\pdfoutput=1
\documentclass[aps,prd,twocolumn,superscriptaddress,longbibliography,letterpaper]{revtex4}
\usepackage{siunitx}
\usepackage{amssymb}
\usepackage{cancel}
\usepackage{color,graphicx}
\usepackage{amsmath}
\usepackage{amsbsy}
\usepackage{amsthm}
\usepackage{bbm}
\usepackage[normalem]{ulem}
\usepackage{epsfig}
\usepackage{lscape}
\usepackage{float}
\usepackage{graphicx}
\usepackage{subfigure}
\usepackage{dcolumn}
\usepackage{bbm}
\usepackage{color,epstopdf}
\usepackage{amscd}
\usepackage{amsfonts}  
\usepackage[]{amsmath}
\usepackage{amssymb}    
\usepackage{mathrsfs}
\usepackage{verbatim}
\usepackage[]{cases}
\usepackage{amsmath}
\usepackage{wasysym}
\usepackage[utf8]{inputenc}
\usepackage[T1]{fontenc}
\usepackage{mathtools}
\usepackage[dvipsnames]{xcolor}
\usepackage{url}

\usepackage[colorlinks=true,linkcolor=blue,urlcolor=blue,citecolor=blue,pdfusetitle]{hyperref}
\usepackage{soul}

\usepackage{braket} %
\usepackage{bbm} %
\usepackage{colonequals} %
\usepackage[scr=boondoxo]{mathalfa} %

\usepackage{qcircuit} %
\usepackage{simplewick} %

\newcommand{\al}{\alpha}
\newcommand{\be}{\beta}
\newcommand{\ga}{\gamma}
\newcommand{\de}{\delta}
\newcommand{\ep}{\epsilon}

\newcommand{\et}{\eta}
\newcommand{\tht}{\theta}  %

\newcommand{\la}{\lambda}
\newcommand{\rh}{\rho}
\newcommand{\ta}{\tau}

\newcommand{\ph}{\phi}
\newcommand{\ch}{\chi}

\newcommand{\om}{\omega}

\newcommand{\Tht}{\Theta}  %
\newcommand{\La}{\Lambda}

\newcommand{\Om}{\Omega}

\newcommand{\vtht}{\vartheta} %

\newcommand{\vph}{\varphi}

\newcommand{\cellcolour}{\cellcolour} %
\newcommand{\eqn}[1]{\begin{equation}#1\end{equation}} %
\newcommand{\eqns}[1]{\begin{equation*}#1\end{equation*}} %
\newcommand{\aln}[1]{\begin{align}#1\end{align}} %
\newcommand{\gthr}[1]{\begin{gather}#1\end{gather}} %
\newcommand{\subeqn}[1]{\begin{subequations}#1\end{subequations}} %

\newcommand{\pmx}[1]{\begin{pmatrix}#1\end{pmatrix}} %
\newcommand{\iu}{\mathrm{i}} %
\newcommand{\ec}{\mathrm{e}} %

\newcommand{\lb}{\left}  %
\newcommand{\rb}{\right}  %
\newcommand{\ct}{\dagger}  %
\newcommand{\infi}{\infty} %
\newcommand{\mc}{\mathcal} %
\newcommand{\nn}{\nonumber} %
\newcommand{\ti}{\widetilde} %
\newcommand{\ce}{\colonequals} %
\newcommand{\bb}{\mathbb}

\newcommand{\id}{\mathbbm{1}}
\newcommand{\tp}{\otimes} %
\newcommand{\bd}{\boldsymbol}

\newcommand{\ms}{\mathscr} %
\newcommand{\qq}{\qquad\qquad} %
\newcommand{\qqqq}{\qquad\qquad\qquad\qquad} %
\newcommand{\tx}{\text} %

\newcommand{\sipl}{\hat{\sigma}^+} %
\newcommand{\simi}{\hat{\sigma}^-} %
\newcommand{\oa}{\hat{a}} %
\newcommand{\oad}{\hat{a}^{\dagger}} %
\newcommand{\Abs}[1]{\left|{#1}\right|} %

\DeclareMathOperator{\Tr}{Tr}

\DeclareMathOperator{\erf}{erf}

\DeclareMathOperator{\erfi}{erfi}

\makeatletter
\newcommand{\vast}{\bBigg@{3}}	
\newcommand{\Vast}{\bBigg@{4}}
\makeatother

\renewcommand{\Re}{\operatorname{Re}}
\renewcommand{\Im}{\operatorname{Im}}

\graphicspath{
  {"./NewFigures/"}
}

\allowdisplaybreaks

\begin{document}

%\title{2+1 Spherical EH with $\delta$-switching (title is a WIP)}
%\title{Detecting UV cutoffs on a sphere using particle detectors}
\title{Locally detecting UV cutoffs on a sphere with particle detectors}

\author{Ahmed Shalabi}
\email[]{ashalabi@uwaterloo.ca}
\affiliation{Department of Physics and Astronomy, University of Waterloo, Waterloo, Ontario, Canada, N2L 3G1}
\author{Laura J. Henderson}
\email[]{l7henderson@uwaterloo.ca}
\affiliation{Centre for Engineered Quantum Systems, School of Mathematics and Physics,
The University of Queensland, St. Lucia, Queensland 4072, Australia}
\affiliation{Department of Physics and Astronomy, University of Waterloo, Waterloo, Ontario, Canada, N2L 3G1}
%\affiliation{Institute for Quantum Computing, University of Waterloo, Waterloo, Ontario, Canada, N2L 3G1}
\author{Robert B. Mann}
\affiliation{Department of Physics and Astronomy, University of Waterloo, Waterloo, Ontario, Canada, N2L 3G1}
\affiliation{Institute for Quantum Computing, University of Waterloo, Waterloo, Ontario, Canada, N2L 3G1}
\affiliation{Perimeter Institute, 31 Caroline St. N. Waterloo Ontario, N2L 2Y5, Canada}

\begin{abstract}
 The potential breakdown of the notion of a metric at high energy scales could imply the existence of a fundamental minimal length scale below which distances cannot be resolved. One approach to realizing this minimum length scale is construct a quantum field theory with a bandlimit on the field.  We report on an investigation of the effects of imposing a bandlimit on a field on a curved and compact spacetime and how best to detect such a bandlimit if it exists. To achieve this operationally, we couple two Gaussian-smeared UDW detectors to a scalar field on a $S^2 \times R$ spherical spacetime through delta-switching. The bandlimit is implemented through a cut-off of the allowable angular momentum modes of the field. We observe that a number of features of single detector response in the spherical case are  similar to those in flat spacetime, including  the  dependence on the geometry of the detector,  and that smaller detectors couple more strongly to the field,  leading to an optimal size for bandlimit detection.  We find that in flat spacetime squeezed detectors are more senstive to the bandlimit provided they are larger than the optimal size; however, in spherical spacetime the bandlimit itself determines if squeezing improves the sensitivity. We also explore setups with two detectors, noting that in the spherical case, due to its compact nature, there is a lack of dissipation of any perturbation to the field, which results in locally excited signals being refocused at the poles. Quite strikingly, this feature can be exploited to significantly improve   bandlimit detection via field mediated signalling.  Moreover, we find that squeezing on a sphere introduces extra anisotropies that could be exploited to amplify or weaken the response of the second detector.
\end{abstract}
\maketitle

%Quite strikingly, squeezing increases the response of the detectors in flat space but decreases the analogous response on the spherical spacetime.

\section{Introduction}
% - General introduction and the need for cutoffs

The theories of General relativity and quantum field theory describe all fundamental interactions in nature. Yet they are based on entirely different mathematical structures and are
 empirically applicable over very different energy and length scales. While semi classical descriptions of quantum field theory on curved spacetime exist, a fundamental step towards their unification into some higher energy theory of quantum gravity will entail understanding what happens at   short distance scales. Fluctuations of quantum fields might potentially break down the notion of a  general relativistic metric upon approaching the Planck scale. As such, it is believed that there is a finite minimum length beneath which distances cannot be resolved \cite{Kempf:1994su}.

% Motivating Bandlimitation and Shannon sampling theorem
There are several consistent high energy theories of quantum gravity, each with its own treatment of spacetime. Generally, there are two overarching approaches to dealing with the nature of spacetime in theories of quantum gravity \cite{Hossenfelder:2012lrr}. One is to model spacetime as some sort of discrete structure. This approach is conducive to quantization and would naturally entail some sort of ultraviolet (UV) cutoff. However it comes at the price of a loss of local  Lorentzian symmetries. %\rbm{Which symmetries?} 
The other broad approach, based on continuous structures, does not suffer from these issues.  However understanding the notion of metric breakdown   remains an open problem \cite{Kempf_2018}. %

In an attempt to reconcile these issues  a hybrid proposal \cite{Kempf_2010_ST} treats spacetime as both continuous and discrete, analogous to the way that Shannon's sampling theorem \cite{6773024} regards information as both continuous and discrete.  More concretely, consider a continuous signal modelled by a  function $f(t)$. Shannon's sampling theorem states that if $f(t)$ is bandlimited i.e. contains frequencies in a finite interval $(-\Lambda,\Lambda)$, then taking a discrete set of samples  $\left\{f\left(t_{n}\right)\right\}_{n=-\infty}^{n=\infty}$ is enough to reconstruct the signal via the Shannon sampling formula for all times, provided the samples are taken at intervals $t_{n+1} - t_n = 1/(2 \Lambda)$. This was generalized to physical fields on Lorentzian manifolds,  establishing how this form of bandlimitation on the momentum modes of a field is equivalent to a UV cutoff \cite{PhysRevLett.100.021304}. It is important to note that, unlike quantizing a field on discrete lattice, bandlimitation of a QFT preserves local Euclidean symmetries.  Furthermore, although   this form of bandlimitation is not Lorentz invariant, it can be generalized to a fully covariant cutoff \cite{Kempf:2013jmp,Chatwin-Davies:2017prl}. 

With all of this established, it is of utmost importance to study UV cutoffs at the intersection between quantum field theory and general relativity --  in other words, imposing a cutoff on a quantum field on a curved background. To this end,
we here  study this question on
 an $S^2 \times R$ spacetime and, for comparison, its  $(2+1)$-dimensional Minkowski counterpart. We do so for several reasons.  First, a quantized scalar field on an $S^2 \times R$  background has well defined angular momentum modes. Moreover, this spacetime is compact and bounded, so a quantized scalar field would have a countably infinite number of modes if no UV cutoff existed. Furthermore,  AdS$_3$ is conformal to $S^2 \times R$,  making our results straightforwardly transferrable to that context. AdS spacetimes have been studied extensively in the context of holographic duality and the AdS/CFT correspondence. In addition, the field correlation functions on AdS$_3$ are related to those in BTZ spacetimes via image sums \cite{Henderson:2017yuv}.

% General introduction to RQI/particle detector models
 The most straightforward way to probe quantum fields locally is through particle detectors. First  proposed by Unruh   \cite{Unruh:1976}, the detector is modelled as a two-level system that (linearly) couples to the field. As such, it
serves as a local probe of the field, providing both a concrete notion of  locality 
and an operational definition of a  particle. In other words, "A particle is what a particle detector detects"~\cite{Scully_2008}.

Particle detectors probe and study the semi classical regime of quantum field theory on curved spacetime. They sample the fluctuations and (if more than one detector is employed) correlations of a quantum field by coupling to its momentum modes. By smearing a particle detector over a region of spacetime, we can probe the quantum field in question via local interactions over that region.  Since the field degrees of freedom and the spatial profile of the detector (which quantifies where the detector couples to the field) enter the model at the same level, an investigation of how they interact can yield a better operational understanding of the finite spatial volume of the discrete degrees of freedom.  Such a study  was recently carried out in $(3+1)$-dimensional flat spacetime \cite{Henderson:2020ucx}.  Here we take the next natural step by considering this problem in $S^2 \times R$, with appropriate comparison to $(2+1)$-dimensional flat spacetime.

There are several models for the field-detector coupling. These
  include non linear scalar field coupling \cite{physrevd.103.125021}, field derivative coupling \cite{Ju_rez_Aubry_2014,Ju_rez_Aubry_2018}, fermionic couplings \cite{PhysRevD.102.093003,Perche_2022} and
  delocalized matter \cite{stritzelberger2019coherent,PhysRevD.103.016007}. However a simple linear coupling \cite{Pozas-Kerstjens:2015gta} is an appropriate approximation to the full light-matter interaction if angular momentum exchange is negligible \cite{Pozas-Kerstjens:2016rsh,PhysRevA.103.013703}. We shall only consider this coupling in our investigation.

% Entanglement harvesting, spacetime geometry/topology, signalling and communication
 Particle detector models have found many applications in the study of quantum information in both flat and curved spacetimes. These include studies of the entanglement structure of quantum fields using the entanglement harvesting protocol \cite{Salton:2014jaa,Pozas-Kerstjens:2015gta}, the Unruh effect \cite{Unruh_no_thermality}, Hawking radiation \cite{henderson2019btz}, probing the geometry \cite{PhysRevD.105.066011,PhysRevD.105.125011} and topology \cite{Smith:2016prd} of spacetime, providing a measurement framework for quantum field theory \cite{PhysRevD.105.065003}, and other applications like communication protocols \cite{PhysRevD.101.125005,Tjoa_2022} and thermodynamics \cite{PhysRevA.102.052219}.
%\rbm{Can mention communication protocols}\ahs{done}

 We consider here the question of how to best detect the presence of a cutoff using particle detectors in both flat and curved spacetimes. We utilize particle detectors  by coupling them to vacuum states of quantum fields in each of $(2+1)$-dimensional Minkowski spacetime and on $S^2 \times R$ as a prototypical $(2+1)$-dimensional curved spacetime. We will implement the UV cutoffs via a hard/conventional bandlimitation on the momentum modes of the scalar field.  We will take the field-detector coupling duration to be the shortest length scale in the problem by modelling it as a $\delta$-function. This delta coupling has several advantages. It removes the need for time ordering and allows a full non-perturbative determination of the final detector-field joint state \cite{Simidzija:2017kty}. After the field is traced out, the final state of the detector carries information about the geometry of the underlying spacetime \cite{PhysRevD.105.125011}.  Furthermore, as discussed earlier, although the conventional band\-limit is not covariant, we expect our results to be similar to a full covariant generalization  since the duration of the coupling  we employ is smaller than any other length scales in the problem.  Finally, we will use two detectors, switching on one before the other to study the impact of field mediated signalling on the detection of the bandlimit.

% Organization of the paper
The rest of the paper is organized as follows. In section~\ref{UDWmod} we present the basic formalism of the UDW model in the context of
$\delta$-switching and bandlimited quantum fields for both the flat and spherical cases we consider, and in section \ref{setup}
we describe the spatial profiles of the detectors.  We then present our results for bandlimit detection using a single detector in
 section~\ref{sdres} and for two detectors in section~\ref{sec:signalling}. We present our conclusions in section~\ref{conc} along with a discussion of directions for further work.  A set of appendices contains technical details pertinent to our investigation.

%%%%%%%%%%%%% Section II
\section{The UDW detector model and Dirac $\delta$ switching}
\label{UDWmod}

The   Unruh-DeWitt (UDW) detector~\cite{Unruh:1976,summers_bells_1987,DeWitt:1980hx} 
%serves as a local probe of the quantum scalar field, providing both  a concrete notion of locality and an operational definition of particle: ``a particle is what a particle detector detects''~\cite{Scully_2008}. 
is a 2-level system whose ground and excited states  are respectively given by $\Ket{g}_D$ and $\Ket{e}_D$,   separated by an energy gap  $\Om_D$. We shall  consider two such detectors A and B linearly coupled to a massless scalar field
 such that the initial joint detector-field state is given by
\begin{equation}
    \hat{\rho}_{i} = \ket{g}_A\prescript{}{A}{\bra{g}} \otimes \ket{g}_B\prescript{}{B}{\bra{g}}\otimes \ket{0}_{\phi}\prescript{}{\phi}{\bra{0}}
\end{equation}
or in other words, the field is in the vacuum state and the detectors are in their ground states.
The interaction detector-field  Hamiltonian is 
\begin{equation}
    \hat H_{I,AB}(t) = \hat H_{I,A}(t) + \hat H_{I,B}(t)
\end{equation}
 in the interaction picture, where $\hat H_{I,D}(t)$ is given by
\aln{
  \hat{H}_{I,D}(t) &= \la_D \ch_D(t)\lb(\ec^{\iu\Om_D t}\sipl_D + \ec^{-\iu\Om_D t}\simi_D\rb) \nn\\
  &\qquad \tp \int d^n\bd{x}\ F_D[\bd{x}-\bd{x}_D]\hat{\ph}(\bd{x},t)
  \label{eq:HIntD}
}
with $D \in \{A,B\}$,
where $\lambda_D$ is the field-detector coupling constant and $\chi_D(t)$ is the switching function that controls the duration of the field-detector interaction. The operators $\sipl_D \ce \ket{e}_D\prescript{}{D}{\bra{g}} $ and $\simi_D \ce \ket{g}_D\prescript{}{D}{\bra{e}}$ are the $\tx{SU}(2)$ ladder operators acting on the Hilbert space of detector $D$.
We have introduced a spacial profile $F_D(\bd{x}-\bd{x}_D)$
for each detector, centred around its position $\bd{x}_D$. We interpret this as describing the size and shape of the detector \cite{PhysRevD.87.064038,McKay:2017pra}.  

The time evolution of the full  detector-field system is
\eqn{
    \hat{U}=\mathcal{T} \exp \left[-\mathrm{i} \int_{-\infty}^{\infty} \mathrm{d} t \hat{H}_{\mathrm{I}, {AB}}(t)\right] \label{TimeEvoleqn}}
where $\mathcal{T}$ is the time ordering operator. The final state of the two detector-field system is given by 
\begin{equation}
    \hat{\rho}_f = \hat{U}\hat{\rho}_{i}\hat{U}^{\dagger}
\end{equation} 
in turn yielding the reduced density matrix describing   the final state of the two detector system 
\aln{
  \hat{\rh}_{AB} &\ce \Tr_\ph\big[\hat{\rh}_{f}\big] %=\sum_{m,n}\Tr_\ph\big[\hat{U}^{(m)}\hat{\rh}_{i}\hat{U}^{(n)\ct}\big].
  \label{eq:rhoABSeries}
}
obtained by tracing out the Hilbert space of the field.

The general approach from here would be to solve for the matrix elements of $\hat{\rho}_{AB}$ perturbatively. However it is possible to  solve the two detector density matrix exactly 
\cite{Simidzija:2017kty,Simidzija:2018ddw}
by  using the switching function \eqn{
  \ch_D(t)=\eta_D\de(t-T_D)
}
where $T_D$ is the time at which the interaction takes place.  We shall briefly review this 
`$\delta$-switching' formalism without imposing the original constraint \cite{Simidzija:2017kty} of working in Minkowski space.

We will assume without loss of generality that detector A switches before B ($T_A \leq T_B$). Applying the $\delta$-switching allows us to write the time evolution operator (\ref{TimeEvoleqn}) as
\eqn{
  \hat{U}_{\de} = \exp\big(\hat H_{I,B}(T_B)\big) \exp\big(\hat H_{I,A}(T_A)\big)
}
or alternatively as 
\eqn{
 \hat{U}_{\de} = \exp\big(\hat{\mu}_B(T_B) \otimes \hat{\mc{Y}}_B\big) \exp \big(\hat{\mu}_A(T_A)   \otimes \hat{\mc{Y}}_A\big)
}
where the operator
\eqn{
  \hat{\mu}_D(t) = \ec^{\iu\Om_D\ta_D(t)}\sipl_D+\ec^{-\iu\Om_D\ta_D(t)}\simi_D
}
describes the evolution of the detector and
\eqn{
  \hat{\mc{Y}}_D \ce -\iu\la_D \eta _D\int d^n\bd{x} F_D(\bd{x}-\bd{x}_D)\hat{\ph}(\bd{x},T_D)
  \label{eq:GenY}
}
which is the smeared field operator.

By expanding the Taylor series of the exponential and noting that $\hat{\mu}_D(t)^2 = \id$, we can write the time evolution operator $\hat{U}_{\delta}$ as
\aln{
  \hat{U}_{\de} &= \lb(\id_A\tp\id_B\tp\cosh(\hat{\mc{Y}}_B) + \id_A\tp\hat{\mu}_B(T_B)\tp\sinh(\hat{\mc{Y}}_B)\rb) \nn\\
  &\times \lb(\id_A \tp \id _B \tp \cosh(\hat{\mc{Y}}_A) + \hat{\mu}_A(T_A)\tp\id_B\tp\sinh(\hat{\mc{Y}}_A)\rb)
  \label{eq:TimeEvOpdelta}
}

Moreover, we can rewrite $\hat{U}_{\delta}$ using the complex exponential form of the hyperbolic trigonometric functions utilizing the following definition. Let $j,k \in \{1,-1\}$ and write 
\eqn{
    \hat{X}_{(j,k)} = \frac{1}{4}(e^{\hat{\mc{Y}}_B} + j e^{-\hat{\mc{Y}}_B})(e^{\hat{\mc{Y}}_A} + k e^{-\hat{\mc{Y}}_A})\label{xij}}
which gives
\aln{
\hat{U}_{\delta} &=
\id_{A} \otimes \id_{B} \otimes \hat{X}_{(1,1)}
+ \hat{\mu}_{A}\left(T_{A}\right) \otimes \hat{\mu}_{B}\left(T_{B}\right) \otimes \hat{X}_{(-1,-1)}
  \nn \\
& + \hat{\mu}_{A}\left(T_{A}\right) \otimes \id_{B} \otimes \hat{X}_{(1,-1)}  +\id_{A} \otimes \hat{\mu}_{B}\left(T_{B}\right) \otimes \hat{X}_{(-1,1)} 
}
from \eqref{eq:TimeEvOpdelta}.
The two detector subsystem evolves to the final state 
\aln{
\hat{\rho}_{AB} = \Tr_{\phi}[\hat{U}_{\delta} \hat{\rho}_i \hat{U}_{\delta}^{\dagger}]
}
which in the $\Ket{a}_A \otimes \Ket{b}_B$ basis for $a,b \in \{g,e\}$ contains cross terms of the form $\bra{0}\hat{X}_{(j,k)}\hat{X}_{(l,m)} \ket{0}$. This motivates the following definition:
\aln{
\label{f_jklm}
    f_{jklm} &\coloneqq \left\langle 0\left|\hat{X}_{(j,k)}^{\dagger} \hat{X}_{(l,m)}\right| 0\right\rangle}
where $j,k,l,m \in \{1,-1\}$. Using the Baker-Campbell-Hausdorff (BCH) formula $e^Xe^Y = e^Z$ with $Z$ given by
\eqn{
Z=X+Y+\frac{1}{2}[X, Y]+\frac{1}{12}\big[X,[X, Y]\big]-\frac{1}{12}\big[Y,[X, Y]\big]+\cdots}
and using arguments similar to those in the Minkowski space case \cite{Simidzija:2017kty}, we   rewrite the $f_{jklm}$ matrix elements as 
\aln{
 f_{jklm} &=\frac{1}{16} \left[ (1+j \ell+k m+j k \ell m)+[(1+j \ell)(k+m)] f_{A} \right. \nn \\
 &+\left[(\ell+j k m) e^{2 i \Theta}+(j+k \ell m) e^{-2 i \Theta}\right] f_{B} \nn \\
+ & \left. \left[(j k+\ell m) e^{\omega}+(j m+k \ell) e^{-\omega}\right] f_{A} f_{B}\right] 
}
where  
\eqn{
f_D = \Bra{0}\exp(2\hat{\mc{Y}}_D)\Ket{0}
\label{eq:Genf_D}
}
and the quantities $\Theta$ and $\omega$ are defined as:
\begin{equation}
\begin{aligned}
\Theta &:= -\mathrm{i}\left\langle 0\left|\left[\mc{Y}_{A}, \hat{\mc{Y}}_{B}\right]\right| 0\right\rangle 
= 
\left\langle 0\left|\hat{\Theta} 
\right| 0\right\rangle 
\\
\omega &:= 2\left\langle 0\left|\left\{\hat{\mc{Y}}_{A}, \hat{\mc{Y}}_{B}\right\}\right| 0\right\rangle
= 
\left\langle 0\left|\hat{\omega} 
\right| 0\right\rangle 
\label{eq:GenThetaOmega}
\end{aligned}
\end{equation}
which are respectively the vacuum expectation values of the smeared field commutator and anti-commutator. For two regions in a spacetime, these operators encode the signalling through the field and the correlations in the field between the two regions. 

The smeared field commutator $\Theta$ is non zero when two smeared detectors are in casual contact, in which case communication between them is possible. It is zero when the detectors are spacelike separated.  It is important to note that for two detectors that are initially separable there is no dependence of the final state on $\omega$ after the interaction. This  is to be expected since $\omega$ encodes the amount of correlations between two detectors --  if they are uncorrelated initially then they cannot harvest entanglement from the field via delta coupling \cite{Simidzija:2017kty,Yoshimura_2021}. The role played by the anti-commutator
$\omega$ in detector correlations with delta coupling was recently studied \cite{Yoshimura_2021}.  We will analyze in section~\ref{sdresflat} the relevance of $\Theta$ in  signalling.
%\rbm{We need a consistent notation for the hats on $\Theta$ and $\omega$. Is what I have done above OK?} \ahs{Yes, I personally like this since the vacuum expectation becomes a number anyway. Fixed}

The final reduced density matrix $\hat{\rho}_{AB}$ can then be written in terms of the $f_{jk\ell m}$ matrix elements as
\eqn{
  \hat{\rh}_{AB} = \pmx{
    \rh_{11} & 0 & 0 & \rh_{14} \\
    0 & \rh_{22} & \rh_{23} & 0 \\
    0 & \rh_{23}^* & \rh_{33} & 0 \\
    \rh_{14}^* & 0 & 0 & \rh_{44}
  }
  \label{eq:GeneralrhoAB}
}
where the non zero $\rho_{ij}$ matrix elements are given by
\subeqn{
  \aln{
    \rh_{11} &= \frac{1}{4} \Big(1+f_A+f_B\cos(2\Theta)+f_Af_B\cosh(\om)\Big) \\
    \rh_{14} &= \frac{1}{4} \ec^{-\iu(\Om_AT_A+\Om_BT_B)} f_B \Big(\iu\sin(2\Theta)+f_A\sinh(\om)\Big) \\
    \rh_{22} &= \frac{1}{4} \Big(1+f_A-f_B\cos(2\Theta)-f_Af_B\cosh(\om)\Big) \\
    \rh_{23} &= -\frac{1}{4} \ec^{-\iu(\Om_AT_A-\Om_BT_B)} f_B\Big(\iu \sin(2\Theta)+f_A\sinh(\om)\Big) \\
    \rh_{33} &= \frac{1}{4} \Big(1-f_A+f_B\cos(2\Theta) - f_Af_B\cosh(\om)\Big) \\
    \rh_{44} &= \frac{1}{4} \Big(1-f_A-f_B\cos(2\Theta)+f_Af_B\cosh(\om)\Big)
  }
}
We can also trace out the detectors individually to obtain the following density operators
\eqn{
\hat{\rho}_A = \Tr_B[\hat{\rho}_{AB}] = \frac{1}{2}
    \begin{pmatrix}
    1 + f_A & 0 \\
    0   & 1 - f_A
    \end{pmatrix} \label{rho_A}}
and
\eqn{
\hat{\rho}_B = \Tr_A[\hat{\rho}_{AB}] = \frac{1}{2}
    \begin{pmatrix}
    1 + f_B\cos(2\Theta) & 0 \\
    0   & 1 - f_B\cos(2\Theta)
    \end{pmatrix} \label{rho_B}}
    
Note that the dynamics of detector B is modified by the commutator of the field. This is a consequence of the fact that detector B interacts with an evolved state of the field subsequent to its interaction with the first detector. Finally, we can read off the transition probabilities for the first and second detector to be
\eqn{
    P_A = \frac{1}{2}(1 - f_A) \quad  \text{and} \quad   P_B = \frac{1}{2}\big(1 - f_B\cos(2\Theta)\big)\label{PD}}
    
The dynamics and the response of $\delta$-coupled detectors can be extended to those   on any curved spacetime by quantizing the scalar field on the background spacetime. The Klein-Gordon equation in curved spacetime is
\eqn{\left(\frac{1}{\sqrt{|g|}} \frac{\partial}{\partial x^{\mu}} g^{\mu \nu} \sqrt{|g|} \frac{\partial}{\partial x^{\nu}}\right) \hat{\phi}(x, t)=0}
We can solve the Klein-Gordon equation by assuming the following  ansatz for the scalar field operator
\eqn{
\hat{\phi}(x,t) = \sum_k 
\left[u_k(x,t)\hat{a}_k + u_k(x,t)^*\hat{a}_k^{\ct}\right]  }
where the functions $u_k(x,t)$ are solutions to the Klein-Gordon equation, and $\hat{a}_k$, $\hat{a}_k^{\ct}$ are the raising and lowering operators of the scalar field. If the spacetime is globally hyperbolic, then a set of modes $u_k(x,t)$ exists. If we can quantize the scalar  field  on some background spacetime, the task would be to derive the smeared field operator $\hat{\mc{Y}}_D$ and from it the expressions for $f_D$ and $\Theta$ that define the response of the two detectors to the coupling. Moreover, we can particularize those expressions to the shape and localization of the detectors on the background geometry.
%\rbm{Not clear how the next paragraph relates to non-uniqueness of the modes.}\ahs{You are right it does not, trimmed it down}
\medskip

\subsection{Flat spacetime}
Here we briefly summarize the recent $(3+1)$-dimensional  flat space analysis \cite{Henderson:2020ucx} in a $(2+1)$-dimensional context for ease of comparison with the spherical case.  We decompose the scalar field into plane-wave modes as
\eqn{
  \hat{\ph}(\bd{x},t) = \frac{1}{(2\pi)^{n/2}} \int \frac{d^2\bd{k}}{\sqrt{2\Abs{\bd{k}}}}\ \lb[\ec^{\iu(\Abs{\bd{k}}t-\bd{k}\cdot\bd{x})} \oad_{\bd{k}} +\tx{H.c.}\rb]
  \label{eq:FlatSpaceField}
}
where $\hat{a}_{\bd{k}}^\dagger$, $\hat{a}_{\bd{k}}$ are  creation and annihilation operators that obey the canonical commutation relations
\eqn{
    \left[\hat{a}_{\bd{k}},\hat{a}^\dagger_{\bd{k}'}\right] = \delta^{(2)}(\bd{k}-\bd{k}').
}

After writing down the field, we can now calculate the density matrix describing the joint state of the two detector system \eqref{eq:GeneralrhoAB}.  In the case of $(2+1)$-dimensional Minkowski space, the smeared field operator is 
\eqn{
  \hat{\mathcal{Y}}_D = -\iu\lambda_D\eta_D \int \frac{d^2\bd{k}}{\sqrt{2|\bd{k}|}} \Big(\ti{F}_D(\bd{k}) \ec^{-\iu(|\bd{k}|T_D-\bd{k}\cdot{x}_D)}\oa_{\bd{k}} + \tx{H.c.}\Big)
  \label{eq:FlatY}
}
where
\eqn{
  \ti{F}_D(\bd{k}) = \frac{1}{2\pi} \int d^2\bd{x}\ F_D(\bd{x}) \ec^{\iu\bd{k}\cdot\bd{x}}
}
is the Fourier transform of the spacial profile.

It is straightforward to calculate the  matrix element functions \eqref{eq:Genf_D} and \eqref{eq:GenThetaOmega} as \cite{Simidzija:2017kty}
\begin{widetext}
\aln{
    f_D &= \exp\lb(-\la_D^2\et_D^2\int_{\Abs{\bd{k}}<\La} \frac{d^2\bd{k}}{\Abs{\bd{k}}}\Abs{\ti{F}_D(\bd{k})}^2\rb) \nn\\
    \Tht &= -\frac{\iu\la_A\la_B\et_A\et_B}{2} \int_{\Abs{\bd{k}}<\La} \frac{d^2\bd{k}}{\Abs{\bd{k}}}\Big(\ti{F}_A^*(\bd{k})\ti{F}_B(\bd{k})\ec^{-\iu\Abs{\bd{k}}(T_B-T_A)}\ec^{\iu\bd{k}\cdot(\bd{x}_B-\bd{x}_A)} - \tx{H.c.}\Big) \nn\\
    \om &= \la_A\la_B\et_A\et_B \int_{\Abs{\bd{k}}<\La} \frac{d^2\bd{k}}{\Abs{\bd{k}}}\Big(\ti{F}_A^*(\bd{k})\ti{F}_B(\bd{k})\ec^{-\iu\Abs{\bd{k}}(T_B-T_A)}\ec^{\iu\bd{k}\cdot(\bd{x}_B-\bd{x}_A)} + \tx{H.c.}\Big)
    \label{eq:MatrixFunctionsFlat}
}
\end{widetext}

Expanding the field in the plane-wave modes of  Eq.\ \eqref{eq:FlatSpaceField}, we are able to easily introduce a hard momentum cutoff by removing modes where $\Abs{\bd{k}}>\La$ \cite{Henderson:2020ucx}.  We note that this cutoff is not Lorentz invariant, but expect that results will be similar to the case of the full covariant cutoff, since the switching time of the detector is shorter than any other scale in the problem \cite{Kempf:2013jmp,Chatwin-Davies:2017prl}.

%%%%%%%%%%%%%%%%%%% 2.2 %%%%%%%%%%%%%%%%%%%%
\subsection{Spherical Spacetime}

To generalize $\delta$-switching to the spherical spacetime we will need to quantize the scalar field $\hat{\phi}$ and derive from it the smeared field operator $\hat{\mc{Y}}_D$ for each detector. We shall then obtain expressions for $f_D$ that define the transition probability $P_D$ for each detector. We shall also need an expression for the commutator $\Theta$ of the smeared field operators to obtain the transition probability of the second detector after it has interacted with the evolved state of the field.

Quantizing a conformally coupled scalar field on $\mathbb{R} \times \mathbb{S}^2$ was done in \cite{Lifschytz_1994}. From the metric
\eqn{ds^2 = -dt^2 +d\tht^2 + \sin^2(\theta)d\varphi^2}
where $- \infty < t < \infty$ , $0 < \theta < \pi$ and $0 < \varphi < 2\pi$, the Klein-Gordon equation then becomes
\eqn{
\square \psi-\frac{1}{8} R \psi=0}
whose solutions are given by
\eqn{\label{psilm}
\psi_{\ell m} = \frac{1}{\sqrt{2\ell+1}}
e^{-i(\ell + \frac{1}{2}) t}Y_{\ell m}(\theta,\varphi)}
where \eqn{Y_{\ell m}(\theta, \varphi)=(-1)^{m} N_{\ell m} P_{\ell}^{m}(\cos \theta) e^{i m \varphi}
\label{eq:SphHarm}}
are the spherical harmonics basis functions, \ $P_{\ell}^m$ are Associated Legendre Polynomials and 
\eqn{
N_{\ell m} \equiv \sqrt{\frac{(2 \ell+1)}{4 \pi} \frac{(\ell-m) !}{(\ell+m) !}}
}  
%\rbm{We should explicitly write these down somewhere. And we have used $\omega$ above to denote the anti-commutator.  I think we should use a different notation for it -- perhaps $\varpi$?}\ahs{Probably could just change every occurrence of $\omega$ to $\ell + \frac{1}{2}$ - Done}
This allows us to expand the scalar field $\hat{\phi}$ in the modes $\psi_{\ell m}$ as follows
\begin{equation}
    \hat{\phi} = \sum_{\ell,m}\psi_{\ell m} \hat{a}_{\ell m} + \psi_{\ell m}^* \hat{a}_{\ell m}^{\dagger}
\end{equation} 
where $ \hat{a}_{\ell m}$ and $\hat{a}^{\dagger}_{\ell m}$ are creation and annihilation operators such that $\hat{a}^{\dagger}_{\ell m} \ket{0} = \ket{\ell,m}$, and where
\eqn{
\label{SphericalCommRelations}
[\hat{a}^{\dagger}_{ij},\hat{a}_{\ell m}] = \delta_{ i \ell }\delta_{j m}}

That is, $ \hat{a}_{\ell m}$ and $\hat{a}^{\dagger}_{\ell m}$ raise and lower the angular momentum of the scalar field. We can expand any function on  $\mathbb{S}^2$ in terms of spherical harmonics as:
\begin{equation}\label{fexp}
h( \tht, \varphi)=\sum_{l, m} h_{l m} Y_{l m}(\tht, \varphi)
\end{equation}
and using   
the orthogonality condition  
    \begin{equation}
\int \int \mathrm{d} \tht \mathrm{d} \varphi\  \sin(\tht)Y_{\ell m}(\tht, \varphi)Y_{pq}^*(\tht, \varphi) =\delta_{mq} \delta_{pl}\end{equation}
we can compute  the coefficients  
\begin{equation}
h_{l m} = \int \int \mathrm{d} \tht \mathrm{d} \varphi \sin (\tht) h(\tht, \varphi) Y_{\ell m}^{*}(\tht, \varphi)\; .
\end{equation}
in \eqref{fexp}.

Expanding the smeared field operator in terms of the scalar field modes we get
\begin{equation}\hat{\mc{Y}}_D =-\iu\lambda_D\eta_{D} \int \mathrm{d} \Omega F_{D}\left(\bd{\theta}-\bd{\theta}_{D}\right) \left(\sum_{\ell, m} \psi_{\ell m} \hat{a}_{\ell m}+\psi_{\ell m}\hat{a}_{\ell m}^{\dagger}\right),
\end{equation}
where $\int \mathrm{d} \Omega \coloneqq \int \int \mathrm{d} \tht \mathrm{d} \varphi \sin(\theta)$ for a profile $F(\bd{\tht})$ centered at $\bd{\tht}_D=(\tht_D,\vph_D)$. Expressing the spacial localization function  in the spherical harmonics basis, with $F_D = \sum_{p,q} f_{p q} Y_{p q}$
we obtain 
\begin{equation}\hat{\mc{Y}}_D =-\iu\lambda_D\eta_{D}\int \mathrm{d} \Omega \sum_{p,q} f_{pq} Y_{pq}(\theta, \varphi) \sum_{l, m}\left( \psi_{\ell m} \hat{a}_{\ell m}+\psi_{\ell m} \hat{a}_{l m}^{\dagger}\right)
\end{equation}
 for the smeared field operator.  
Writing $\hat{\mc{Y}}_D \coloneqq \hat{\ms{y}} + \hat{\ms{y}}^*$
we have
\begin{align}
    \hat{\ms{y}}^* &=-\iu\lambda_D\eta_{D} \int \mathrm{d} \Omega \sum_{p,q} f_{pq} Y_{pq}(\theta, \varphi)\sum_{l, m}\psi_{\ell m} \hat{a}_{l m}^{\dagger} \nn\\
   &=-\iu\lambda_D\eta_{D} \sum_{l, m} \sum_{p,q} \frac{e^{i(\ell + \frac{1}{2}) T_D}f_{pq}}{\sqrt{(2\ell+1)}} \int \mathrm{d} \Omega   Y_{pq}(\theta, \varphi)Y_{l m}^{*} \hat{a}_{l m}^{\dagger} \nn\\
   &=-\iu\lambda_D\eta_{D} \sum_{l, m}\frac{1}{\sqrt{(2\ell+1)}}e^{i(\ell + \frac{1}{2}) T_D}f_{\ell m}\hat{a}_{l m}^{\dagger}
\end{align}
using \eqref{psilm} and the orthogonality condition of the spherical harmonics. 

For the $\hat{\ms{y}}$ term a similar argument gives
\begin{equation}
    \hat{\ms{y}} = +\iu\lambda_D\eta_{D} \sum_{l, m}\frac{1}{\sqrt{2\ell+1}}e^{-i(\ell + \frac{1}{2}) T_D}f_{\ell m}^*\hat{a}_{l m}
\end{equation}{}
so that 
\begin{align}
    \hat{\mc{Y}}_D &= \sum_{l, m} \frac{-\iu\lambda_D\eta_{D}}{\sqrt{(2\ell+1)}} \left(  e^{\iu(\ell + \frac{1}{2}) t_D}f_{\ell m}\hat{a}_{l m}^{\dagger} +  e^{-\iu(\ell + \frac{1}{2}) t_D}f_{\ell m}^*\hat{a}_{l m}\right)
    \nn\\
    &=\sum_{\ell,m} \left(\alpha_{\ell m} \hat{a}_{\ell m}^{\dagger}-\alpha_{\ell m}^{*} \hat{a}_{\ell m}\right)
\label{YD}    
\end{align} 
with $\alpha_{\ell m}$ defined as
\begin{equation}
    \alpha_{\ell m} \ce -\frac{\iu\lambda_D\eta_{D}}{\sqrt{2\ell+1}}e^{i(\ell + \frac{1}{2}) T_D}f_{\ell m}.
\end{equation}

Analogous to the case in Minkowski space \cite{Simidzija:2017kty}, the exponential of the smeared field operator acts on the vacuum state, resulting in a state that corresponds to the phase space displacement of the vacuum state 
\eqn{
    \hat{\mc{D}}_{\ell m}\Ket{0} = \ec^{\hat{\mc{Y}}_D}\Ket{0} = \Ket{\al_{\ell m}}
}
due to the interaction with the detector.
The displaced state is indeed a coherent state, since it is an eigenstate of the annihilation operator
\eqn{
    \oa_{ij}\Ket{\al_{\ell m}} = \al_{ij}\Ket{\al_{\ell m}}
}
 as shown in appendix A. To compute $f_D = \bra{0}\exp^{2\hat {\mc{Y}}_D}\ket{0}$  we begin by computing  the Taylor series for the exponential of the smeared field operator. The computation follows as a special case of the proof presented in appendix A of \cite{Simidzija:2017kty} due to the linearity of sums. 
\begin{widetext}
\aln{
    f_D &= \bra{0}\exp{\sum_{\ell,m} 2(\alpha_{\ell m} \hat{a}_{\ell m}^{\dagger} - \alpha_{\ell m}^* \hat{a}_{\ell m})}\ket{0} \nn\\
    &= \Braket{0|0} + \sum_{\ell,m} 2\bra{0}\cancel{\alpha_{\ell m}\hat{a}_{\ell m}^{\dagger}} - \cancel{\alpha_{\ell m}^*\hat{a}_{\ell m}}\ket{0} + \frac{2}{2!}\sum_{\ell,m}\sum_{\ell',m'}\bra{0}(\cancel{\alpha_{\ell m}\hat{a}_{\ell m}^{\dagger}} - \alpha_{\ell m}^*\hat{a}_{\ell m})(\alpha_{\ell' m'}\hat{a}_{\ell' m'}^{\dagger} - \cancel{\alpha_{\ell' m'}^*\hat{a}_{\ell' m'})}\ket{0}  + .... \nn\\
    &= 1 + 0 -\sum_{\ell,m}\sum_{\ell',m'} \braket{\ell',m'|\ell,m}\alpha_{\ell'm'}\alpha_{\ell m}^* + ...}

The orthogonality relation implies that $\Braket{\ell,m|\ell',m'} = \delta_{\ell \ell'}\delta_{mm'}$. Moreover,
each term in the sum containing an odd number of creation and annihilation operators vanishes by Wick's theorem, i.e
\begin{equation}\left\langle 0\left|\left(\alpha_{j} \hat{a}_{j}^{\dagger}-\alpha_{j}^{*} \hat{a}_{j}\right)^{(2n+1)}\right| 0\right\rangle=0\end{equation}
in general. The even terms   recombine \cite{Simidzija:2017kty}, yielding 
%\rbm{Is the expression below correct, given the factor of $1/2$ above?}\ahs{fixed}
\eqn{
    f_D = \exp\left(-\sum_{\ell = 0}^{\infty}\sum_{m= -\ell}^{\ell} |\alpha_{\ell m}^D|^2\right)
    \label{f_D}}
Finally, using \eqref{YD} and the commutation relations \eqref{SphericalCommRelations},  we find
%\rbm{Do we need all the steps below?}\ahs{Good point, removed intermediate step and referenced the commutation relations}
\aln{
  \Theta& =-\iu\lb[\hat{\mc{Y}}_A,\hat{\mc{Y}}_B\rb] 
  = \iu\la_A\la_B\et_A\et_B \Bigg[ \sum_{\ell,m}\frac{1}{\sqrt{2\ell+1}} \sum_{i,j} \frac{1}{\sqrt{2i+1}} {f_{\ell,m}^A}^* f_{i,j}^B \ec^{-\iu(\ell+1/2)T_A} \ec^{\iu(i+1/2)T_B} \lb(\oa_{\ell,m}\oad_{i,j} - \oad_{i,j}\oa_{\ell,m}\rb) \nn\\
  &\qqqq \qquad\qquad\qquad\qquad  + f_{\ell,m}^A {f_{i,j}^B}^* \ec^{\iu(\ell+1/2)T_A} \ec^{-\iu(i+1/2)T_B} \lb(\oad_{\ell,m}\oa_{i,j} - \oa_{i,j}\oad_{\ell,m}\rb) \Bigg] \nn\\
  &= \iu \la_A\la_B\et_A\et_B \sum_{\ell,m}\frac{1}{2\ell+1} \lb[ {f_{\ell,m}^{A*}}  f_{\ell,m}^B \ec^{\iu(\ell+1/2)(T_B-T_A)} - f_{\ell,m}^A {f_{\ell,m}^{B^*}} \ec^{-\iu(\ell+1/2)(T_B-T_A)}\rb].
  \label{eq:SphereCommNoLocation}}
\end{widetext}

\begin{figure*}
    \includegraphics[width=0.48\linewidth]{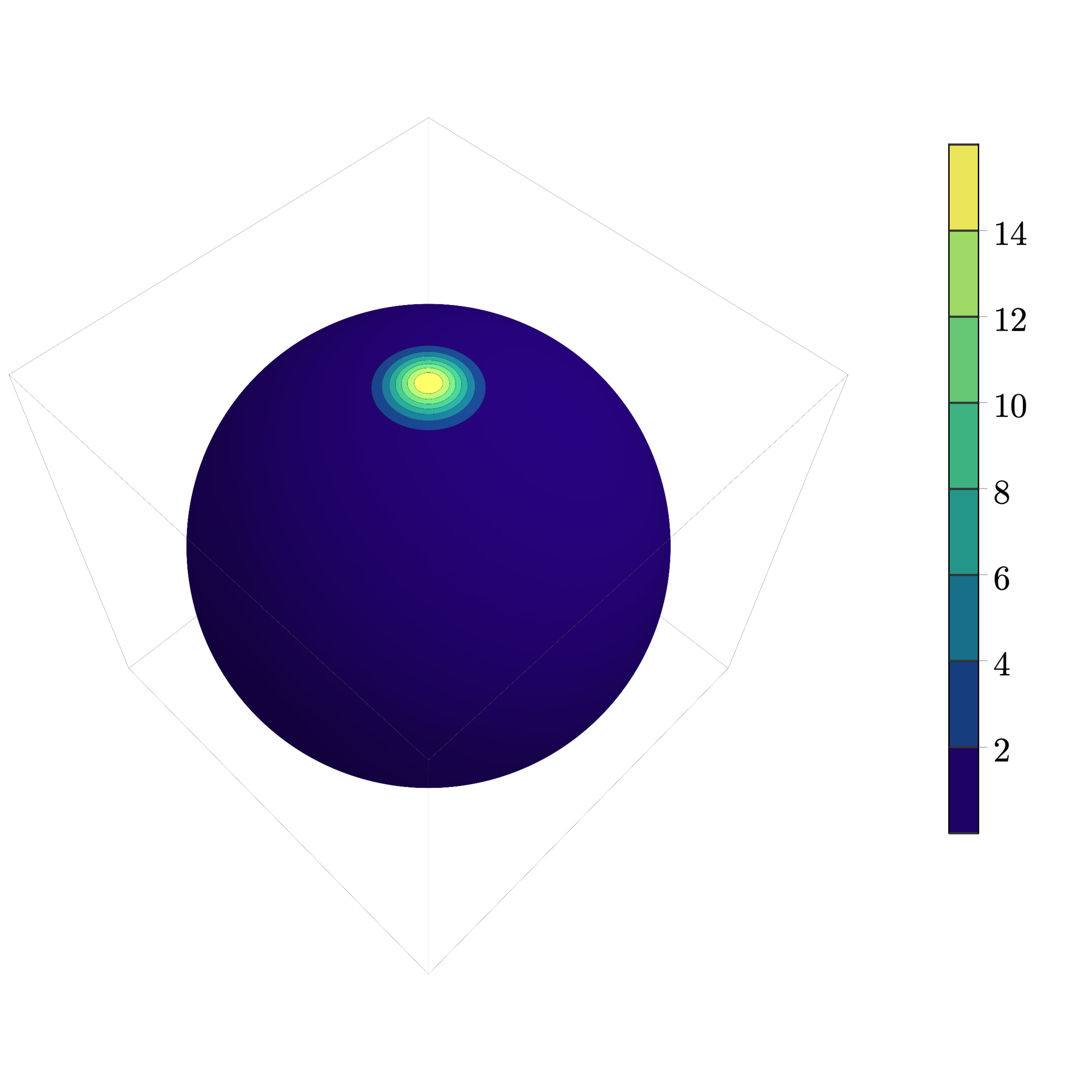}
    \hspace{0.5cm}
    \includegraphics[width=0.48\linewidth]{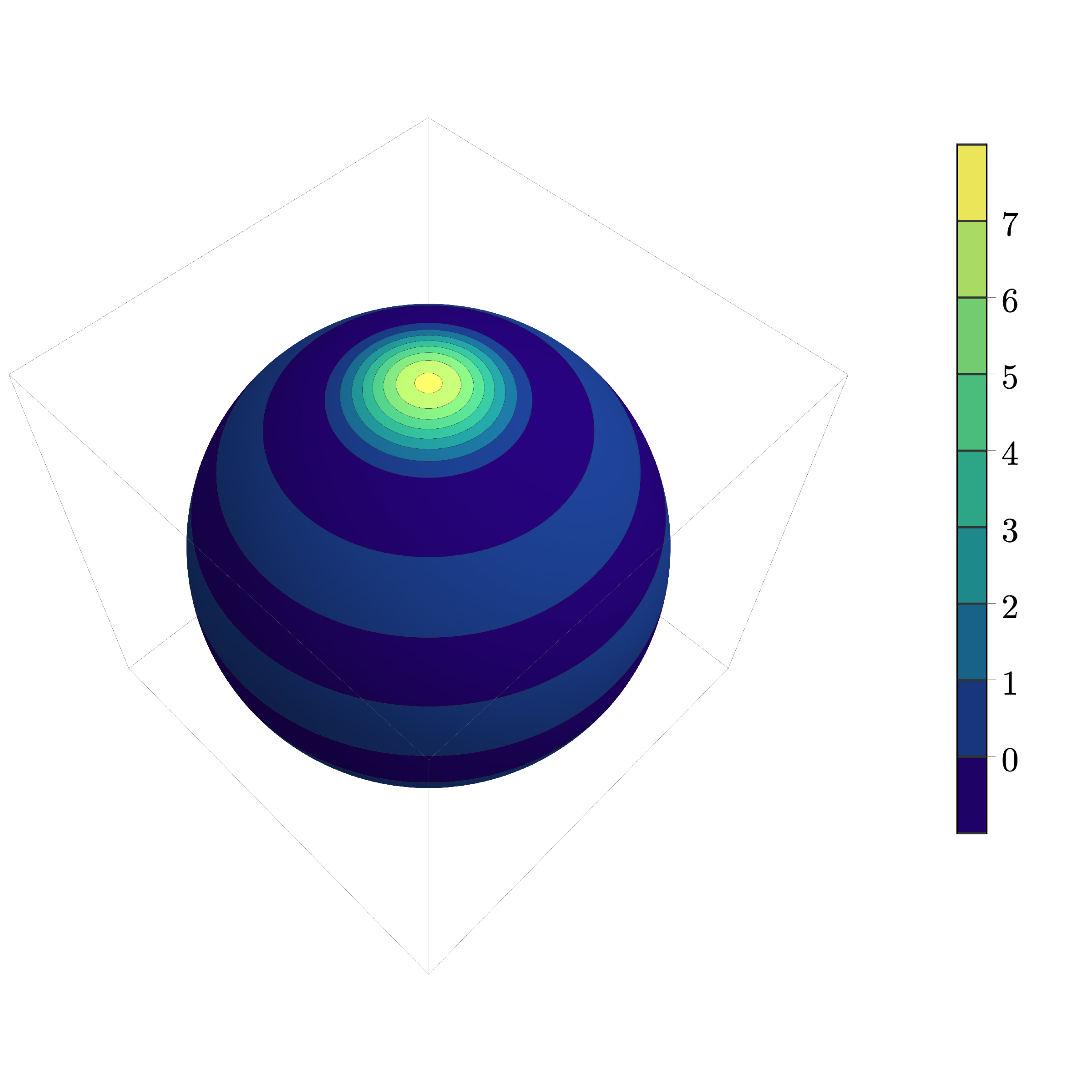}
    \caption{%
        \ \textit{Left}: The spherical analogue of a Gaussian distribution, the Fisher-Bingham Five-Parameter (FB5) distribution, with a size of $\kappa=100$  and zero squeezing $(\beta=0)$ [Eq.~\eqref{FBSymmetric}]. \textit{Right}: The same FB5 distribution with $\kappa=100$ and $\beta=0$, approximately reconstructed in the sphereical harmonic basis [Eq.~\eqref{fexp}] with a cutoff of $\ell_{\text{Max}}=10$.  When the cutoff is not present, the distribution is highly localized at the north pole, but when the cutoff is present, the distribution has support  over the entire sphere.
    }
    \label{fig:ProfileRecon}%
\end{figure*}

Similar to the Minkowski spacetime bandlimit, the hard cutoff is  implemented by removing all modes with $\ell > \ell_{max}$. This is a straightforward extension of the flat space cutoff to $S^2 \times R$, since it is a cutoff of the conjugate momentum degrees of freedom to $\theta$ and $\varphi$. The same argument as to why we would not expect artifacts due to the non covariant nature of the cutoff apply here as well. Finally, as was discussed in \cite{Henderson:2020ucx}, the impact of the bandlimit can also viewed as a non local profile in the absence of a bandlimit since  if we were to expand the profile in spherical harmonics
\aln{
    \sum_{\ell = 0}^{\ell_{max}} \sum_{m=-\ell}^{m = \ell} f_{\ell m}Y_{\ell m} &= \sum_{\ell = 0}^{\infty}\sum_{m=-\ell}^{m = \ell} h_{\ell}{f}_{\ell m}Y_{\ell m} \\ \nn 
    &= \sum_{\ell = 0}^{\infty} \sum_{m=-\ell}^{m = \ell} \Tilde{f}_{\ell m}Y_{\ell m}
}    
where $h_{\ell}$ is 1 for $\ell \leq \ell_{max}$ and 0 otherwise.  The $\Tilde{f}_{\ell m}$ coeeficients correspond to the equivalent non local profile in the absence of a bandlimit. Figure \ref{fig:ProfileRecon}, shows that the bandlimit leads to a detector profile that is highly nonlocal.
%\ahs{Still need to explain why this is non local} 

%%%%%% Section III
\section{Setup}
\label{setup}

Throughout  the rest of this paper, we will take the two UDW detectors to couple to the field with the same interaction strength $\la_A=\la_B=\la$ and to have  identical switching functions $\et_A=\et_B=\et$.  We will also take the spacial profile of the detectors to be two-dimensional normal or Gaussian distribution profiles, henceforth referred to as Gaussian detectors.  We choose Gaussian detectors to maintain consistency with many previous studies in Relativistic Quantum Information; our analysis can be extended to other types of spacial profile straightforwardly.

\subsection{Flat Spacetime}

In $(2+1)$-dimensional Minkowski spacetime, 
a Gaussian detector has the spacial profile
\eqn{
    F(x,y) = \frac{1}{2\pi ab}\exp\lb(-\frac{x^2}{2a^2}\rb)\exp\lb(-\frac{y^2}{2b^2}\rb)
} 
where $a$ describes the width in the $x$-direction and $b$ describes the width in the $y$-direction.  We note that decreasing the values of $a$ and $b$ decreases the size of the detector.  It is convenient to re-write the profile in polar coordinates as
\eqn{
    F(r,\vartheta) = \frac{\sqrt{1-\ep^2}}{2\pi b^2}\exp\lb(-\frac{r^2}{2b^2}\big(1-\ep^2\cos^2(\vtht)\big)\rb)
    \label{eq:FDflat}
}
where
\eqn{
    \epsilon\ce\sqrt{1-\frac{b^2}{a^2}}\in[0,1)
    \label{eq:epsilon}
}
describes the eccentricity of the Gaussian, with higher values of $\ep$ denoting larger squeezing.  For completeness,  we  write  down the Fourier transform of the spacial profile:
\eqn{
    \ti{F}(k,\theta) = \frac{1}{2\pi}\exp\lb(-\frac{a^2\Abs{\bd{k}}^2}{2}\big(1-\ep^2\sin^2(\tht)\big)\rb).
    \label{eq:FTDflat}
}
%\rbm{What are $\vartheta$ and $\theta$ in the above?}

By introducing a non-zero eccentricity to the spacial profile, we will be able to rotate the semi-major axes of the detectors away from the $x$-axis of the chosen coordinate system by some angle, $\varphi_D$.  Since the rotation operator commutes with the Fourier transform, the rotation can be easily implemented by taking $\theta\to\theta-\varphi_D$ in equation \eqref{eq:FTDflat}.

With the spacial profile of the two   detectors given, we are able to write down the expressions for the  basic quantities that compose the matrix element functions:
\begin{widetext}
\aln{
    f_D &= \exp\lb(-\frac{\la_D^2\et_D^2}{8\pi^{3/2}a_D}\int_{0}^{2\pi}d\tht\ \frac{\erf\lb(\La a_D\sqrt{1-\ep_D^2\sin^2(\tht)}\rb)}{\sqrt{1-\ep_D^2\sin^2(\tht)}}\rb) \label{eq:fDflat}\\
    \Theta &= \frac{\la_A\la_B\et_A\et_B}{4\sqrt{2}\pi^{3/2}}\int_{0}^{2\pi}d\tht\ \frac{\exp\lb(-\frac{[T-S\cos(\tht)]^2}{2\mathcal{A}(\tht)}\rb)}{\sqrt{\mathcal{A}(\tht)}} \Bigg\{\Re\lb[\erfi\lb(\frac{T-S\cos(\tht)}{\sqrt{2\mathcal{A}(\tht)}}+\frac{\iu\La \sqrt{\mathcal{A}(\tht)}}{\sqrt{2}}\rb)\rb] -  \erfi\lb(\frac{T-S\cos(\tht)}{\sqrt{2\mathcal{A}(\tht)}}\rb)\Bigg\} \label{eq:Thetaflat}\\
    \omega &= \frac{\la_A\la_B\et_A\et_B}{(2\pi)^{3/2}}\int_{0}^{2\pi}d\tht\ \frac{\exp\lb(-\frac{[T-S\cos(\tht)]^2}{2\mathcal{A}(\tht)}\rb)}{\sqrt{\mathcal{A}(\tht)}} \Im\lb[\erfi\lb(\frac{T-S\cos(\tht)}{\sqrt{2\mathcal{A}(\tht)}}+\frac{\iu\La \sqrt{\mathcal{A}(\tht)}}{\sqrt{2}}\rb)\rb] \label{eq:omegaflat}
}
\end{widetext}
where we define $T \ce T_B-T_A$, $S\ce\Abs{\bd{x}_B-\bd{x}_A}$, and
\eqns{
    \mathcal{A}(\tht)\ce a_A^2\big(1-\ep_A^2\sin^2(\tht-\varphi_A)\big)+a_B^2\big(1-\ep_B^2\sin^2(\tht-\varphi_B)\big)
}
and we evaluate final integral over $\tht$ numerically in Mathematica using the \texttt{DoubleExponential} method and a working precision and accuracy of 20.

%\ahs{Needs to be finished}

\subsection{Spherical Spacetime}

%\ljh{Double check the conventions}

We will   consider   Gaussian detectors  on the $(2+1)$ dimensional surface of the sphere. Without loss of generality  the first detector, A, can be centered at the north pole. The centre of the position of the second detector, B, is encoded through an arbitrary polar rotation $\theta$ and an azimuthal rotation $\varphi$ relative to the north pole. To consider elliptical (or squeezed) detectors we introduce an additional parameter $\gamma$ that describes the rotation of a profile about its  semi major axis. 
%\rbm{Don't we have an analogous parameter in flat space?  It wasn't discussed in the previous subsection.}\ljh{Change $\varphi_D \to \gamma_D$ in the flat space case to better match this notation.  It won't be perfect since rotation doesn't commute with an expansion into spherical modes.}
To study the response of the detectors we will need to calculate the spherical harmonics coefficients $f_{\ell m}^D$ and how they transform under an arbitrary rotation on the surface of the sphere.

 A spherical analog of a Gaussian profile 
 is the Fisher-Bingham Five-Parameter (FB5) Distribution \cite{alem2015spherical}, given by:
\eqn{
\label{FBgen}
f(\widehat{\boldsymbol{x}} ; \kappa, \widehat{\boldsymbol{\mu}}, \beta, \boldsymbol{A})=\frac{1}{C(\kappa, \beta)} e^{\kappa \widehat{\boldsymbol{\mu}}^{T} \widehat{\boldsymbol{x}}+\widehat{\boldsymbol{x}}^{T} \beta \boldsymbol{A} \widehat{\boldsymbol{x}}}}
where 
\eqn{\label{Cdef}
C(\kappa , \beta)=2 \pi \sum_{r=0}^{\infty} \frac{\Gamma(r+1 / 2)}{\Gamma(r+1)} \beta^{2 r}(\kappa / 2)^{-2 r-1 / 2} I_{2 r+1 / 2}(\kappa)}
%\rbm{Is the normalization $C(\kappa, \beta \boldsymbol{A})$ or $C(\kappa, \beta)$?}\ahs{It's $C(\kappa, \beta)$, fixed}
is the normalization constant $C(\kappa, \beta)$ 
and where $I_r(x)$ is the modified Bessel function of the first kind of order $r$. Moreover $\boldsymbol{A}$ is a $3 \times 3$ symmetric matrix defined as
\begin{equation}
\boldsymbol{A}=\left(\widehat{\boldsymbol{\eta}}_{1} \hat{\boldsymbol{\eta}}_{1}^{T}-\widehat{\boldsymbol{\eta}}_{2} \hat{\boldsymbol{\eta}}_{2}^{T}\right)
\end{equation}
where the unit vectors $\hat{\boldsymbol{\mu}},\hat{\boldsymbol{\eta_1}},\hat{\boldsymbol{\eta_2}}$ respectively correspond to the mean centre, semi major and semi minor axes of the FB5 distribution. The parameter $\kappa \geq 0$ quantifies the spatial concentration around the center $\hat{\boldsymbol{\mu}}$ and the parameter $\beta \leq \frac{\kappa}{2} $ quantifies the ellipticity of the distribution. The higher the values of $\kappa$ and $\beta$ are, the more concentrated and elliptical the FB5 distribution.

The rotation group SO(3) is the continuous group of all rotations around an origin in three dimensional Euclidean space that preserve the inner product in $\mathbb{R}^3$ under composition. Any arbitrary rotation $R(\alpha, \beta, \gamma)$ (like that on the surface of a sphere) can be characterized by the Euler angles $(\al,\be,\ga)$ in the `$zyz$' convention:  rotate first by an angle $\ga$ around the $z$-axis, then rotate by an angle $\be$ around the $y$-axis, and finally rotate by an angle $\al$ around the $z$-axis, yielding
\eqn{R(\alpha, \beta, \gamma)=e^{-\iu\alpha L_{z}} e^{-\iu \beta L_{y}} e^{-\iu \gamma L_{z}}}
where $L_i$ is the rotation generator    in the $i$-th direction. These satisfy  the commutation relation
\eqn{
\left[L_{i}, L_{j}\right]=\iu \sum_{k=1}^{3} \epsilon_{i j k} L_{k} \quad \forall i, j \in\{1,2,3\}
}

The matrix elements of the rotation operator $R$ in the $\Ket{l,m}$ basis are given by the Wigner-$D$ matrix 
\eqn{
\Bra{\ell,m}R(\alpha,\beta,\gamma)\Ket{\ell',m'} = \delta_{\ell \ell'}D^{\ell}_{m m'}(\alpha,\beta,\gamma)
}
and using the fact that $\ket{\ell,m}$ are eigenstates of the $\hat{L}_z$ operator we can write  
\aln{
D^{\ell}_{m m'}(\alpha,\beta,\gamma) &=\bra{\ell, m} R(\alpha, \beta, \gamma)\ket{\ell, m^{\prime}} \nn \\
&=\bra{\ell, m}e^{-\iu \alpha L_{z}} e^{-\iu \beta L_{y}} e^{-\iu \gamma L_{z}}\ket{\ell, m^{\prime}} \nn\\
&=e^{-\iu m \alpha}\bra{\ell, m}e^{-\iu \beta L_{y}}\ket{\ell, m^{\prime}} e^{-\iu m'\gamma} \nn \\
&=e^{-\iu m \alpha} d^{\ell}_{m m^{\prime}}(\beta) e^{-\iu m'\gamma} }
where $d^{\ell}_{m m^{\prime}}$ are the Wigner-$d$ matrix elements \cite{Wignerd}. The Wigner-$d$ matrix elements allow us to transform the spherical harmonics coefficients $f_{\ell m}^R$ under an Euler rotation since
\aln{
f_{\ell m}^R &= \bra{\theta, \varphi} f R(\alpha, \beta , \gamma) \Ket{\ell ,m} \nn \\
&= \sum_{\ell',m'} \bra{\theta, \varphi} f \Ket{\ell' ,m'} \Bra{\ell', m'}R(\alpha, \beta , \gamma)\ket{\ell ,m} \nn \\
&= \sum_{\ell',m'} \bra{\theta, \varphi} f \Ket{\ell' ,m'} \delta_{\ell \ell'}D^{\ell}_{m'  m}(\alpha,\beta,\gamma) \nn \\
%&= \sum_{m'} D^{\ell}_{m'  m}(\alpha,\beta,\gamma)  \bra{\theta, \varphi} f \Ket{\ell ,m'} \nn \\
&= \sum_{m'=-\ell}^\ell D^{\ell}_{m'  m} (\alpha,\beta,\gamma) f_{\ell m'}
}

To derive these coefficients for an arbitrary FB5 distribution we follow the approach presented in \cite{alem2015spherical}, where we begin with an FB5 distribution centered at the north pole with the mean, semi-major and semi-minor axes aligned as follows
\eqn{
\widehat{\boldsymbol{\mu}}^{0}=\left[\begin{array}{lll}
0 & 0 & 1
\end{array}\right]^{T} \quad \widehat{\boldsymbol{\eta}}_{1}^{0}=\left[\begin{array}{lll}
1 & 0 & 0
\end{array}\right]^{T} \quad \widehat{\boldsymbol{\eta}}_{2}^{0}=\left[\begin{array}{lll}
0 & 1 & 0
\end{array}\right]^{T}}
respectively. 
We will refer to this as the standard FB distribution  $g(\widehat{x} ; \kappa, \beta)$, with the general FB5 distribution (\ref{FBgen})   related to the standard one via
\eqn{
f(\widehat{\boldsymbol{x}} ; \kappa, \beta, \widehat{\boldsymbol{\mu}}, \boldsymbol{A}) = R(\alpha,\beta,\gamma)g(\widehat{x} ; \kappa, \beta)}

This allows us to calculate the spherical harmonics coefficients $g_{\ell m}$ for the standard FB distribution and automatically obtain the spherical harmonics $f_{\ell m}$ for the general FB5 distribution through the transformation
\eqn{
  f_{\ell,m} = \sum_{m'=-\ell}^{\ell}D_{m',m}^\ell(\al,\be,\ga)g_{\ell,m'}
  \label{eq:WignerDTransf}
}
where $D_{m,m'}^\ell$ is the Wigner-$d$ function of degree $\ell$ and \{$\alpha , \beta , \gamma $\} are the Euler angles. 
Under these assumptions the standard FB distribution  is   given by
\eqn{g(\widehat{x} ; \kappa, \beta) = \frac{1}{C(\kappa, \beta)} e^{\kappa \cos \theta+\beta \sin ^{2} \theta \cos 2 \varphi}
\label{FB dist}}

Consider first that detector A  is unsqueezed ( $\beta = 0$). Its probability distribution function $g$ is then
\eqn{\label{FBSymmetric}
g = \frac{\sqrt{\kappa} e^{\kappa \cos{\theta}}}{ (2 \pi)^{3/2} I_{1 / 2}(\kappa)}} 
using \eqref{Cdef}.
We can derive an analytic expression for the spherical harmonics coefficients  
$g_{\ell m} = \Braket{g(\kappa,\beta,\theta),Y_{\ell}^m}$  
\aln{
g_{\ell m} =\delta_{m, 0} \sqrt{\frac{2 \ell+1}{4 \pi}} \frac{I_{\ell+1 / 2}(\kappa)}{I_{1 / 2}(\kappa)} \equiv \delta_{m, 0} g_{\ell}
\label{glmUnsqueezed}}
as shown in Appendix B.

%Note that the modes converge as $\ell$ increases \red{(double check this)}
%\eqn{
%\lim _{\kappa \rightarrow \infty} \frac{I_{\ell+1 / 2}(\kappa)}{I_{1 / 2}(\kappa) \kappa}=0}

If the detector is squeezed ($\beta \neq 0 $)
then these coefficients become 
%\rbm{Where is $N_{\ell m}$ defined?}\ahs{Fixed-Defined in the previous section now, add reference to equation}
\eqn{
g_{\ell m}= 2 \pi \frac{N_{\ell m}}{C(\kappa, \beta)} \int_{-1}^{1} e^{\kappa z} P_{\ell}^{m}(z) I_{m / 2}\left(\beta \left( 1 - z^2 \right) \right)dz 
\label{glmSqueezed}
}
also shown in Appendix B, where 
$P_{\ell}^{m}(z)$ is an associated Legendre function.
Note that the integral above is 0 whenever $m$ is odd.
%\rbm{Why?} \ahs{We know this numerically, need an analytic explanation} 
The coefficients in \eqref{glmSqueezed} can be evaluated numerically for a given $\kappa$ and $\beta$.

We shall   place  detector A at the north pole so that   it is described by the standard FB distribution whose   spherical harmonics coefficients $g_{\ell m}$ are given by \eqref{glmUnsqueezed} if unsqueezed and by \eqref{glmSqueezed} if squeezed.   The second detector, B,  is placed anywhere on the sphere via  an Euler rotation, and its spherical harmonics coefficients are given by  \eqref{eq:WignerDTransf}. The integrals are evaluated using the same method \texttt{DoubleExponential} as discussed at the end of the last subsection.

With this established, in Appendix B we show that the local coupling term for each detector can be simplified to
\eqn{
f_D = \exp\left[-\sum_{\ell = 0}^{\infty} \sum_{m= -\ell}^{\ell}\frac{\lambda_D^2 \eta_D^2}{2\ell+1}|g_{\ell m}^D|^2\right]
%\qquad  \text{and} \qquad f_B = \exp\left[-\frac{1}{2}\sum_{\ell = 0}^{ \infty} \sum_{m= -\ell}^{\ell}\frac{\lambda_B^2 \eta_B^2}{2(\ell+1/2)}|\sum_{m_1=-\ell}^{ \ell} D_{m,m_1}^{\ell}(\alpha,\beta,\gamma) g_{\ell m_1}^B|^2\right]
}

This is to be expected since the spacetime has constant spatial curvature everywhere, so we would not expect the detector localization to depend on where the detector coupled to the field. Furthermore, the smeared field commutator becomes,
\begin{widetext}
\aln{
  \Theta &= \iu\la_A\la_B\et_A\et_B \sum_{\ell,m}\frac{1}{2(\ell+1/2)}  \vast[ g_{\ell,m}^{A^*} \lb(\sum_{m_{1}=-\ell}^{\ell} D_{m,m_1}^\ell(\al_B,\be_B,\ga_B) g_{\ell,m_1}^B\rb) \ec^{\iu(\ell+1/2)(T_B-T_A)} \nn\\
  &\qquad\qquad\qquad\qquad\qquad\qquad\qquad\qquad  - g_{\ell,m}^A \lb(\sum_{m_{1}=-\ell}^{\ell} D_{m,m_1}^\ell(\al_B,\be_B,\ga_B) g_{\ell,m_1}^B\rb)^* \ec^{-\iu(\ell+1/2)(T_B-T_A)}\vast] \label{eq:SphereCommWignerD}
}
after substituting  \eqref{eq:WignerDTransf} into   \eqref{eq:SphereCommNoLocation}. In Appendix D we show that if both detectors are regular
and the second detector is centered at an angle $\alpha = \theta$, then $\Theta $ can be simplified to
\eqn{
  \Theta = \iu\lambda_A\lambda_B\eta_A\eta_B \sum_{\ell=0}^{\infty} \frac{1}{2\ell+1} \Bigg[ {f_\ell^A}^* f_{\ell}^B P_{\ell}[\cos(\theta)] e^{\iu(\ell+1/2)(T_B-T_A)} - f_{\ell}^A {f_{\ell}^B}^*P_{\ell}[\cos(\theta)] e^{-\iu(\ell+1/2)(T_B-T_A)}\Bigg].
\label{eq:ThetaRegularComm}}
\end{widetext}

%%%%%%%%%%%%%%%%%%%%%%%%%%%%%%%% Results %%%%%%%%%%%%%%%%%%%%%%%%%%%%%
\section{Single detector Results}
\label{sdres}

We consider first the response of a single detector to a bandlimited field in both the Minkowski and spherical spacetimes.  Note that a bandlimit is expressed as a cutoff $\Lambda$ in momentum in flat spacetime, whereas it is expressed as a maximum value of $\ell$ 
(denoted as $\ell_{max}$)
in the  spherical spacetime.
 In the spherical case we are unable to numerically sum over all values of $\ell$, and so will always have to impose a cutoff at some maximal value.

\subsection{Flat Spacetime}
\label{sdresflat}
 
The transition probability of a single UDW detector with a spacial profile given by Eq.\ \eqref{eq:fDflat}   depends on three parameters:  its size, $a$, its eccentricity, $\ep$, and the bandlimit of the scalar field $\La$.

\begin{figure}[H]
    \centering
    \includegraphics[width=0.99\linewidth]{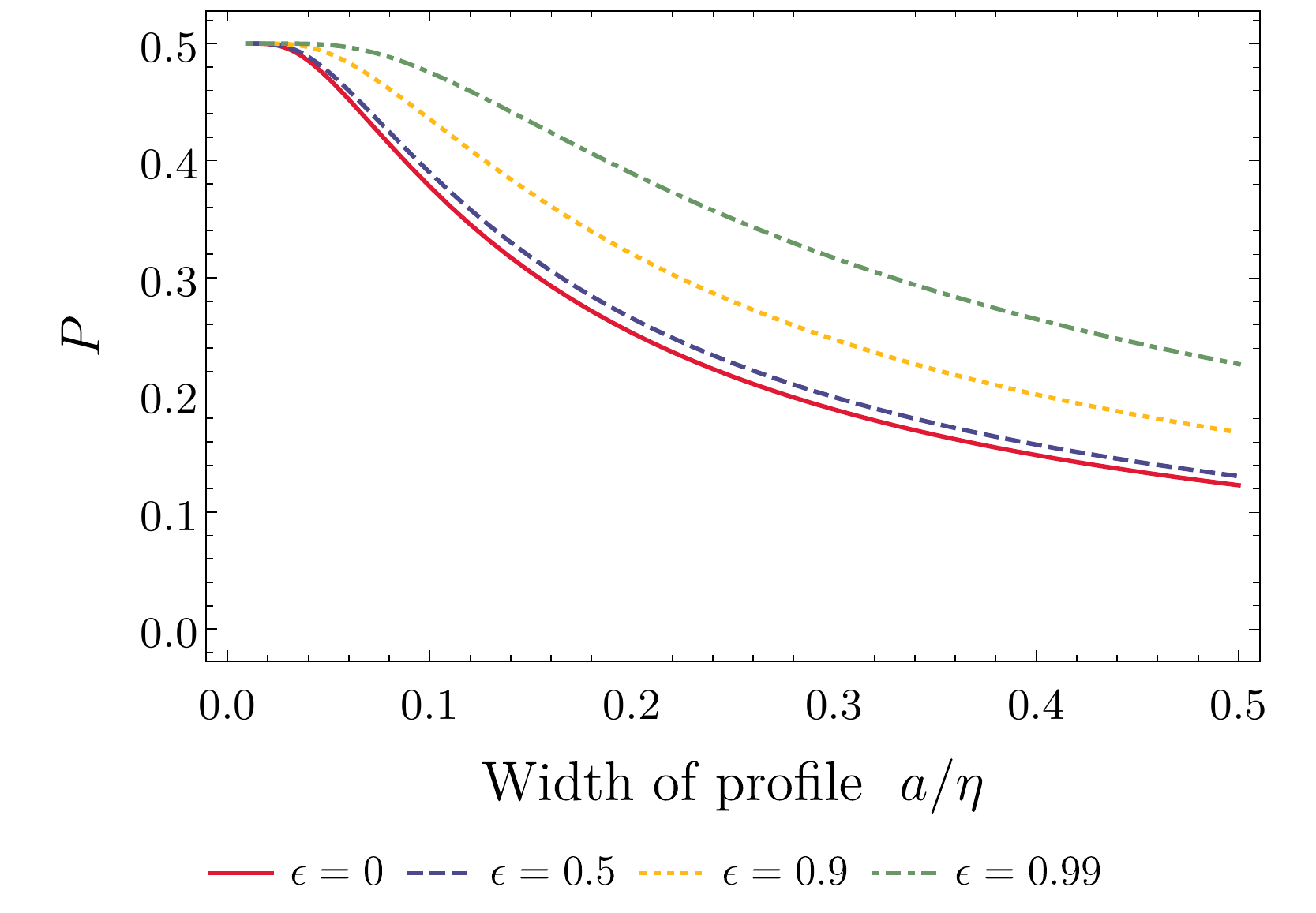} 
    \caption{%
        The transition probability $P$ of a single detector in flat spacetime that interacts with a scalar field with no bandlimit as a function of its size for various values of the eccentricity.   The transition probability increases for decreasing detector size and increasing eccentricity. The dimensionless coupling constant is $\lambda\eta^{1/2}=1$.
    }
    \label{fig:PAvsaNoBL}%
\end{figure}

In figure \ref{fig:PAvsaNoBL}, we plot the transition probability of a single UDW detector as a function of its size and eccentricity that couples to a field with no bandlimit $(\La\to\infi)$.  We find that the transition probability of the detector increases as the overall size of the detector   decreases for a given eccentricity,
 approaching a value of $0.5$ in the pointlike limit.   This is commensurate with previous results on bandlimited detectors \cite{Henderson:2020ucx}, where it was shown that for a detector with an unsqueezed Gaussian spacial profile, smaller detectors have higher transition probabilities due to an increased sensitivity to high momentum field modes. We see that this behaviour holds for squeezed detectors as well; detectors with a smaller overall size have a higher transition probability.

Conversely, for a given size, as the eccentricity  increases, the transition probability also increases. We can understand this by noting that squeezing reduces the length scale of the profile in one direction, say the $x$-direction.  This will increase the sensitivity of the detector to field modes with a large momentum in the  $x$-direction, yielding a larger transition probability.

\begin{figure}[H]
    \centering
    \includegraphics[width=0.99\linewidth]{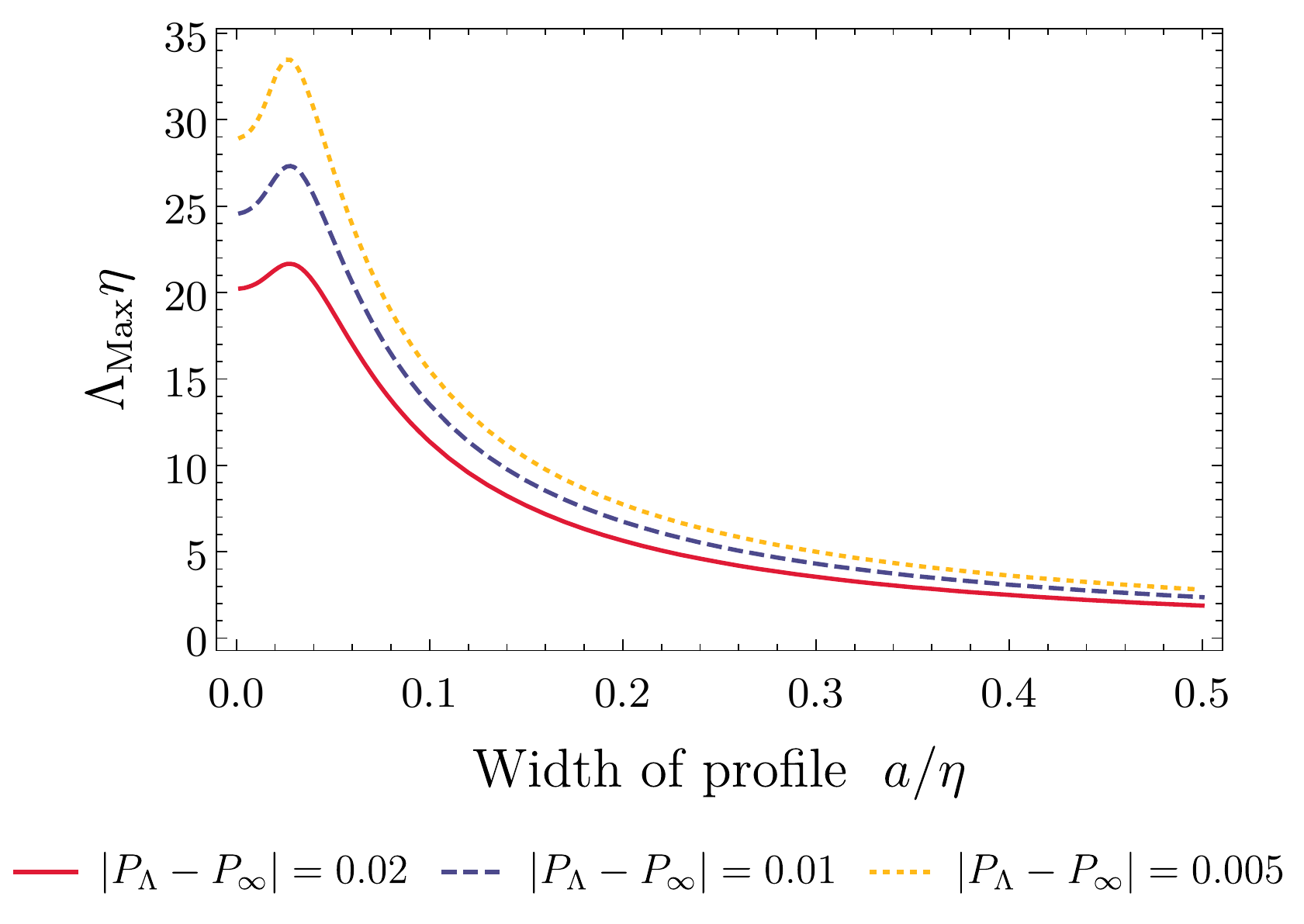}
    \caption{%
        The value of the maximum bandlimit $\La_\tx{max}$ such that the difference between the transition probability in the bandlimtied field and the non-bandlimited field is equal to a specified (arbitrary) tolerance as a function of the size of the unsqueezed $(\epsilon=0)$ spatial profile  in flat spacetime.  The dimensionless coupling constant is $\lambda\eta^{1/2}=1$.
    }
    \label{fig:BestSizeNoBL}%
\end{figure}

In order to explore the effect of detector squeezing on bandlimit detection, for a baseline comparison we first consider the case of an unsqueezed detector.  In figure \ref{fig:BestSizeNoBL}, we plot the value of the bandlimit $\La_{\tx{Max}}$ for which the transition probability of an unsqueezed detector is within a chosen tolerance away from the corresponding non-bandlimited value 
($\Lambda \to \infty$)
as a function of detector size.  We choose this value as a measure of bandlimit detection for two reasons.  First, it provides a precise definition of when the bandlimited transition probability is ``close'' to its asymptotic value.  Second, this value provides an operational notion of bandlimit detection.  In an experimental setup, one can only distinguish between two measures to within some tolerance. Hence given a detector size and tolerance, there is a maximum value of field bandlimit  that can be distinguished from the case  with  no bandlimit.

We find that, regardless of the chosen tolerance, small detectors are able to detect a higher bandlimit than larger detectors. However, once the detector's size is smaller than a (tolerance dependent) critical size,   we find that within this regime larger detectors are able to detect larger bandlimit than smaller detectors.  These results match what was presented for $(3+1)$-dimensional Minkowski space \cite{Henderson:2020ucx}, suggesting dimensional independence. 

\begin{figure*}
    \includegraphics[width=0.48\linewidth]{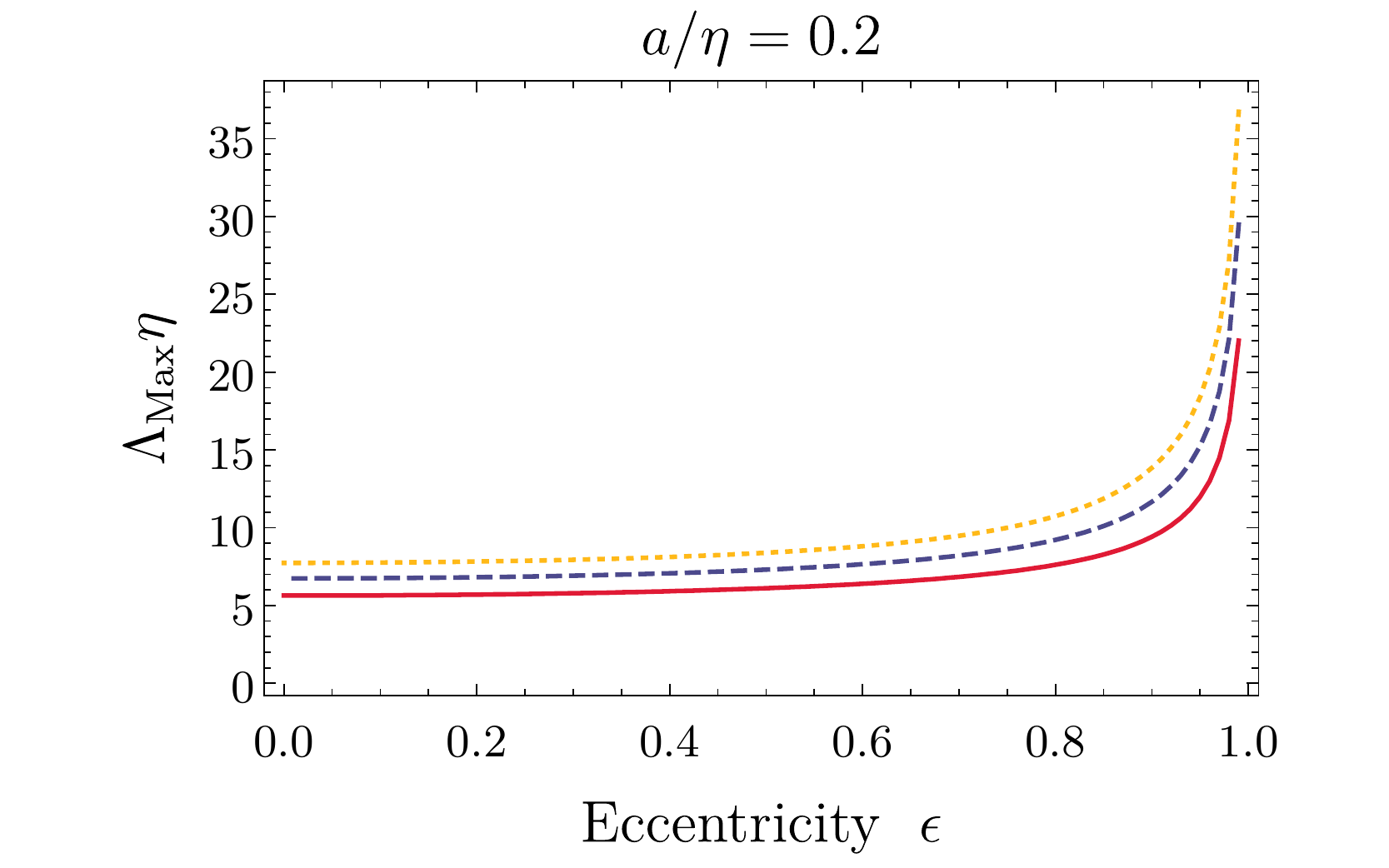}
    \hspace{0.5cm}
    \includegraphics[width=0.48\linewidth]{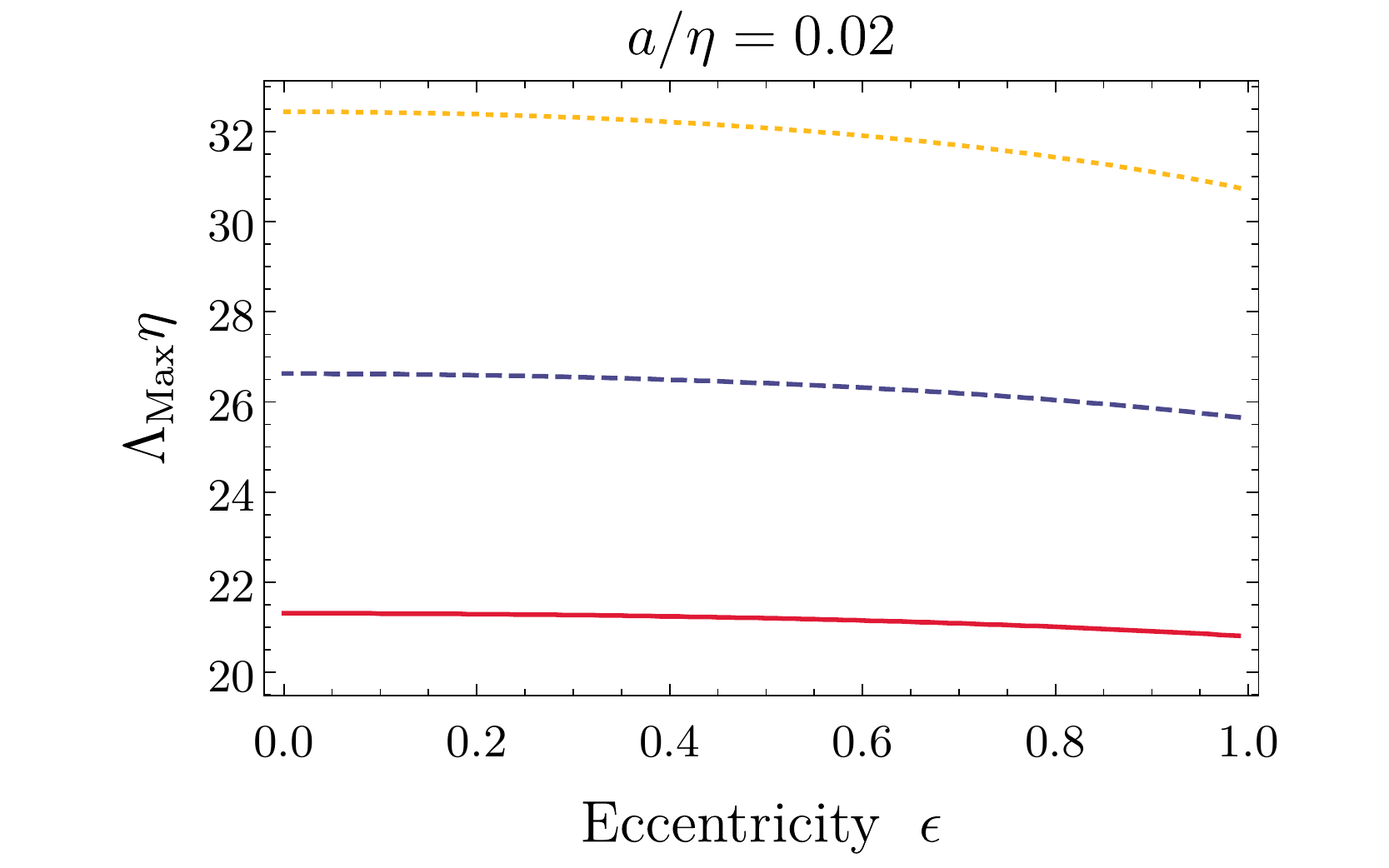}\\
    \includegraphics[width=0.7\linewidth]{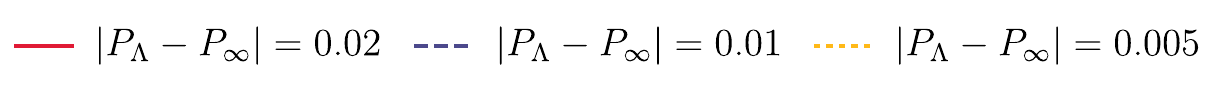}
    \caption{The value  in flat spacetime of the maximum bandlimit $\La_\tx{max}$ such that the difference between the transition probability in the bandlimited field and the non-bandlimited field is equal to a specified (arbitrary) tolerance as a function of the eccentricity of the spacial profile for a detector size of \textit{left:} $a_A=0.2\et$ and \textit{right:} $a_A=0.02\et$.  The dimensionless coupling constant is $\lambda\eta^{1/2}=1$.
    }
    \label{fig:BestSqeezing}%
\end{figure*}

In figure \ref{fig:BestSqeezing}, we explore the impact of squeezing on bandlimit detection, and find that squeezing does not necessarily increase sensitivity to a bandlimit.  Instead, we find the effect of squeezing depends on the overall size of  the detector.  If the detector is larger than the optimal size found in figure \ref{fig:BestSizeNoBL}, then increasing the squeezing increases its sensitivity   to the bandlimit, whereas if the detector is near or smaller than the optimal size, increased squeezing decreases its sensitivity  to the bandlimit.  Since squeezing reduces the length scale of the detector along one direction, this further suggests there is an overall optimal detector scale for bandlimit detection.
%\rbm{Is there any way that we can get figure 3 to appear before figure 4?}\ljh{Moving the figures will the last step once all the text is completed.}

\subsection{Spherical Spacetime}

\begin{figure}[t]
\centering
  \includegraphics[width=0.99\linewidth]{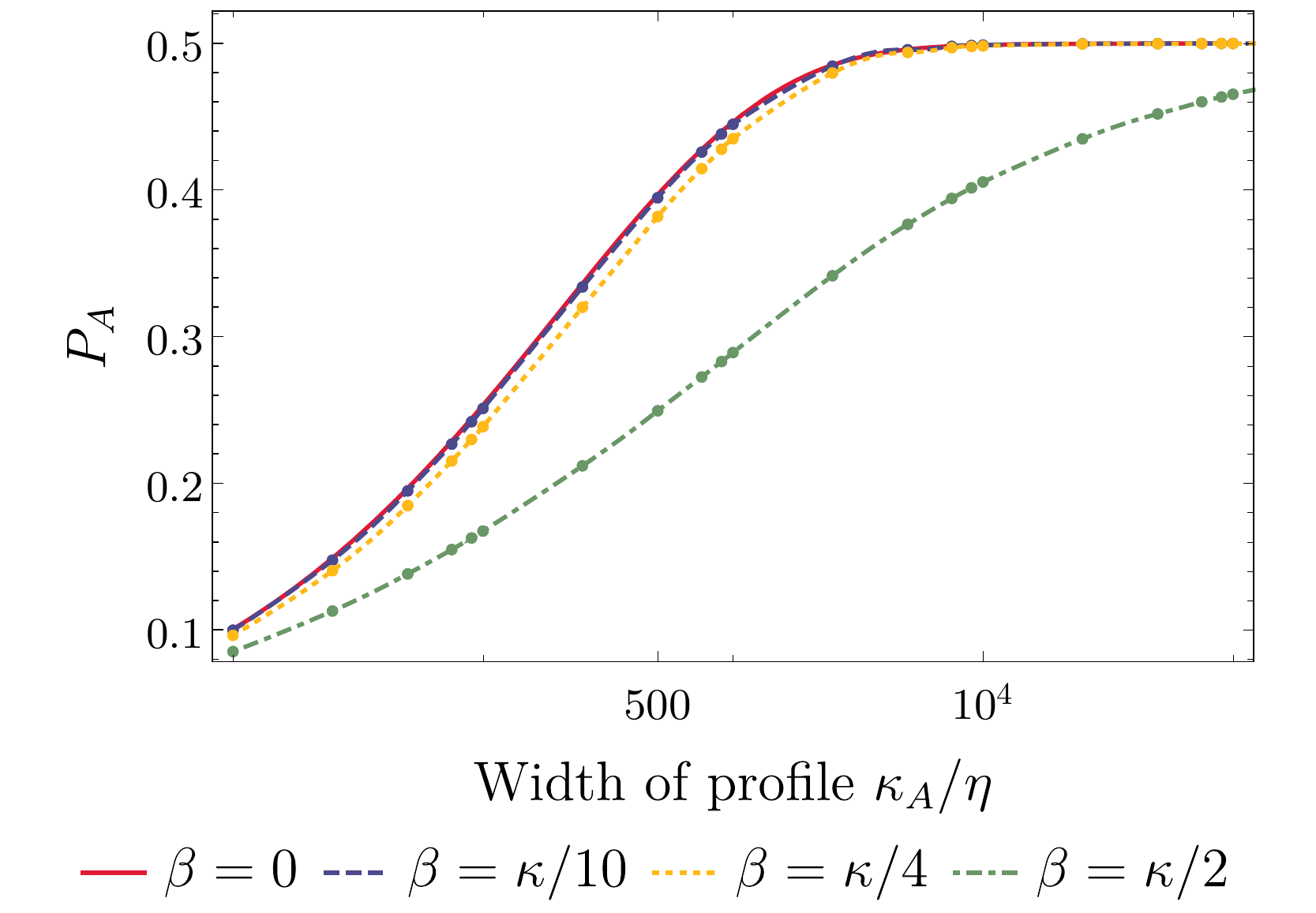}
  \caption{The transition probability $P$ of a single detector on a sphere that interacts with a scalar field with a bandlimit $\ell_{max} = 100$ as a function of its size for various values of the squeezing parameter $\beta$. The transition probability increases for decreasing detector size and 
  decreasing squeezing, the latter in contrast to the flat spacetime case. The dimensionless coupling constant is $\lambda\eta^{1/2}=1$.}
  \label{fig:1detOnSphere}
\end{figure}

Turning to the spherical case, in figure \ref{fig:1detOnSphere} we analyze the transition probability of a single detector as a function of its size $\kappa$ for various values of the squeezing parameter $\beta$. The value of the bandlimit, $\ell_{max}$, was set to $100$, which is sufficient for the results to converge to a tolerance of $0.001$, much smaller then the resolution of the figure. As in the flat spacetime setting, we see that the transition probability increases with decreasing  detector size (increasing $\kappa$), approaching $1/2$ in the pointlike limit $\kappa\to\infty$. This is because smaller detectors have larger spherical harmonics coefficients, and so end up coupling to more field modes, analogous to the  Fourier transform of the Gaussian shape in flat space. In addition, we see that contrary to flat space, squeezing decreases the response of a detector.

It is important to note that squeezing is implemented differently in the FB5 distribution [Eq.~\eqref{FB dist}] than in a 2-dimensional Gaussian distribution [Eq.~\eqref{eq:FDflat}].  While in both cases, increasing the squeezing parameter decreases the length of the semi-minor axis and increases the length of the semi-major axis, the two distributions renormalize the function differently.  The former case reduces the overall height of the distribution while the latter increases the overall height of the distribution.  The mathematical consequence of this is that cross-sections of the squeezed FB5 distribution are no longer Gaussians, whereas cross-sections of the squeezed 2-dimensional Gaussian distribution remain Gaussians.  Functionally, this means that when a detector on a sphere is squeezed, it is more sensitive to field modes with higher momentum perpendicular to the direction of squeezing, but couple to them less strongly overall as compared to an unsqueezed detector. 

%In the former, an increase in the squeezing parameter $\beta$ increases the length of the semi-major axis of the leaving the semi-minor axis unchanged, with a value of $\beta=0$ corresponding to a circle.  In the latter case, an increase in the squeezing parameter $\epsilon$ [Eq.~\eqref{eq:epsilon}] increases the semi-major axis of the profile, but also decreases the length of the semi-minor axis.  Functionally, this means that when a detector on a sphere is squeezed, it's smallest length scale (the semi-minor axis) does not change, but in flat space squeezing decrease the smallest length scale of the detector.}
%
%We need to be careful when comparing spherical squeezing  to the the flat space squeezing (which scales both axes and then renormalizes). The former does not lead to a significantly smaller length scale while the latter does. \rbm{What does that mean?  A length scale of what?}
%This may explain why the transition probability decreases with squeezing in the spherical case contrary to what we see in flat space. \rbm{Where is the figure illustrating this?}
%As a consequence of all of this we will not use squeezed detectors when trying to analyze the sensitivity of our detectors to a bandlimit.

%BL Detection
 To formalize the notion of optimal detection of a band\-limit we analyze the absolute difference between the transition probability of a bandlimited detector with one that is not bandlimited. In the absence of closed form expressions of the series sums that define the response of the detector, we take an $\ell_{max}$ of 350 to be the value at which we truncate the series and call that "infinity". This value was chosen since the results converge up to machine precision. This $\Abs{P_{A,\infty} - P_{A,\ell_{max}}}$ value will be the criterion for optimal bandlimit detection. The higher this absolute difference,  the more sensitive a detector is (in some configuration) to the value of the bandlimit should it exist. Moreover, this absolute difference or tolerance criterion can be understood in an operational sense. That is, if we can resolve the transition probabilities of the detectors in an experimental setup up to an accuracy or a tolerance lower than the absolute difference between the non bandlimited and the bandlimited response, then the bandlimit is detectable.

\begin{figure}
    \centering
    \includegraphics[width=0.99\linewidth]{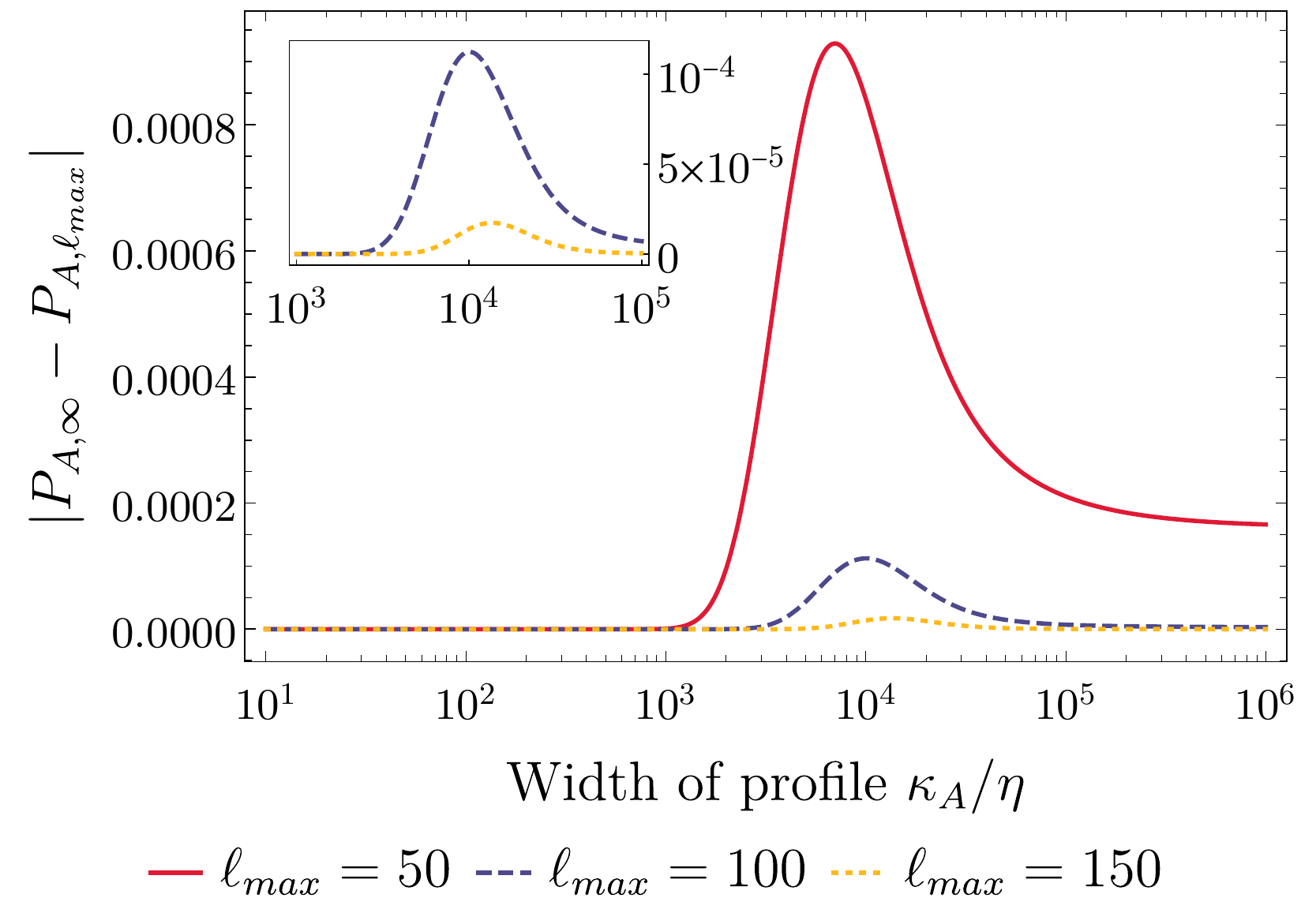} 
    \caption{The value of the absolute difference bandlimit detection criterion of a single regular detector $(\beta=0)$ on the sphere as a function of the detector size for several values of the bandlimit $\ell_{max}$. The dimensionless coupling constant is $\lambda\eta^{1/2}=1$.}
    \label{fig:Optimal-BL-D1-S2}%
\end{figure}

 In figure \ref{fig:Optimal-BL-D1-S2}, we plot the absolute difference between the non bandlimited and bandlimited response as a function of the detector size for several values of $\ell_{max}$. For clarity,   we have made the plot of the tolerance  continuous though the $\ell_{max}$ bandlimit is discrete. In figure \ref{fig:Optimal-BL-D1-S2}, we see that regardless of the chosen tolerance, for any bandlimit, there is a size where the detectors are optimally sensitive to the presence of a bandlimit. In other words  bandlimit detection is optimal for small but not pointlike detectors exactly like detectors in flat spacetime.

 The interplay between the size of the detector and optimal bandlimit detection can be understood as follows.  First, the existence of an optimal size for bandlimit detection can be explained by recasting the detection criterion differently. Ultimately, what we are investigating is the contribution of each field mode to the response of the detector. In light of that, the absolute difference would be dictated by how the length scale of the detector determines its coupling to the field. Larger detectors couple weakly so they are only sensitive to a very small bandlimit. On the other hand, pointlike detectors couple to every field mode, as demonstrated by equation \eqref{f_D} in the limit as $\kappa\to \infty$. The optimal length scale is a balance between both effects: we want a sensitive detector that can couple to many of the field modes but not to  be so small (near pointlike)  that it is too sensitive, making its response saturate to $1/2$. In other words, we do not want it  to become a maximally mixed state where we cannot resolve the impact of the cutoff on its response.

\begin{figure*}
    \includegraphics[width=0.48\linewidth]{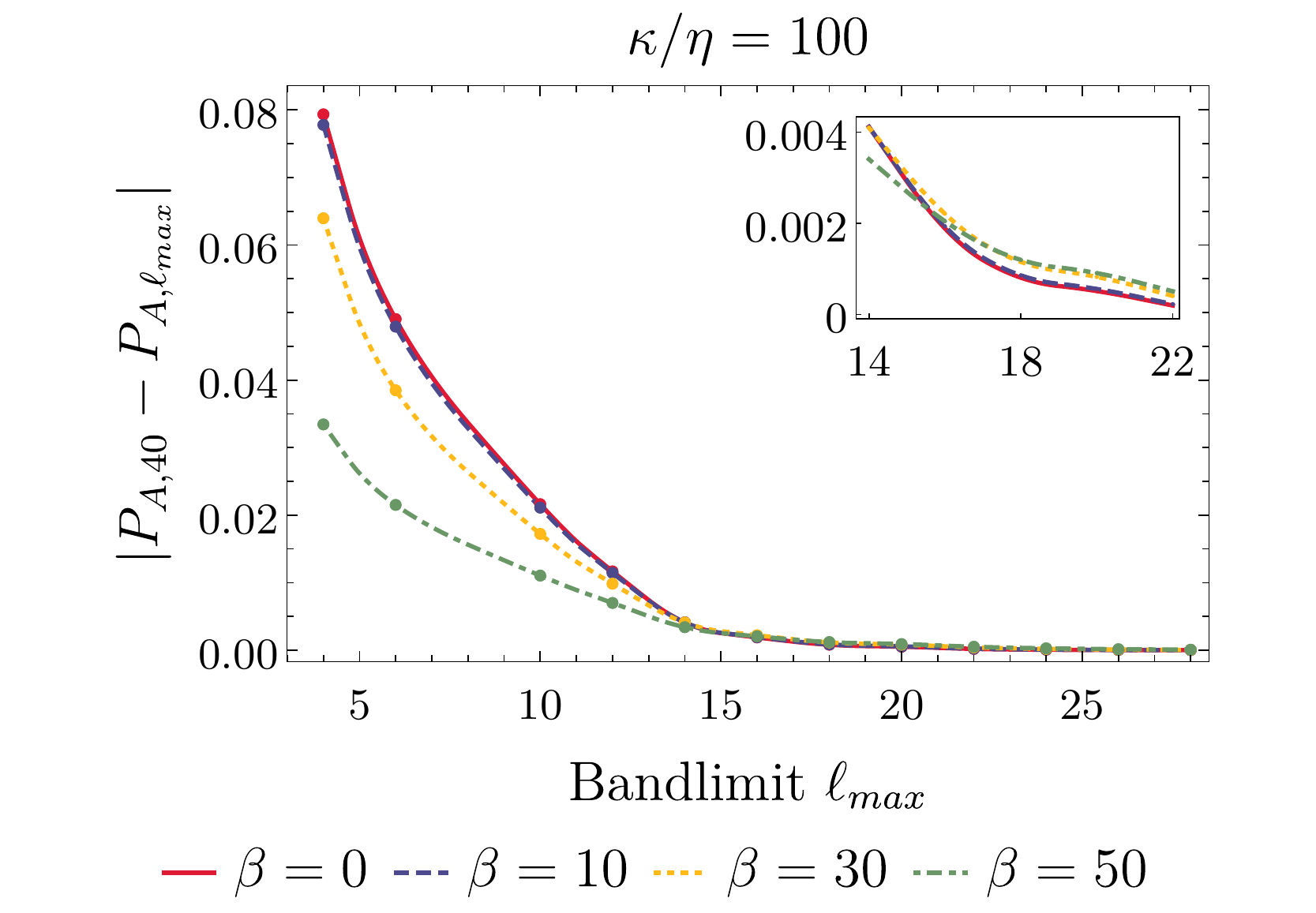}
    \hspace{0.5cm}
    \includegraphics[width=0.48\linewidth]{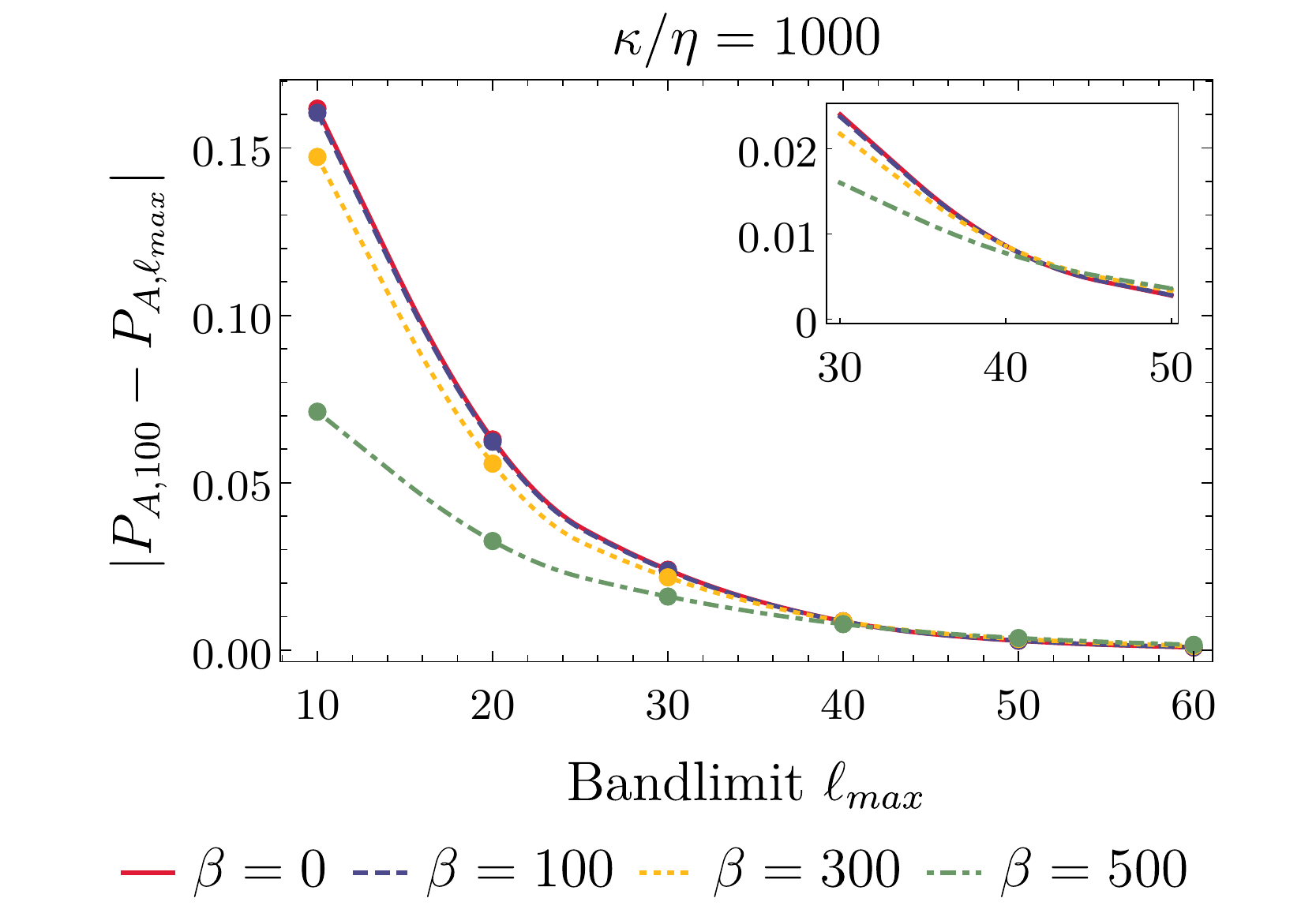}
    \caption{The absolute difference bandlimit detection criterion for squeezed detectors as a function of the cutoff $\ell_{max}$ for several values of the squeezing parameter. Note that $\ell_{max}$ is a discrete parameter, so the interpolation done between the data points in the figures is done for purely illustrative purposes. The detector size on the \textit{left} is $\kappa = 100 \eta$ and on the \textit{right} is $\kappa = 1000 \eta$.The dimensionless coupling constant is $\lambda\eta^{1/2}=1$.}
    \label{fig:BestSqeezing_sphere}%
\end{figure*}

 Finally, we consider the impact of squeezing on bandlimit detection. In figure \ref{fig:BestSqeezing_sphere}a, we pick a large non-optimal size of $\kappa=100 \eta$ and analyze the bandlimit detection criterion as a function of the cutoff $\ell_{max}$ for several values of the squeezing parameter $\beta$. Due to numerical considerations, the values of $\ell_{\infty}$ chosen were based on the values calculated where the results have converged in figure \ref{fig:Optimal-BL-D1-S2} as opposed to some arbitrary tolerance. Contrary to the behavior in flat space shown in figure \ref{fig:BestSqeezing}, squeezing decreases the sensitivity to the bandlimit. However, this is only true up to a certain value of $\ell_{max}$, where we observe a nominal increase in bandlimit detection. This pattern holds for the smaller $k=1000\eta$ detector, which is closer to the optimal size as discussed for figure \ref{fig:Optimal-BL-D1-S2}. This implies that the optimality of squeezing in this context is dependent on the value of the bandlimit itself. This is in notable contrast to flat space, where the optimality (or lack thereof) of squeezing depended primarily on the size of the detector. Ultimately,   optimal bandlimit detection is not just set by the best scale for  probing  field fluctuations, but also by the geometry of the background spacetime (encoded in the fluctuations) and the geometry of the detectors.

\section{Results for two detectors }

\subsection{Signalling and the structure of the smeared field commutator}
\label{sec:signalling}
In the  two detector setup, if both detectors are casually connected, the second detector will interact with an evolved state of the field locally due to the interaction of the first detector with the field.  The response of the second detector will then depend (in part) on the field mediated signalling from the first detector. To that end, let us consider the two point Wightman distribution $W$ for a scalar field $\hat{\phi}$. The Wightman distribution can be split into a real and an imaginary part as
\begin{equation}
W\left(\mathrm{x}, \mathrm{x}^{\prime}\right):=\frac{1}{2}\left(W^{+}\left(\mathrm{x}, \mathrm{x}^{\prime}\right)+W^{-}\left(\mathrm{x}, \mathrm{x}^{\prime}\right)\right)
\end{equation}
where the real and imaginary contributions are defined as
\begin{equation}
\begin{aligned}
&W^{+}\left(\mathrm{x}, \mathrm{x}^{\prime}\right)=\left\langle 0\left|\left\{\hat{\phi}(\mathrm{x}), \hat{\phi}\left(\mathrm{x}^{\prime}\right)\right\}\right| 0\right\rangle \\
&W^{-}\left(\mathrm{x}, \mathrm{x}^{\prime}\right)=\left\langle 0\left|\left[\hat{\phi}(\mathrm{x}), \hat{\phi}\left(\mathrm{x}^{\prime}\right)\right]\right| 0\right\rangle.
\end{aligned}
\end{equation}
and so  $W^+$ and $W^-$ are the vacuum expectation values of the field anti-commutator and commutator respectively. The operators $\omega$ and $\Theta$ defined in equation \eqref{eq:GenThetaOmega} are smeared versions of $W^+$ and $W^-$ due to the  smearing of the detectors  over a compact region of spacetime. Studying the structure of the field commutator is of utmost importance when considering two detectors in causal contact, since it encodes causal signalling through the field between the spacetime regions where the detectors are.

In particular, the strong Huygen's principle states that the solutions of a hyperbolic second-order linear partial differential equation have support only along the null direction. In general however, this does not hold for a massless scalar field on a curved background and only holds in $n+1$ flat spacetime for odd dimensional $n\geq 3$ \cite{Hadamard1942ThePO}. To illustrate this, consider the Green's function for the massless scalar field in $2+1$ Minkowski spacetime,
\begin{align}
  G(\boldsymbol{x}, t,\boldsymbol{x}^{\prime}, t^{\prime})&= \frac{\mathrm{i}}{2 \pi} \frac{\operatorname{sgn}\left(t^{\prime}-t\right)H\left(\left(t-t^{\prime}\right)^{2}-\left|\boldsymbol{x}-\boldsymbol{x}^{\prime}\right|^{2}\right) 
}{\sqrt{\left(t-t^{\prime}\right)^{2}-\left|\boldsymbol{x}-\boldsymbol{x}^{\prime}\right|^{2}}}  
\end{align}
where $H(x)$ is the Heaviside function. It is important to note that due to the $ \operatorname{sgn}\left(t^{\prime}-t\right)/\sqrt{\left(t-t^{\prime}\right)^{2}-\left|\boldsymbol{x}-\boldsymbol{x}^{\prime}\right|^{2}}$ prefactor, part of the signal travels along timelike geodesics inside the lightcone. On the spherical spacetime,  for  two events separated by a time $T$ at two arbitrary positions, the Green's function can be expanded in terms of spherical harmonics modes as \cite{Lifschytz_1994}
\eqn{
\label{GModes}
G = \frac{1}{4 \pi} e^{-\iu T/2} \sum_{\ell=0}^{\infty} e^{-\iu \ell T} P_{\ell}(\cos(\alpha))
}
where $T$ is the difference in time between two events and $\alpha$ is the total angle between ($\theta,\varphi$) and ($\theta^{\prime},\varphi^{\prime}$). Then using the generating function of the Legendre polynomials: $\sum_{n=0}^{\infty} P_{n}(x) z^{n}=\left(1-2 x z+z^{2}\right)^{-\frac{1}{2}}$ for $-1<x<1$ and $|z|<1$, we can write $G$ as
\begin{align}
\label{FullG}
    G(\theta,\varphi,\theta^{\prime},\varphi^{\prime})&=\frac{1}{4 \sqrt{2} \pi}\bigg[\cos (T -\iu \epsilon)-\cos \theta \cos \theta^{\prime} \nonumber \\
    & - \sin \theta \sin \theta^{\prime} \cos(\varphi-\varphi^{\prime})^{-\frac{1}{2}}\bigg]
\end{align}
where $\iu\varepsilon$ is a regularization term that is  taken to 0 at the end of any calculation. Two salient features are implicit in
\eqref{FullG}.  One is that $G$ has support inside the entirety of the lightcone; the other is that it diverges along the null $\theta=T$ geodesics. The former is a general feature of Green's functions of scalar fields on curved spacetimes.

% Notes about the structure of Theta and relating it to the Green's function
The dependence of any field-mediated signalling in the response of the second detector is encoded in the smeared field commutator. In the absence of a bandlimit, it is expected that the smeared field commutator becomes the imaginary part of the Green's function $\Theta \sim \Im[G]$.  To see this, consider the pointlike limit of both detectors $\lim_{\kappa_D \to \infty} f^D_{\ell} = \sqrt{\frac{2 \ell +1}{4 \pi}}$. Inserting this into  \eqref{eq:ThetaRegularComm} yields
\eqn{
  \Theta = \frac{i\lambda_A\lambda_B\eta_A\eta_B}{4 \pi} \sum_{\ell=0}^{\infty} P_{\ell}\left(\cos(\theta)\right) \sin{\left((\ell+1/2)T\right)} \sim \Im[G]
  \label{smearcomm}
  }
where $G$ is given by  \eqref{GModes}. 

Now, let us try to relate $\Theta$ to the Green's function $G$ to understand the role of the detector smearing on the production and detection of the signal through the scalar field. We consider the case of two symmetric detectors in the absence of a bandlimit so that the smeared field commutator is given by   \eqref{eq:ThetaRegularComm}.  Structurally, this is nothing other than multiplying each mode of the Green's function by the  product of the modes $f_{\ell}^A f_{\ell}^B$ (since the mode coefficients are real). The convolution of two functions $f,g$ defined on $S^2$, where one of them is azimuthaly symmetric is defined as \cite{Roddy_2021_SphConv}
\eqn{
(f \circledast g)_{\ell m}=\sqrt{\frac{4 \pi}{2\ell +1}} f_{\ell m} g^*_{\ell 0}}
in the harmonic basis. Equation \eqref{GModes} is a spherical convolution between the two functions $(f^A f^B) \circledast G$ defined on $S^2$, where $f^A f^B$ is the product of the Gaussian smearing functions of the detectors A and B and $G$ is the Green's function. This implies that if the field is bandlimited and   coupled to smeared detectors,   signalling from the first detector to the second is encoded in a bandlimited convolution of the product of the detector sizes with the Green's function $G$. This modulates the individual modes of $G$ and controls the strength of the signal inside the lightcone.  For small first detectors, most of the contribution of the signal will be along the lightcone and for large first detectors, the signal trails inside the lightcone.

% Detector sizes and limit of acausal switches
With all of this established, we now  analyze the structure of the smeared commutator. First  we note that it is symmetric under the exchange of detector sizes. This is due to the fact that the localization of the detectors  does not depend on their positions in a stationary spacetime \cite{PhysRevD.101.125005}.  In addition, if we switch the second detector "off time" i.e $\theta \neq T$,
so that the peak switching time of the second detector is not coordinated with the peak of the signal from the first detector,
the modes of the smeared commutator add up to 0 and the $\cos(2\Theta)$ contribution to the second detector's response vanishes. In other words, if the second detector is switched on too early or too late, thereby `missing' the signal from the first detector, it effectively couples to the vacuum and microcasuality is maintained.

\begin{figure}[]
\centering
  \includegraphics[width=0.99\linewidth]{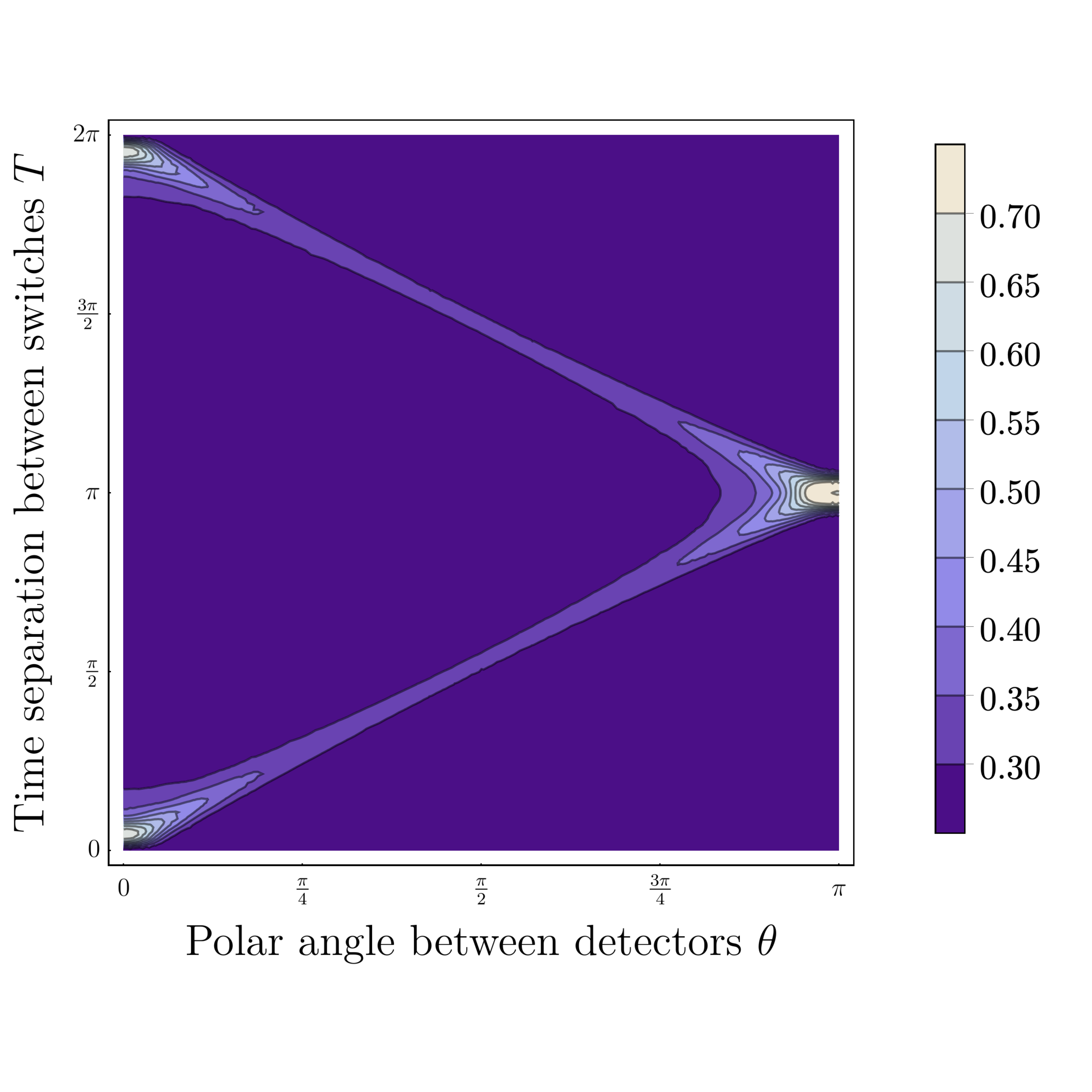}
  \caption{The response of the second detector on the spherical spacetime as a function of spacetime separation between coupling the first and second detectors. Both detectors are equally sized with size $\kappa= 100 \eta$. The dimensionless coupling constant is $\lambda\eta^{1/2}=1$.}
  \label{fig:Lightcone}
\end{figure}

Finally, the smeared commutator is reflection symmetric in time about $t=\pi$, i.e $\Theta(t,\theta,\varphi) = \Theta(t-\pi,\theta,\varphi)$. This property is inherited entirely from the Green's function \eqref{FullG}. This would indicate that any signal from detector A and the response of detector B would also have to follow this reflection symmetry in time. We can make all these features concrete by treating the the second detector as a proxy to the signal from the first. In figure \ref{fig:Lightcone}, we plot the transition probability of a regular detector B of size $\kappa =100$ as a function of the polar angle $\theta$ where it is centered and $T \coloneqq T_B - T_A$, or in other words  where and when it  coupled to the field relative to the first detector. The first detector also has a size of $\kappa=100$ and was coupled to the field at $T_A=0$. First, we see that the signal is concentrated along the lightcone $(\theta=T)$ from the north to the south pole because of the higher excitation probability of detector B. In addition, if the second detector misses the signal from the first detector when switched "off time", then its excitation probability is that of a detector that coupled to the vacuum as discussed earlier, shown as   the dark blue shading in the figure.  Finally, we can see the time symmetry of the signal manifest in the response of the second detector through the reflection along the line $T=\pi$ in the figure.

%%%%%%% Section IV B)
\subsection{Response of two detectors and bandlimit detection on $R \times S^2$}

We now turn our attention to bandlimit detection with two detectors. Can we exploit the field-mediated signalling to find an optimal length scale or an optimal spacetime separation on the spherical spacetime for the detection of the bandlimit?

 In flat space we find that, although the smeared field commutator $\Theta$ depends heavily on the bandlimit, the transition probability of the second detector does not.  Consequently we do not get an large gain in bandlimit detection by employing two detectors. However, when the detectors are on a sphere the signal from detector $A$ does not dissipate (unlike in flat space) and instead becomes focused at the antipodal region. By placing the second detector at that location, we can maximize the effect of the commutator, and hence the sensitivity to the bandlimit.
%\ljh{Moved from the conculsion: We don't consider flat space 2 detector bandlimit detection because just like in $(3+1)$ the effect of the commutator on the transition prob of $B$ is much less important than it's size, so again we can't take advantage of the sensitivity of the commutator to the bandlimit. It's only because the signal is focused at the south pole of a sphere that we are able to boost the relative importance of the commutator and get a much stronger bandlimit detection.  I will add a figure to illustrate this.}

Given the liberty to localize the second detector anywhere on the spherical spacetime, we study the response of the second detector placed  at several distinct latitudes along an arbitrary line of longitude. We will only consider a pair of regular/unsqueezed detectors when analyzing bandlimit detection as per our discussion in section IV. We begin by coordinating the switching time of the second detector with the peak of the signal; in other words we  switch on the second detector on the $\theta = T$ part of the lightcone. We will refer to this as switching "on time". 

\begin{figure}[]
    \centering
    \includegraphics[width=0.49\textwidth]{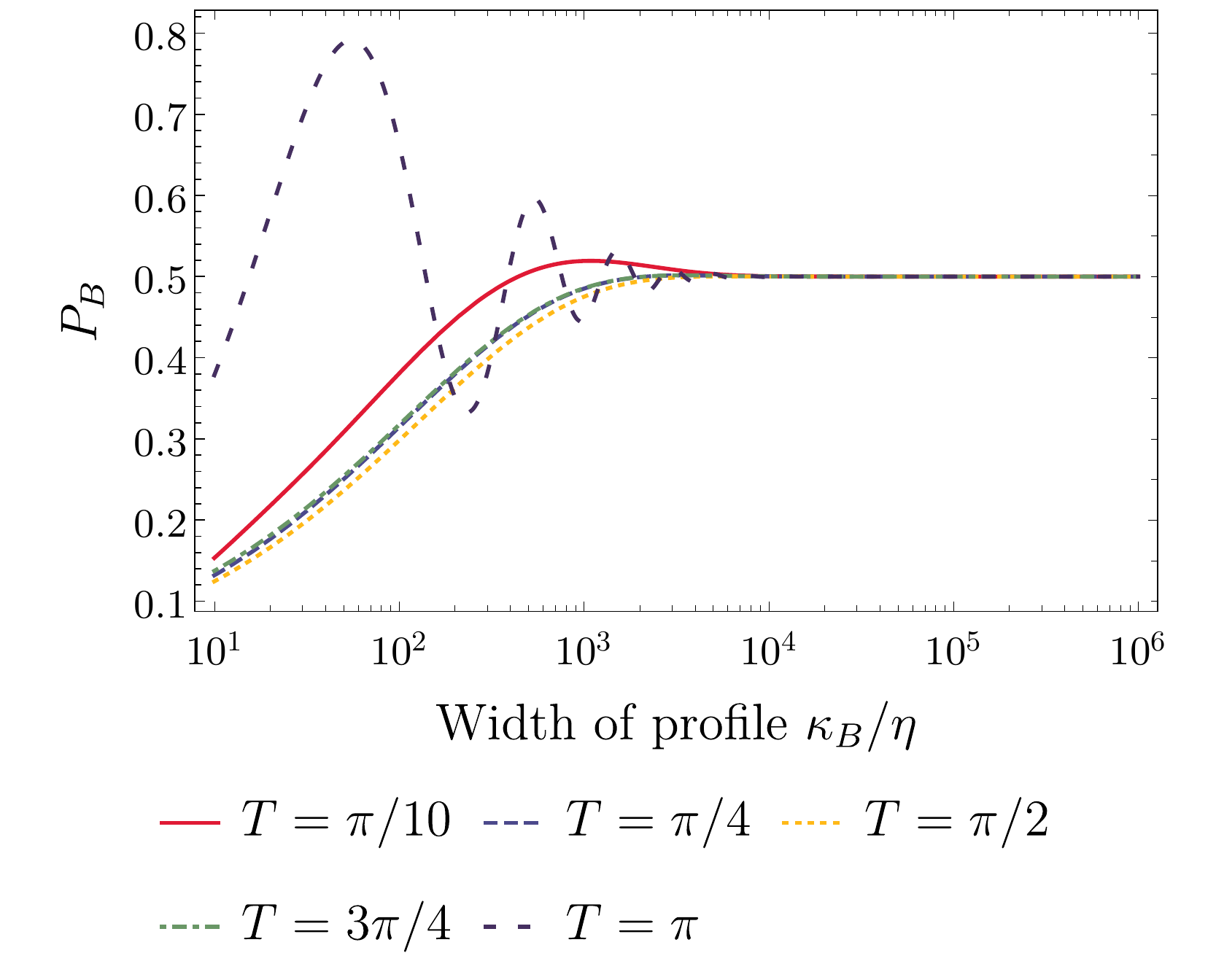}
    \caption{The response of the second detector on the spherical spacetime switched on time as a function of the size of the second detector for various spatial separations between the detectors. The first detector is assumed to be pointlike and the bandlimit is set to $\ell_{max} = 200$. The dimensionless coupling constant is $\lambda\eta^{1/2}=1$.}
    \label{Two-dets-resp}
\end{figure}

\begin{figure*}
    \centering
    \includegraphics[width=0.32\linewidth]{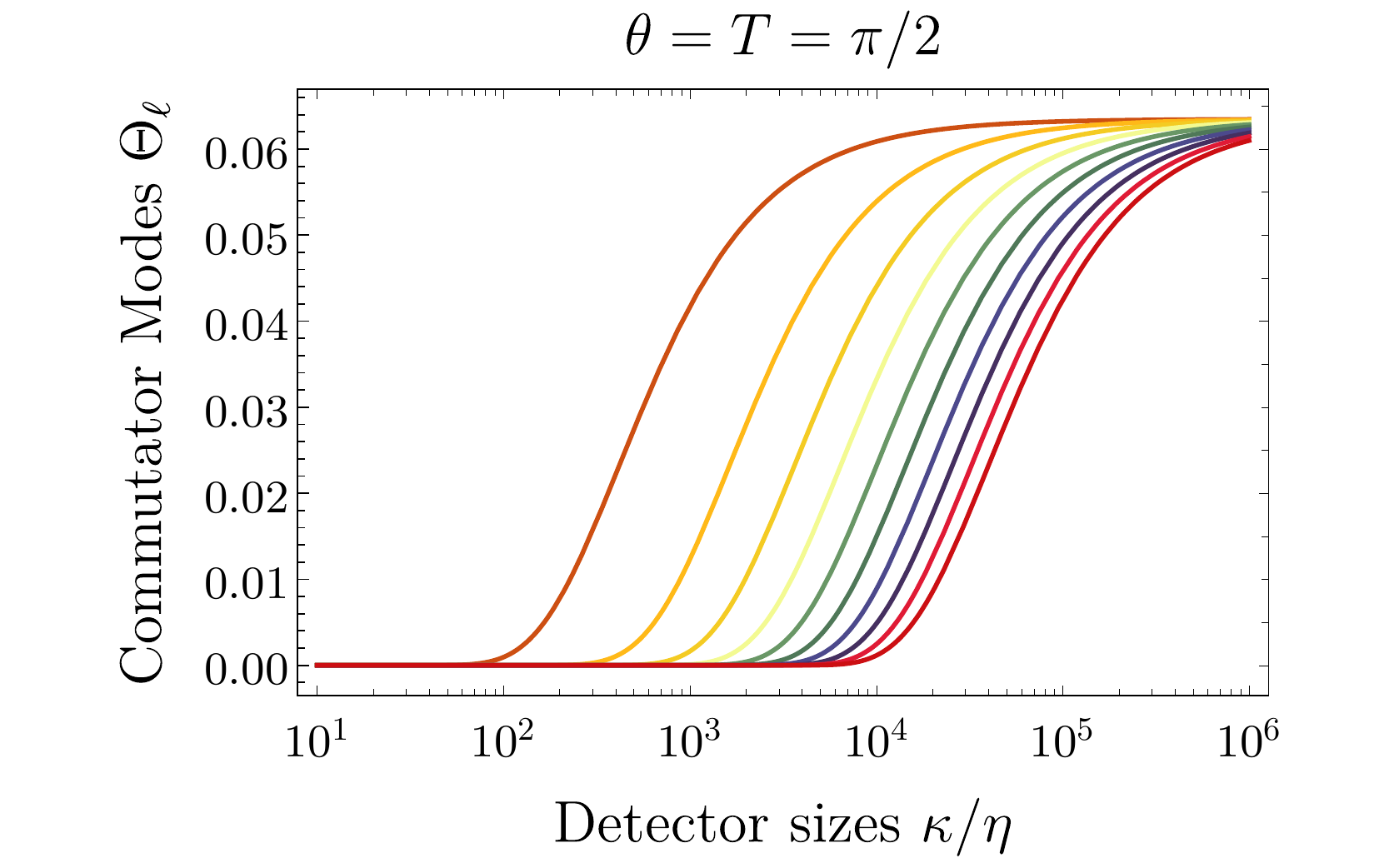}
    \includegraphics[width=0.32\linewidth]{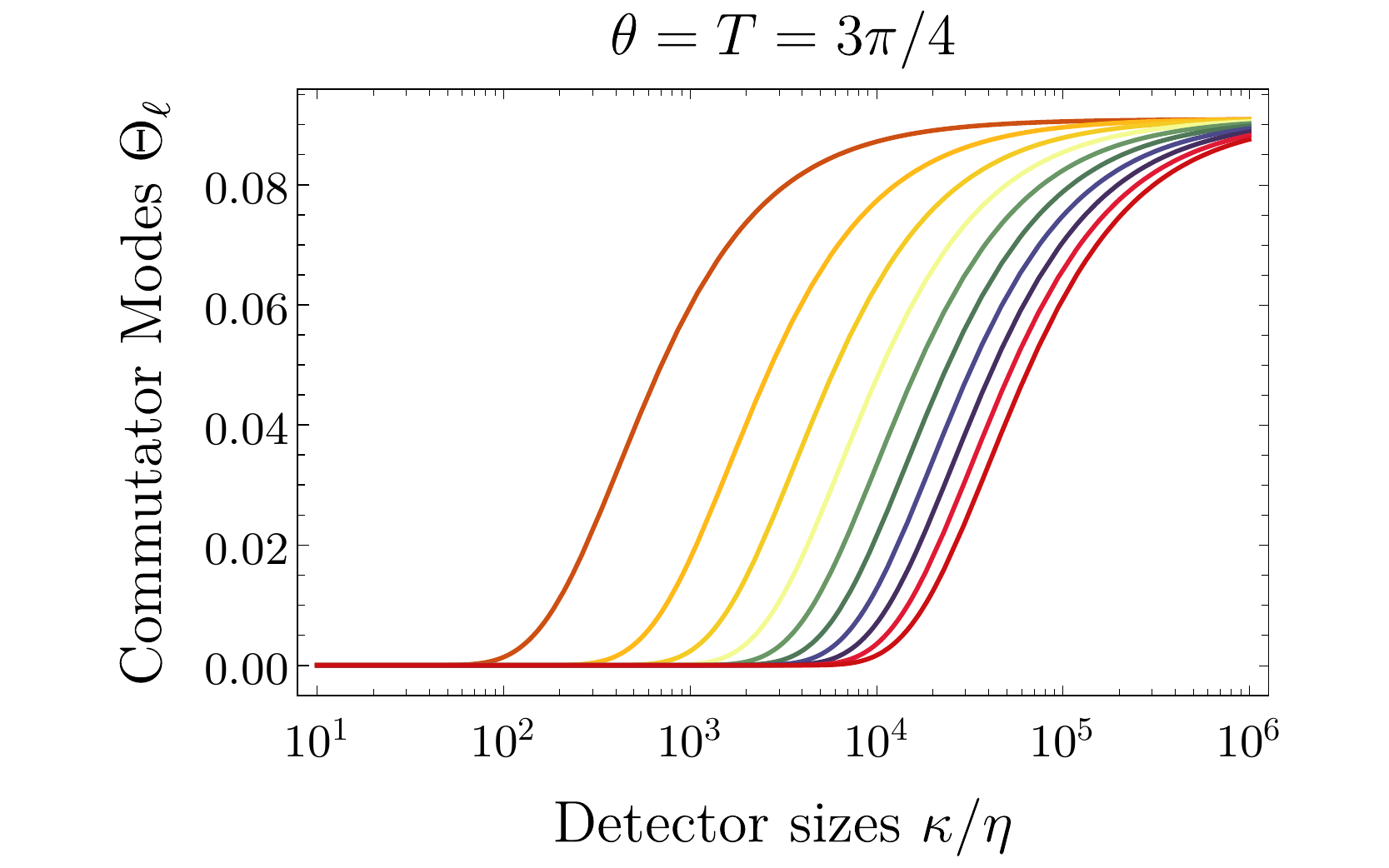}
    \includegraphics[width=0.32\linewidth]{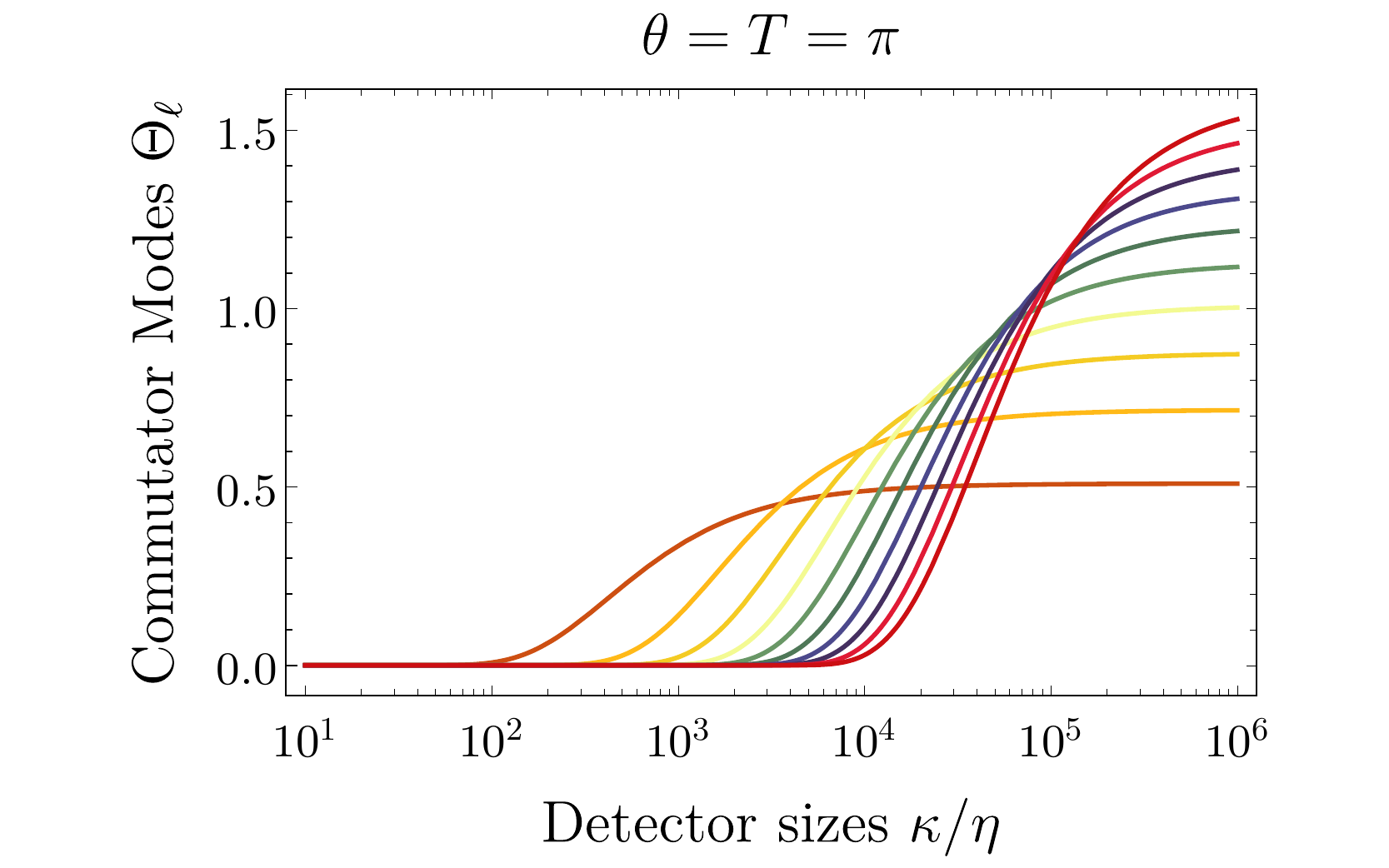}\\
    \includegraphics[width=0.8\linewidth]{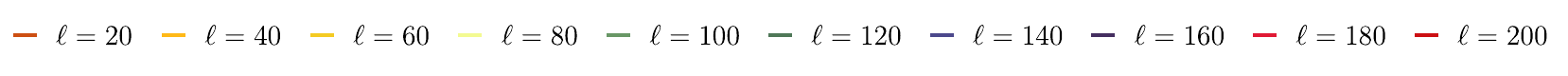}
    \caption{The value of the commutator modes $\Theta_{\ell}$ as a function of the size $\kappa/\eta$ of both detectors at various spatial and switching time separations. The dimensionless coupling constant is $\lambda\eta^{1/2}=1$.}
    \label{fig:comm_modes}%
\end{figure*}

In figure \ref{Two-dets-resp} we depict the response of the second detector B as a function of its size, switched on time at several different values of $T$. Detector A is chosen such that it is point like and we will consider a bandlimit of $\ell_{max} = 200$. As   discussed in section \ref{sec:signalling}, the modes through which the signal propagates depend on the size of the first detector. Consequently, we consider the first detector to be a pointlike so as to not impose an effective cutoff when probing the bandlimit via signalling.
 The response of the second detector is qualitatively similar to that of a single detector, as shown in the $\beta=0$ curve of figure~\ref{fig:1detOnSphere}, for switches on time at $\theta = \frac{\pi}{4},\frac{\pi}{2}$ and $\frac{3\pi}{4}$. We also observe that the response for detectors placed at $\frac{\pi}{4}$ and $\frac{3\pi}{4}$ is identical, as expected due to the symmetry of the spacetime. The slowest increase in response of the second detector is for the $\frac{\pi}{2}$ case. 

The most interesting behavior in the response is for a switch at each pole ($\theta=\pi$).  We see  that it can be larger than $1/2$,  oscillating as a function of the detector size. In the  pointlike (large-$\kappa$) limit, the response converges to $1/2$. This is because the support of the smearing of detector B becomes too small to  resolve the effects of curvature of the spacetime on the field; as expected, its response asymptotes to the pointlike response regardless of where and when it coupled to the field. 

It is worth reiterating that since the spacetime is stationary with constant curvature then the spacial  profiles of the   detectors do not depend on their location in the spacetime.
%That is, our "probe" (the second detector) \blue{of the signal from the first detector are the same regardless of where and when the second detector couples to the field. \rbm{I don't understand this. Fig 5 indicates otherwise.}\ahs{The point being made here is not that the response does not change depending on where the detector is. All whats is being said here is that the detectors are going to be identical everywhere and will couple locally in the same way. Therefore the difference maker is how the signal spreads and focuses.}
This implies that the differences in the behavior of the response of detector B can be attributed to the field mediated signalling. In particular, due to the compact and bounded nature of the spacetime, the various switches in figure \ref{Two-dets-resp} interact with the signal through the field as it spreads from detector A all the way down as it gets focused at the south pole. %\rbm{I don't quite understand what is being said in this paragraph.}\ahs{Hopefully it is more clear now}

%\begin{widetext}
%\begin{figure}[H]
%    \includegraphics[width=0.99\textwidth]{Commutator mode-theta-3.14159-T-3.14159.pdf}
%    \caption{Individual commutator $\ell$-modes $\Theta_{\ell}$ as a \blue{function of the size $\kappa$ of both detectors for values of $\ell$ from $20$  to $200$. The figures were evaluated at $\theta=T=\pi/2$ and at $\theta=T=\pi$ respectively }}
%    \label{fig:comm_modes}%
%\end{figure}
%\end{widetext}

The geometric notion of the field-mediated signal spreading and focusing manifests itself through the smeared field commutator and its modes. To this end, let us analyze the dependence of the smeared commutator on the individual $\ell$ modes. In figure \ref{fig:comm_modes}a,  we plot $\Theta$ as a function of $\kappa_B$ for various values of $\ell$ modes of the smeared commutator $\Theta$,  denoted by $\Theta_{\ell}$,  in increments of $20$ when  the second detector is switched on time at $\pi/2$. This choice  corresponds to a maximally spread  signal. However we also find that the pattern in figure \ref{fig:comm_modes}a holds
as the on time switch increases to $3\pi/4$
(as the signal becomes more focused), as shown in figure \ref{fig:comm_modes}b. We see that  all the commutator modes $\Theta_{\ell}$ asymptote to a value $\tilde{\Theta}(\theta)$ regardless of the $\ell$ mode in the pointlike limit $\kappa_B\to \infty$. Moreover,  higher $\ell$ modes are suppressed at small $\kappa_B$, and only contribute to the smeared commutator $\Theta$ on length scales where the exponential suppression of the local coupling term is strong.  Ultimately, the smeared commutator  ($\Theta(\theta,T) = \sum_{\ell = 0}^{\ell_{max}} \Theta_{\ell}(\theta,T)$) for all $\theta=T$ configurations where the signal is  not entirely focused provides small corrections  to the response of the second detector. Consequently, this provides us with a nominal improvement to bandlimit detection at best. %\rbm{The notion of a commutator $\ell$-mode hasn't been clearly defined. Is it the cumulative contribution of all $\ell$'s up to $\ell$-max, or is it just the contribution of a single $\ell$? I used $\ell$-max, but maybe this is wrong.}\ahs{Fixed, it is the $\ell$ modes not their sum}

\begin{figure*}
    \centering
    \includegraphics[width=0.49\linewidth]{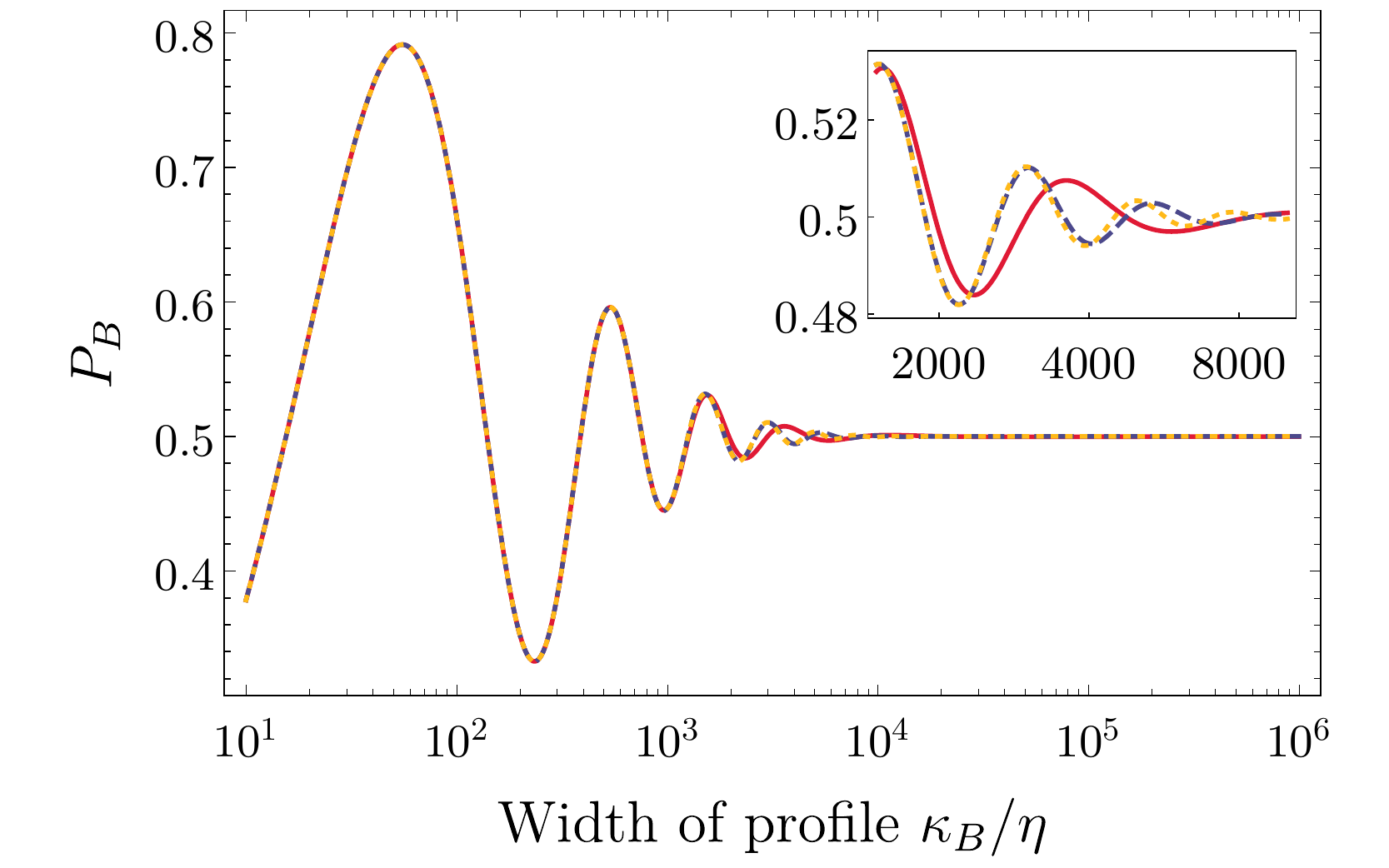}
    \includegraphics[width=0.49\linewidth]{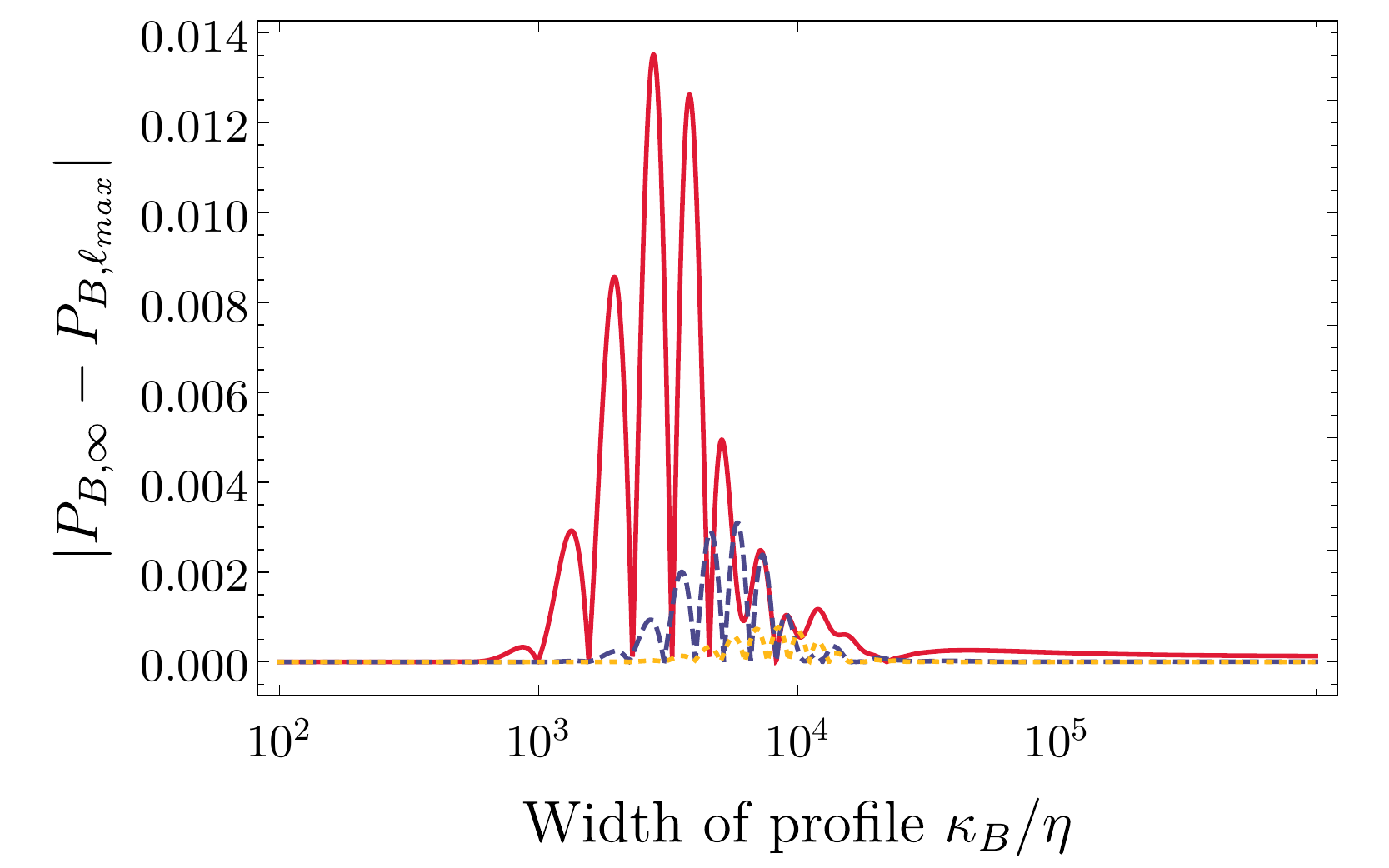}\\
    \includegraphics[width=0.49\linewidth]{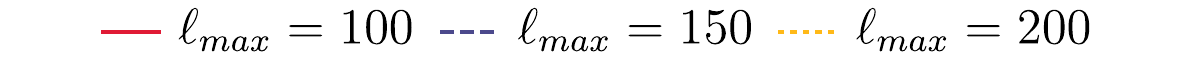}
    \caption{\textit{Left}: The transition probability $P$ of a second regular detector switched on time at the pole  as a function of its size for various values of the bandlimit $\ell_{max}$. \textit{Right}: The absolute difference bandlimit detection criterion for a regular second detector localized at the pole as a function of its size for several values of the bandlimit $\ell_{max}$. For both plots, the dimensionless coupling constant is $\lambda\eta^{1/2}=1$.}
    \label{fig:Two dets}%
\end{figure*}

The situation as the second detector is  switched on time at $\pi$ (at the south pole)  is different. As we can see in figure \ref{fig:comm_modes} c),  the individual contributions to the commutator are two orders of magnitude larger on average than those at maximal spread, shown in \ref{fig:comm_modes}a.  Moreover, the maximum value a mode has as a function of size increases as $\ell$ increases. So the commutator increases rapidly as a function of decreasing size (increasing $\kappa_B$), causing  the signalling part of the response ($\cos(2{\Theta})$)  to oscillate more rapidly. While this accounts for the behavior of the transition probability of the second detector, the rate of   increase in the value of the commutator as a function of  decreasing size  also depends on the value of the bandlimit $\ell_{max}$. In left of figure \ref{fig:Two dets} we plot the response of the detector switched on time at $T = \pi$ against its size.  We see for sufficiently small but not pointlike detectors that oscillations in the response become distinct for different bandlimits. This offset in frequency  could provide a promising candidate for more optimal bandlimit detection, since the response for a given size becomes more dependent on the bandlimit over the this intermediate range of detector sizes.

 With all of this established, we now know
 the optimal location for B that maximizes the  contribution of the signal to B's
 response.
 %the spacetime separation of the detectors that amplifies the signalling contribution to the response of detector B. \rbm{What does that mean? Does it mean we know the optimal location of B relative to A to maximize the signal?}\ahs{We know the optimal location for B to maximize the \textit{contribution} of the signal to the response}
 We now consider the optimal sizes of the detectors.  In the previous subsection, we saw how the size of the first detector dictates the modes through which the signal propagates. Ideally, detector $A$ should be as small as possible, so as to allow detector B full sensitivity to an arbitrary value of the bandlimit. As such, we assume detector A to be pointlike.
 
%  \begin{figure}[H]
%     \centering
%     \includegraphics[width=0.49\textwidth]{Figures- Results - Sphere - 2dets/Two dets-response at pi - pointlike A-Final.pdf}
%     \caption{The transition probability $P$ of a second regular detector switched on time at the pole  as a function of its size for various values of the bandlimit $\ell_{max}$.} 
%     \label{Two dets resp at pi}%
% \end{figure}

% \begin{figure}[H]
%     \centering
%     \includegraphics[width=0.99\linewidth]{Figures- Results - Sphere - 2dets/Optimal size BL abs detection-kappa A-pointlike- theta-T-3.14159.pdf} 
%     \caption{The absolute difference bandlimit detection criterion for a regular second detector localized at the pole as a function of its size for several values of the bandlimit $\ell_{max}$}
%     \label{BL det at the pole}%
% \end{figure}

\begin{figure*}
    \centering
    \includegraphics[width=0.20\linewidth]{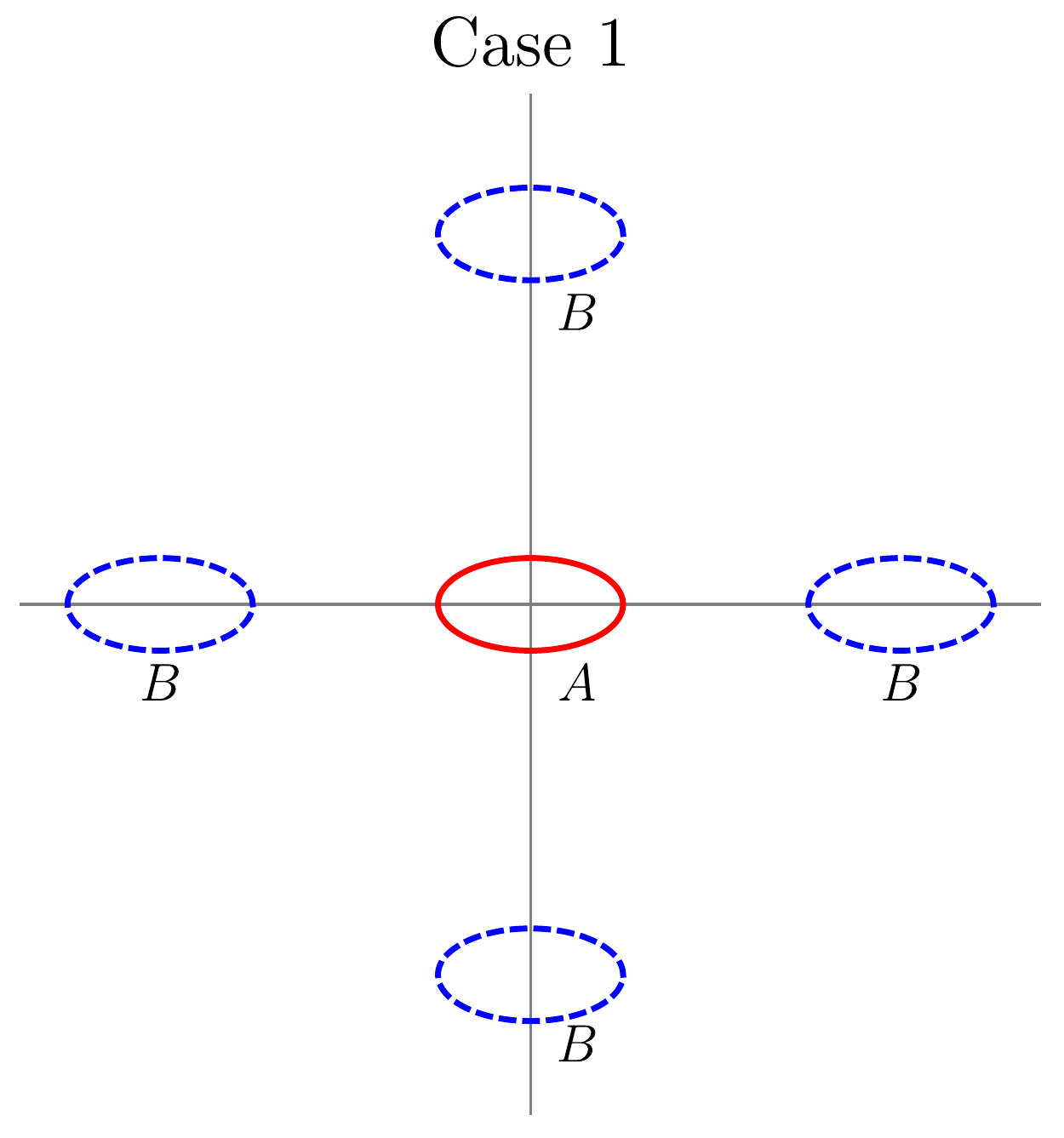} \hspace{5pt}
    \includegraphics[width=0.30\linewidth]{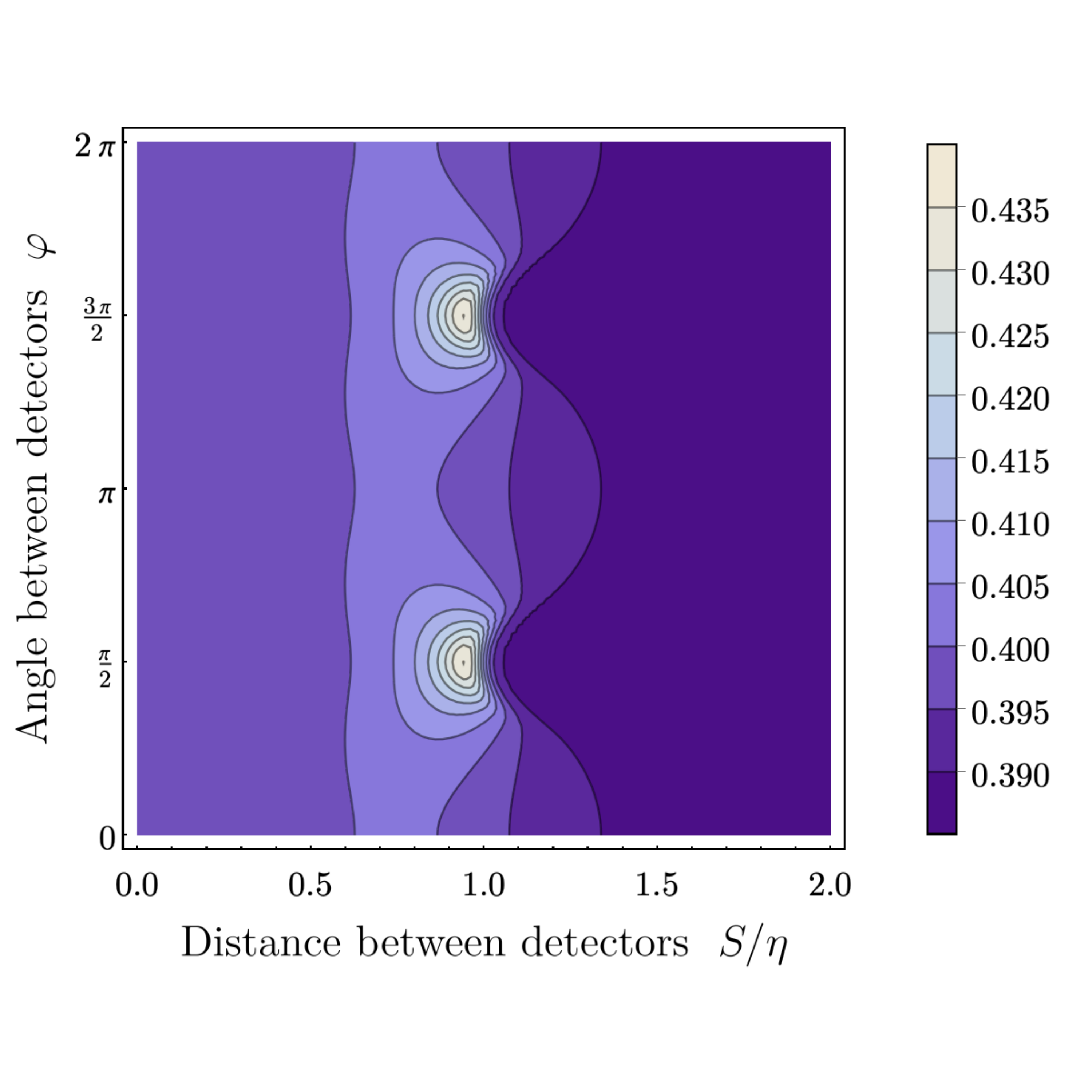} \hspace{5pt}
    \includegraphics[width=0.30\linewidth]{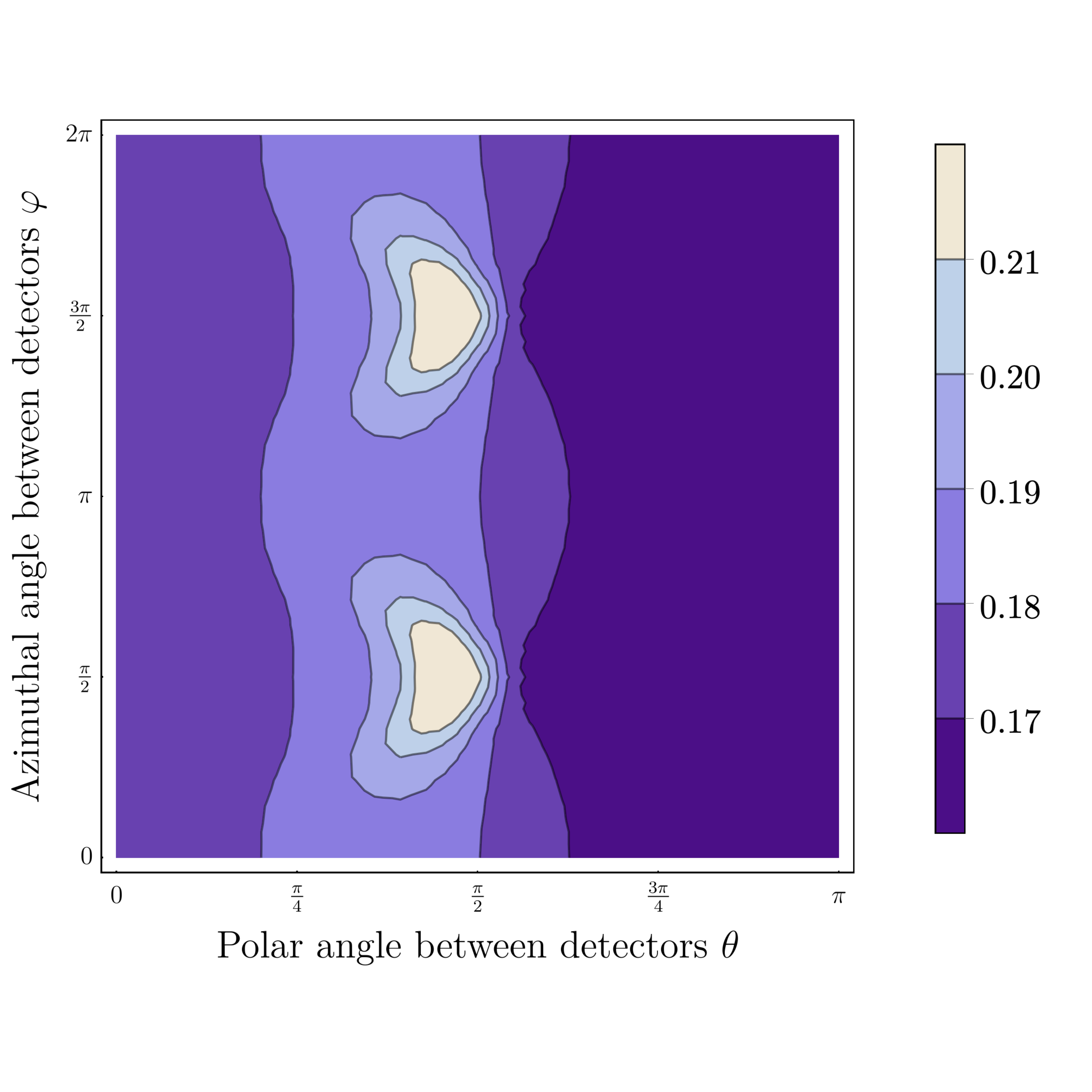}\\
    \includegraphics[width=0.20\linewidth]{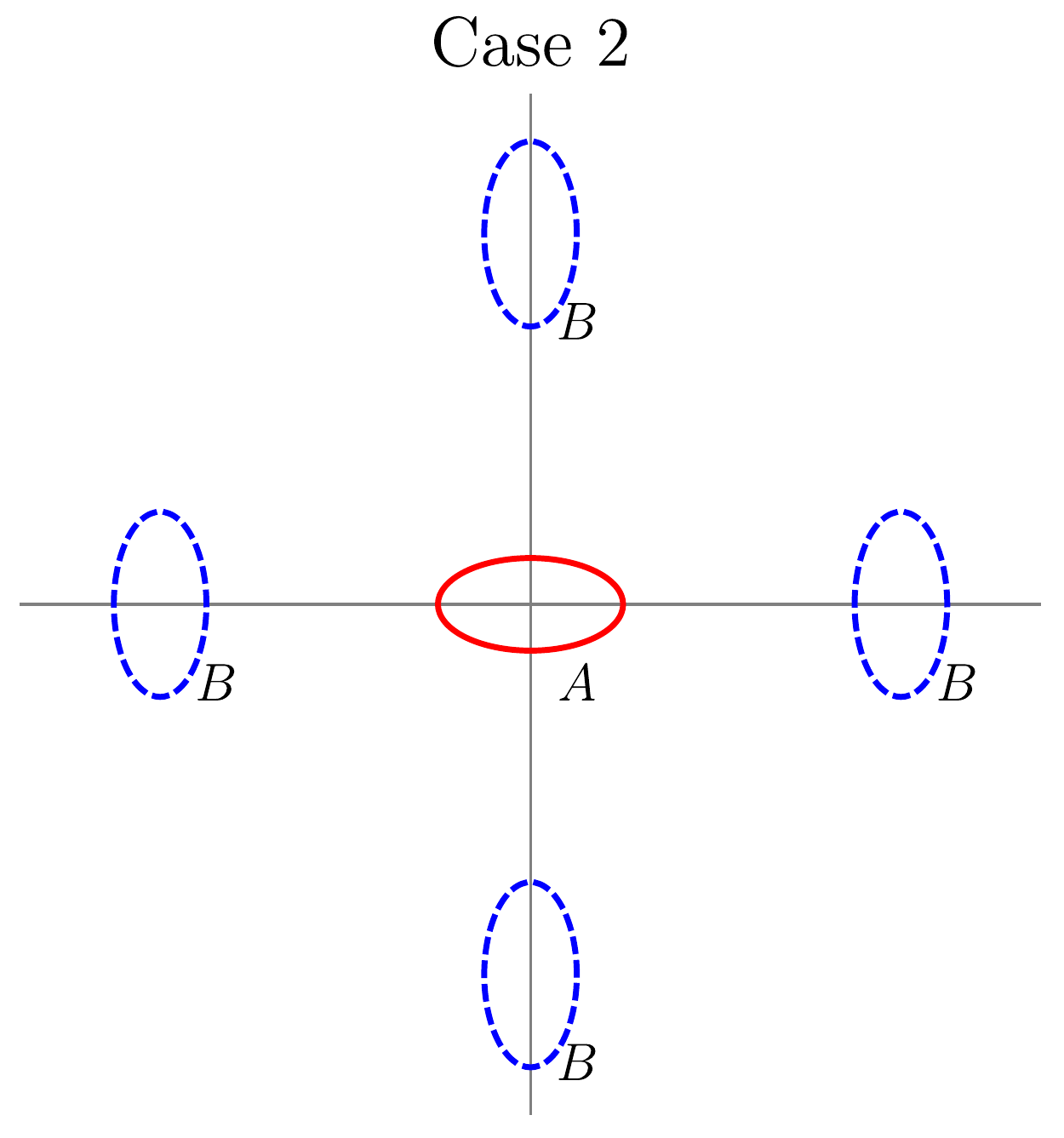} \hspace{5pt}
    \includegraphics[width=0.30\linewidth]{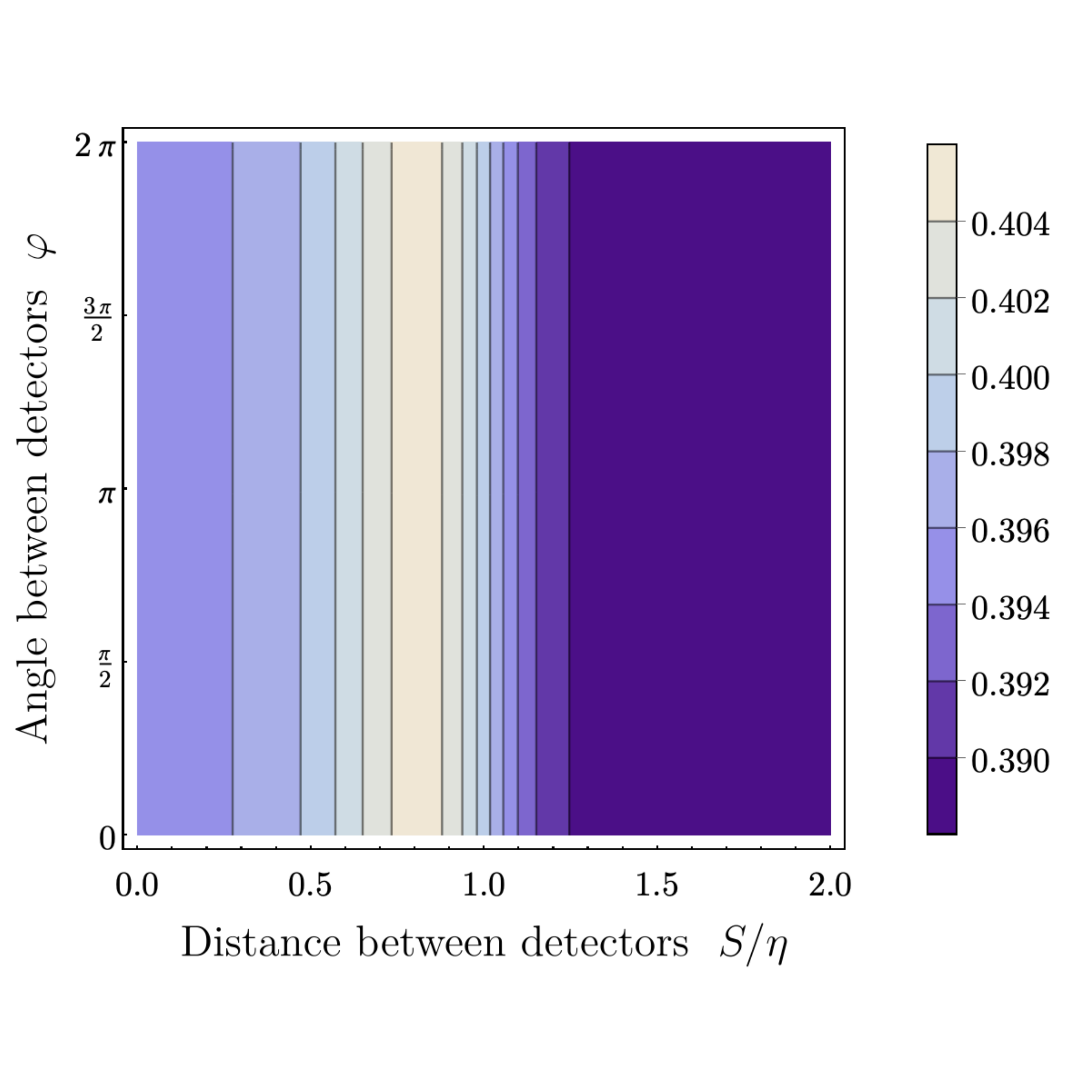} \hspace{5pt}
    \includegraphics[width=0.30\linewidth]{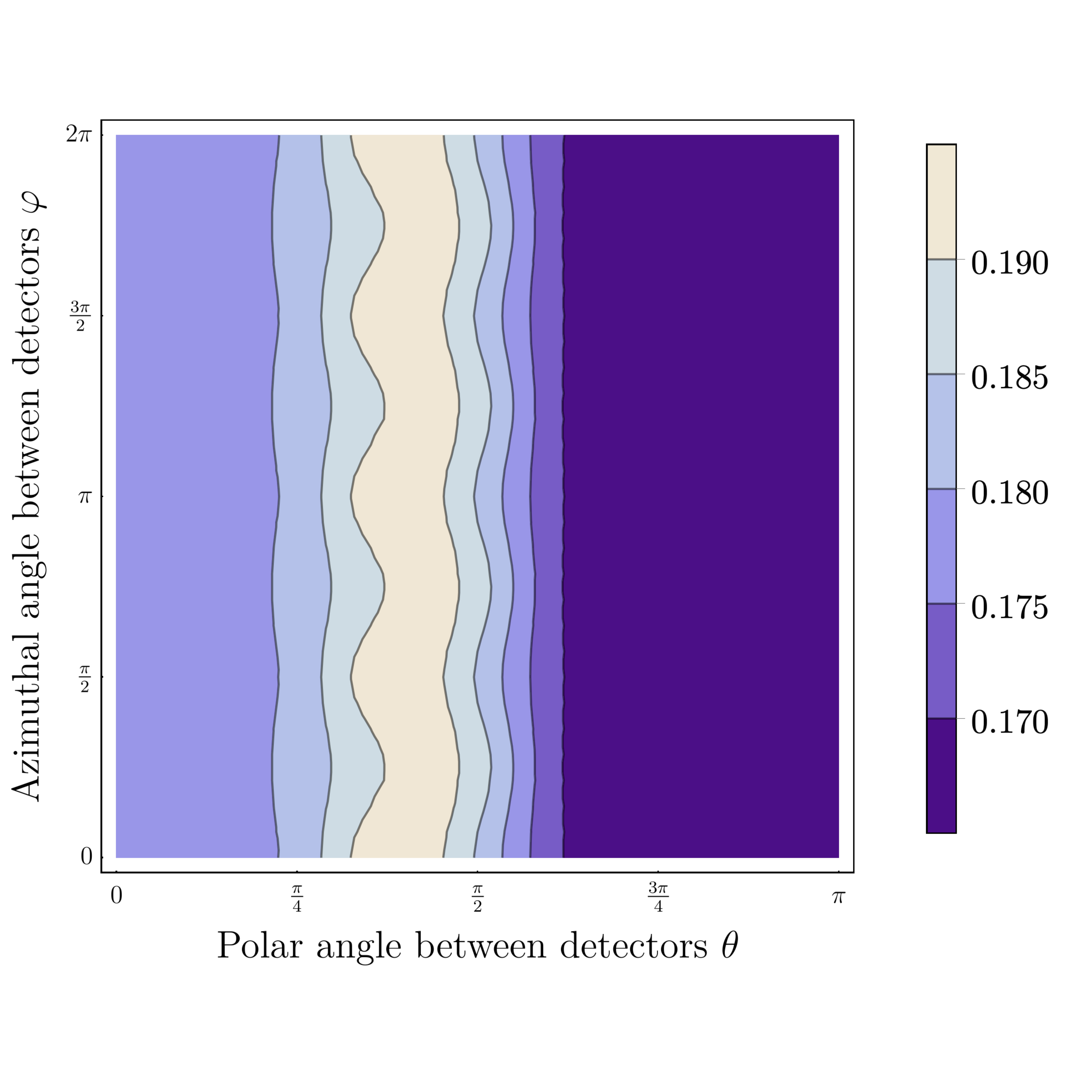}\\
    \includegraphics[width=0.20\linewidth]{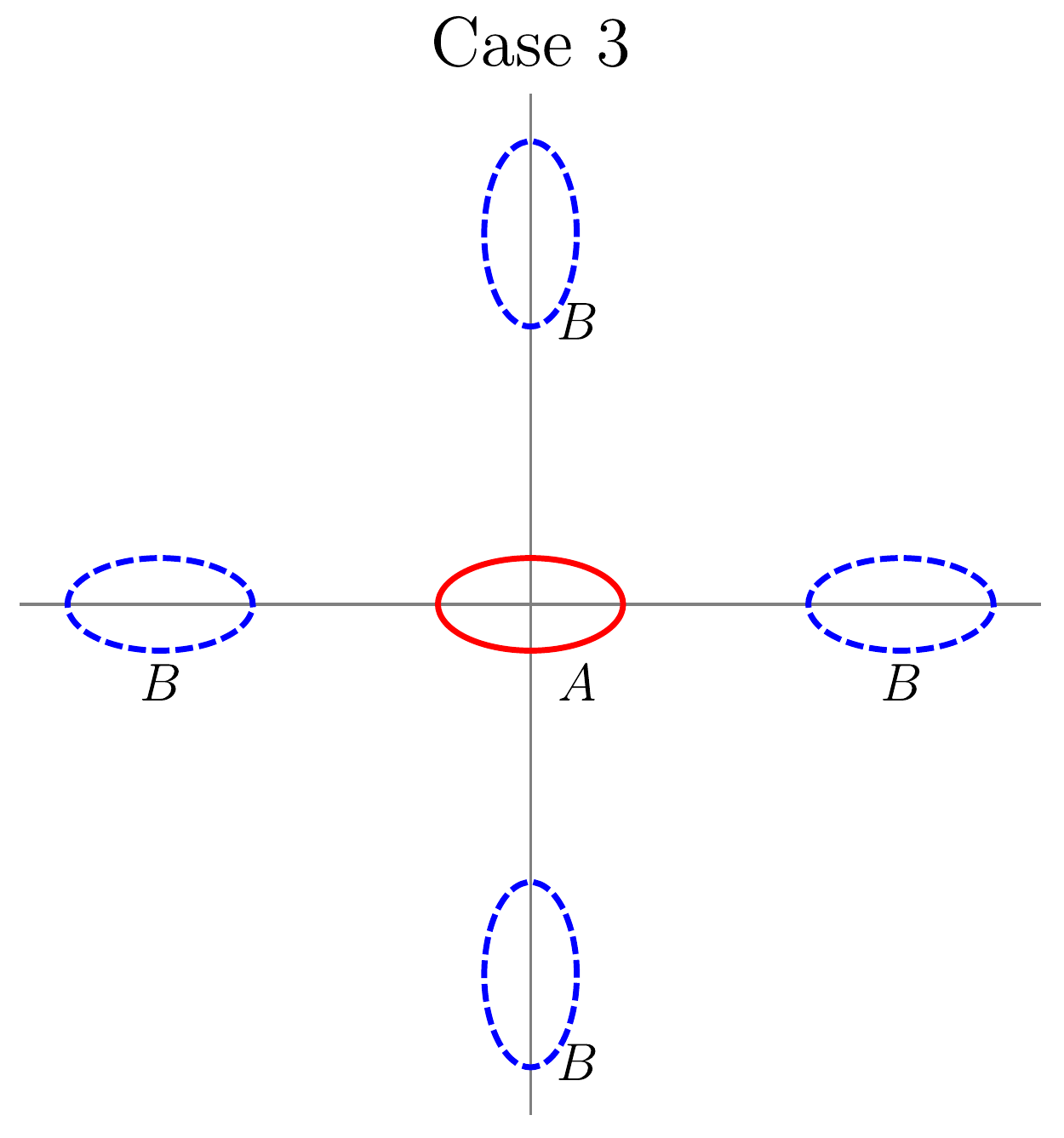} \hspace{5pt}
    \includegraphics[width=0.30\linewidth]{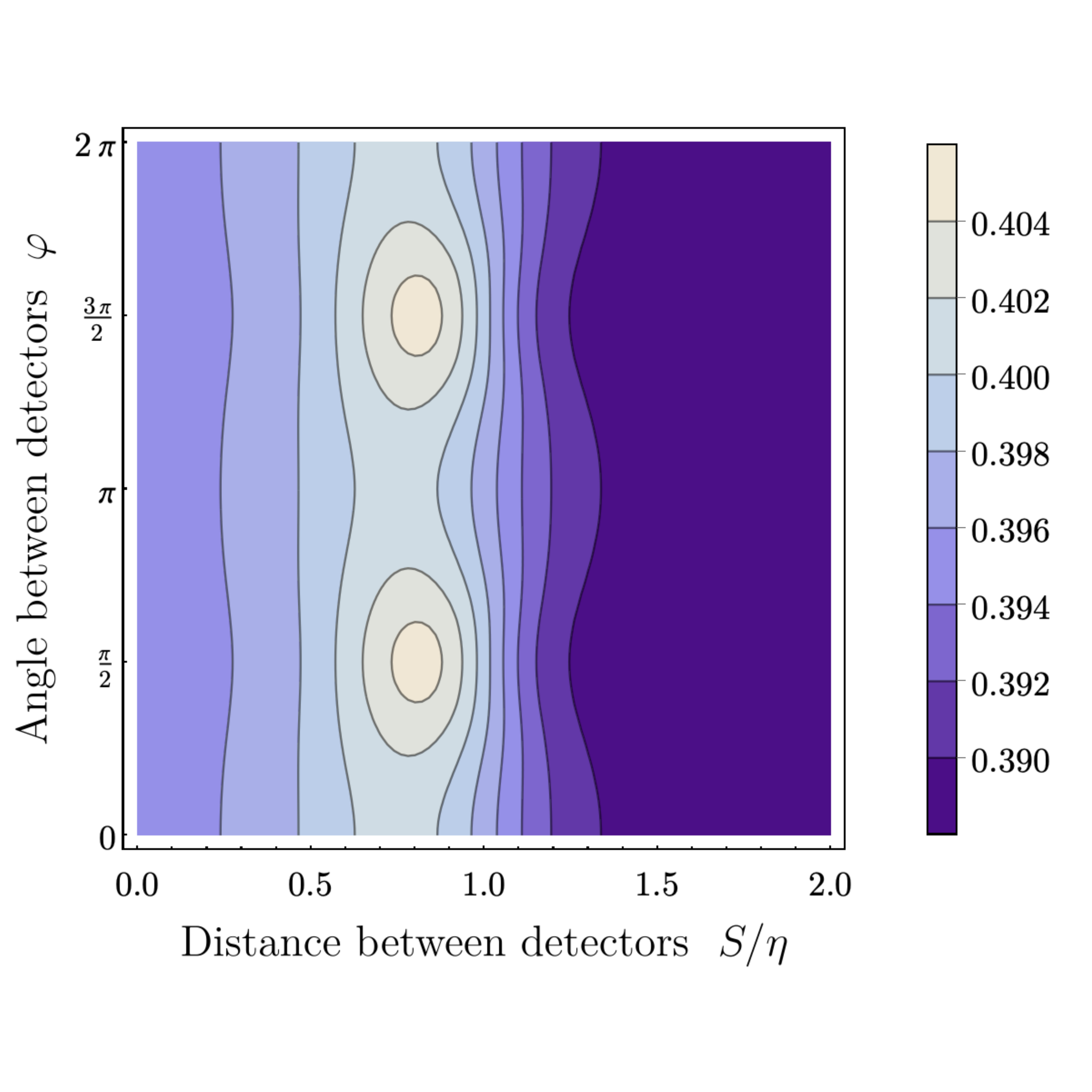} \hspace{5pt}
    \includegraphics[width=0.30\linewidth]{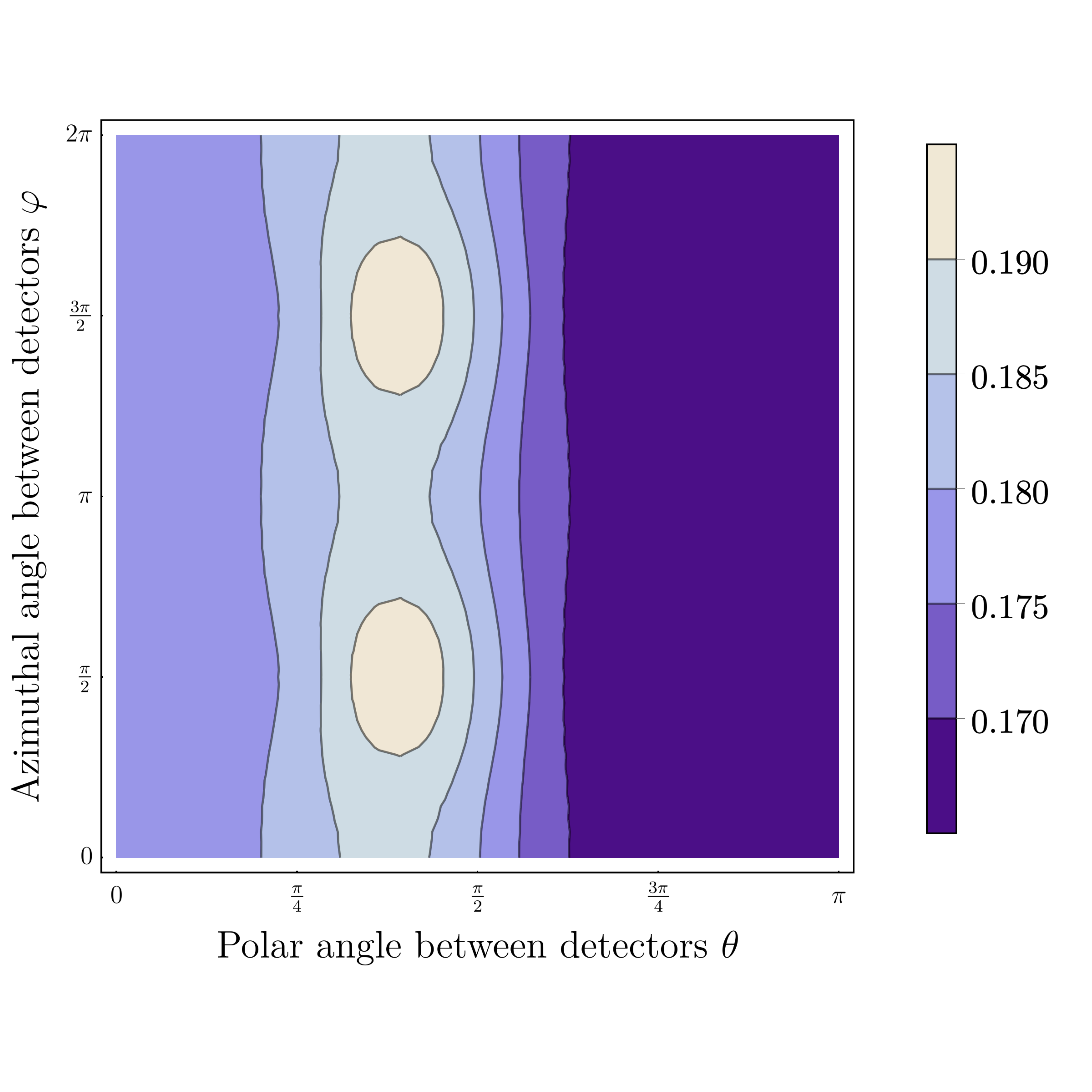}\\
    \includegraphics[width=0.20\linewidth]{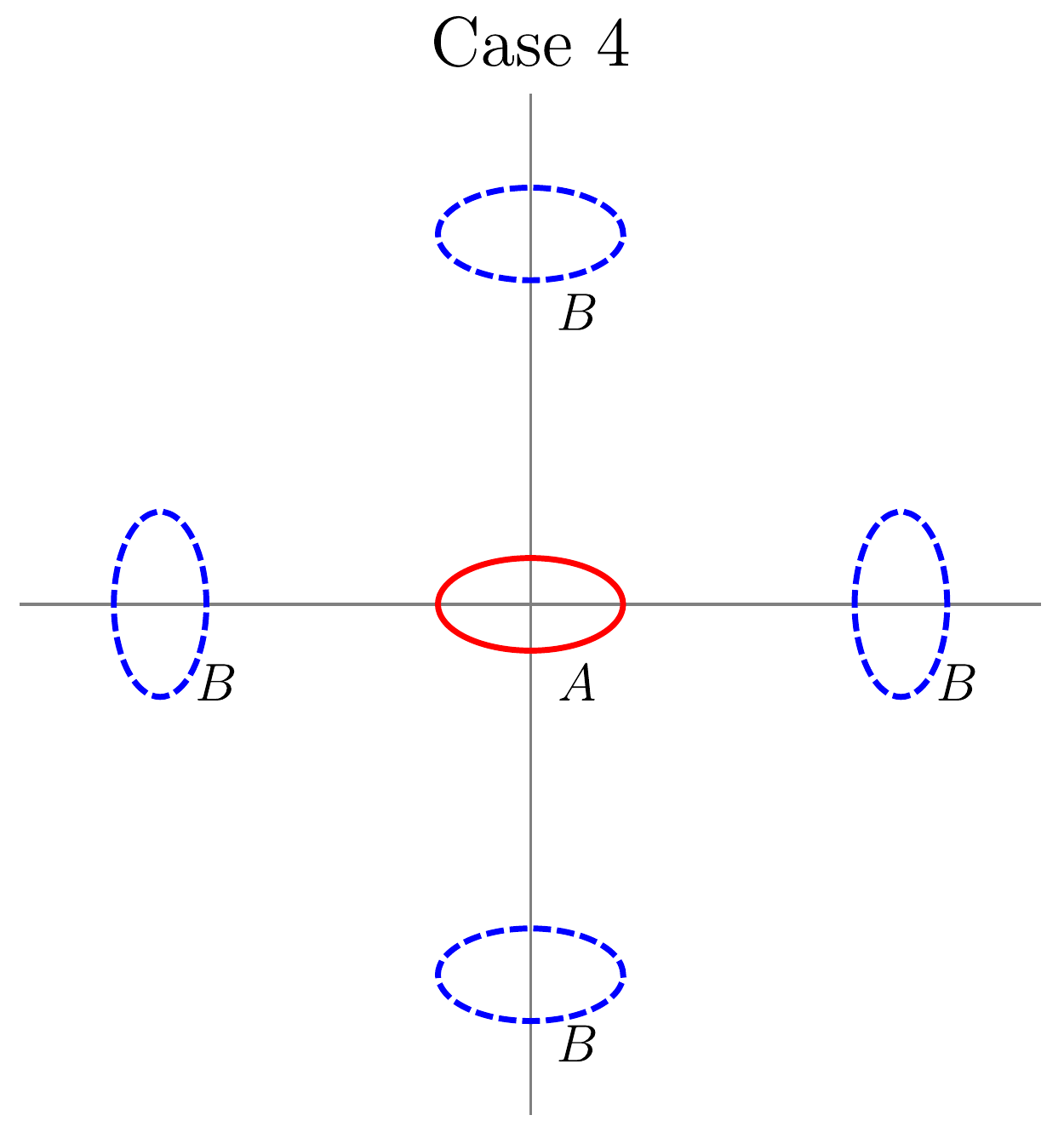} \hspace{5pt}
    \includegraphics[width=0.30\linewidth]{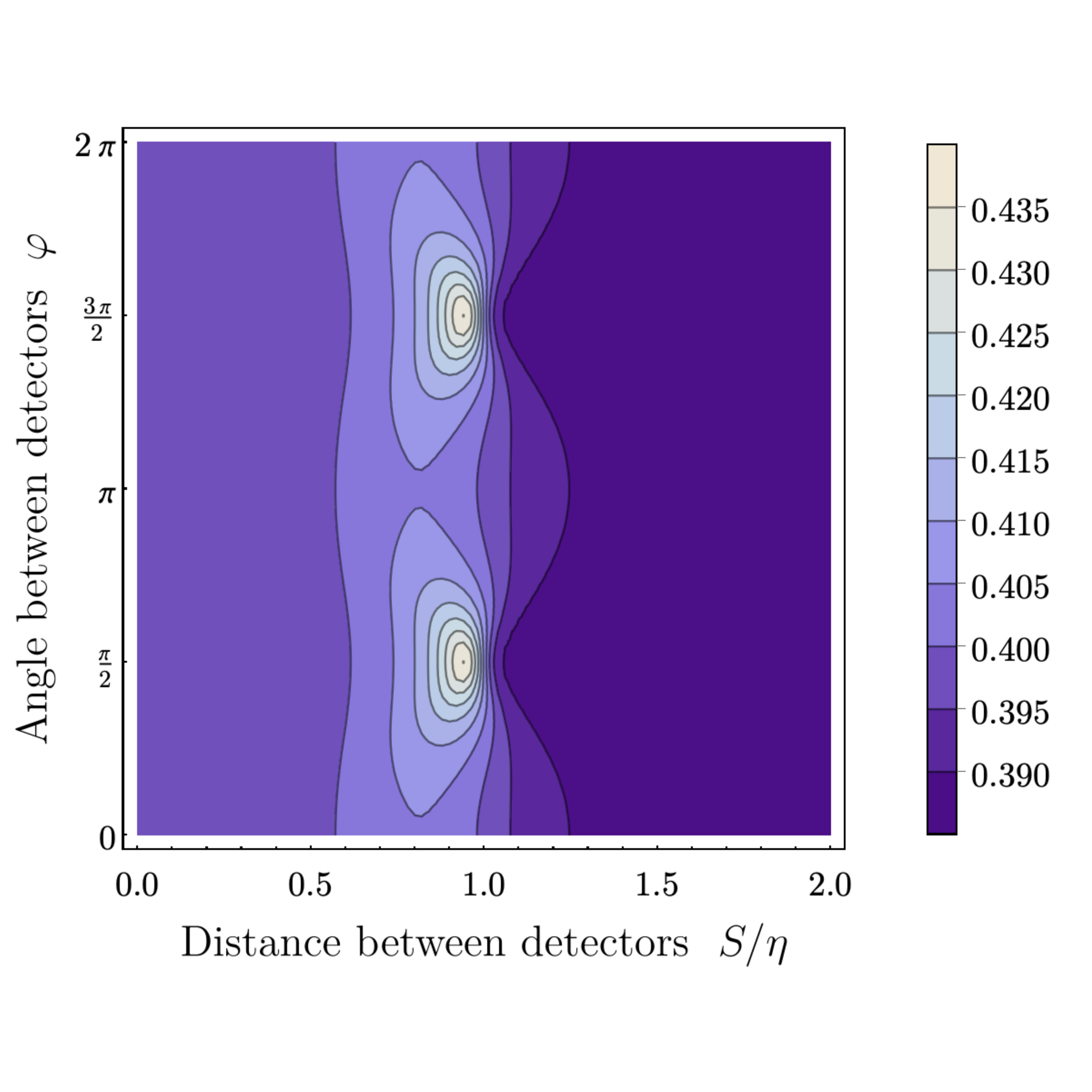} \hspace{5pt}
    \includegraphics[width=0.30\linewidth]{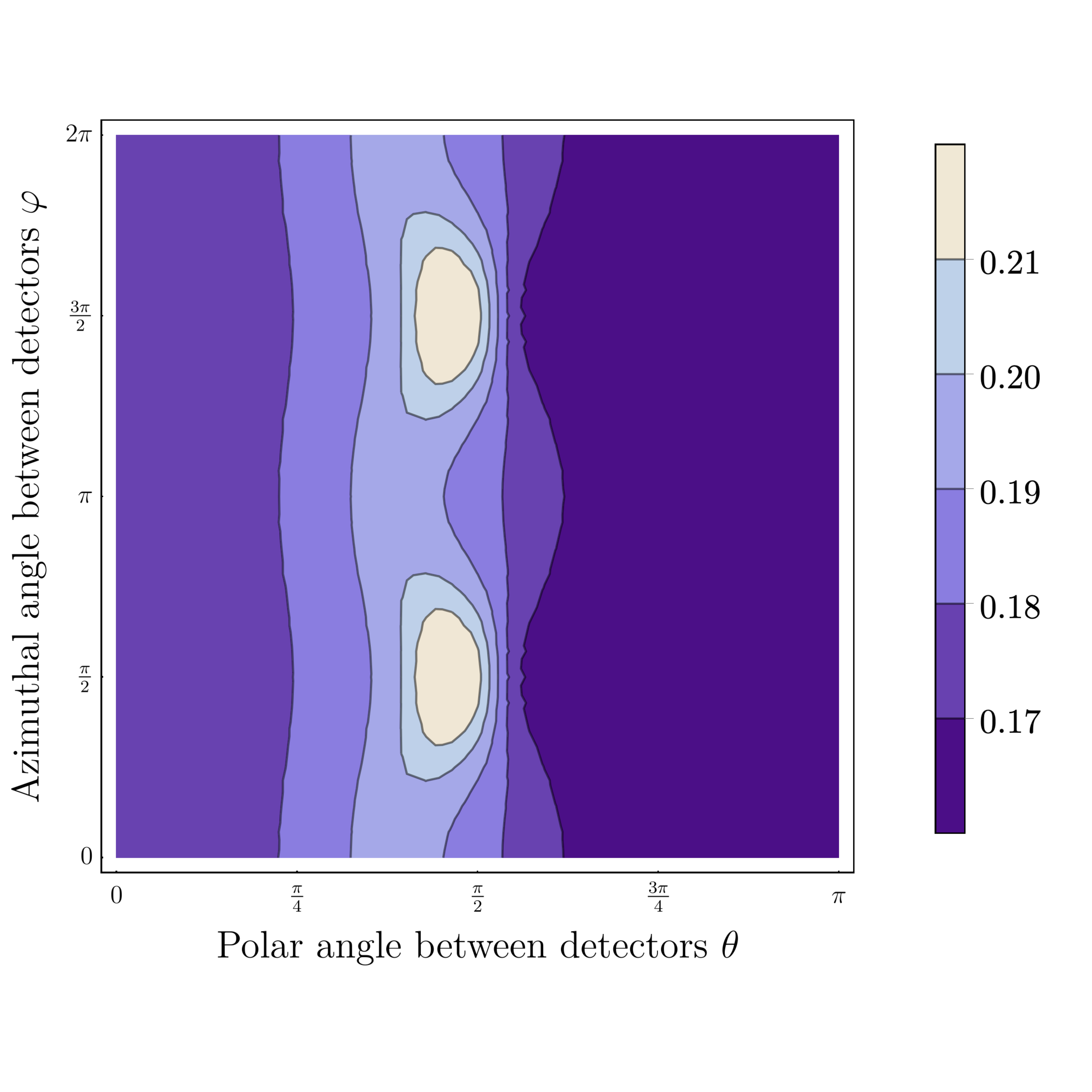}
    \caption{%
        \textit{Left}: A schematic depicting the position of detector $A$ and various positions of detector $B$ for the four types of configurations considered, where each case is a new row.  When the detectors are on the sphere, these figures are interpreted as top-down view of the north pole.
        \textit{Centre}:  The transition probability of detector $B$ in flat space as a function of the separation of the centres of the detectors, $S$, and the angle between detectors, $\vph$.  The time delay is $T=\eta$ and the detectors have a width of $a=0.2\eta$ an eccentricity of $\epsilon=0.99$.
        \textit{Right}:  The transition probability of detector $B$ in spherical spacetime as a function of the polar $\theta$ and azimuthal $\varphi$ separation between the centres of the detectors.  The time delay is $T = \pi/2$ and the detectors have a size $\kappa=100$ and a squeezing of $\beta=50$.
        }
    \label{fig13BothSqueezed}
\end{figure*}

Recall from section IV that the 
large $\kappa$ limit
$$
\lim_{\kappa_A \to \infty} f_{\ell}^A
= \sqrt{\frac{2\ell +1}{4 \pi}}
$$
is the point like limit, with 
the detector modes are given as above.  The smeared commutator in the pointlike limit is
\eqn{
\lim_{\kappa_A \to \infty}{\Theta} = \sum_{\ell=0}^{\ell_{max}} \frac{f_{\ell}^B P_{\ell}(\cos{\theta})}{\sqrt{\pi(2\ell +1)}}\sin((\ell+\frac{1}{2})T).}

In right of figure \ref{fig:Two dets}, we plot for different $\ell_{max}$ the  bandlimit detection parameter for detector $B$ as a function of its size $\kappa_B$ when it  is localized at the south pole  and switched on time ($\theta = T = \pi$), with the first detector pointlike. We highlight several key features. First, the detection parameter is as much as full order of magnitude larger than its single-detector counterpart,   shown in figure \ref{fig:Optimal-BL-D1-S2}. In addition, there is no longer a fixed optimal size for bandlimit detection:  the range of optimal length scales now depends on the value of the bandlimit.  Within this range, the  bandlimit detection parameter oscillates. The origin of both effects is rooted in the reason as to why the response oscillates in the first place.  As $\kappa_B$ increases,  $\Theta$ likewise increases, and so  
$\cos(2\Theta)$ can oscillate over several periods.   % \rbm{And what is that reason?}\ahs{Addressed}. 
This implies that the bandlimit detection criterion using a second detector can be recast as finding the length scale at  which the rate of change of the sum of the $\ell$ modes of the commutator increases the fastest as a function of the detector size.

In flat spacetime  the dominant effect on bandlimit detection is the local coupling of the detector as governed by its size   \cite{Henderson:2020ucx}.
  With two detectors, the smeared commutator is much more sensitive to the bandlimit,  but   gets exponentially suppressed by the local detector coupling. On the spherical spacetime we have engineered a situation where the commutator  grows  rapidly as  the bandlimit increases for all sizes %\rbm{What does that mean?}\ahs{clarified} 
such that the  $\cos(2 {\Theta})$ signalling term rapidly oscillates, counteracting the exponential suppression from the local detector coupling term that happens before the pointlike limit of the detector is reached. In other words, we have engineered a situation where the sensitivity of the commutator to the bandlimit is manifest.

%The notion of optimal bandlimit detection being tied to a length scale where the effects of the background curvature of the spacetime can be "resolved" or felt by the second detector. 

%The focusing of the signal geometrically, can be understood in terms to what happens to the smeared commutator and its modes along different points on the spacetime. Around the southpole for $\theta = T$, the modes are two orders of magnitude bigger than when initialized at the equator, where the signal is the most spread. The commutator becomes unbounded as $\ell_{max}$ increases in this case. The response depends on the cosine of the value of the smeared field commutator, which oscillates 

%This means that the commutator is ultimately even more sensitive to the bandlimit but the issue is that the local length scale of the detector i.e. its size dominates the response. Ultimately, what is necessary for the smeared commutator to manifest its sensitivity to bandlimit detection is to find the spacetime separation such that 

%%%%%%%%%%%%%%%%%%%%% 2 squeezed detectors results %%%%%%%%%%%%%%%%%%%
\subsection{Squeezed Detectors}

To complete the exploration of the parameter space, we now allow both  detectors to be
in squeezed states. They can then be
rotated relative to each other. We shall consider the effects of such rotations on the transition probability of the second detector.

We first identify  four distinct detector  configurations; taken together these  allow for a robust exploration of the parameter space.   Without loss of generality we  orient the coordinate system so that the semi-major axis of detector $A$ is along the $x$-axis in Minkowski space and at $\vph=0$ on a sphere.  Placing the centroid of detector $B$ at a fixed distance from that of $A$, 
its  semi-major axis  
is then taken to be either parallel or perpendicular to that of detector $A$, resulting in the four configurations shown in the left column of
figure \ref{fig13BothSqueezed}. 
  In case 1, the semi-major axis of $B$ is always parallel to the semi-major axis of $A$.  In case 2, the semi-major axis of $B$ is always perpendicular to $A$.  In case 3, the semi-major axis of $B$ always points toward the centre of $A$ and in case 4, the semi-minor axis of $B$ always points toward the centre of $A$.  
 For the $ R \times \mathbb{S}^2$ spacetime  we set the orientation of $B$   at the north pole and then parallel transport it to the coordinates $(\theta_B,\varphi_B)$
 using  Wigner-$d$ functions.

% \begin{figure*}[t]
%     \includegraphics[width=0.24\textwidth]{Figures/C1.pdf}
%     \includegraphics[width=0.24\textwidth]{Figures/C2.pdf}
%     \includegraphics[width=0.24\textwidth]{Figures/C3.pdf}
%     \includegraphics[width=0.24\textwidth]{Figures/C4.pdf}
%     \caption{%
%         The position of detector $A$ and various positions of detector $B$ for the  four types of configurations considered. When the detectors are on the sphere, these figures are interpreted as top-down view of the north pole.
%     }
%     \label{fig:Cases}%
% \end{figure*}

\subsubsection{Flat Spacetime}

 Considering first the situation in flat space, shown in the centre column of figure \ref{fig13BothSqueezed}, we plot the transition probability of detector $B$ as a function of its position.  For convenience, we will define the position of $B$ relative to $A$ in terms of the vector $\bd{S}=(S,\varphi)$, with magnitude, $S$, describing the distance between the centres of the spacial profiles of the two detectors and angle, $\varphi$, relative to the semi-major axis of detector  $A$.  The angle that the semi-major axis of detector $B$ makes with the vector $\bd{S}$ is determined by which of the four configurations is being considered.
 
 We find that in all four cases,  the field mediated signal from detector $A$ leads to an increase in the transition probability of detector $B$ as compared to the case when detector $A$ does not switch.
% \rbm{What does that mean? Increased relative to what?}\ljh{Answered}
 Since both detectors have a compact spacial extent, the signal from $A$ will increase the transition probability of $B$ for separations of their centroids that are greater than the time delay of their switching.  However the maximum increase occurs, for a given value of $\varphi$, when the separation between the detectors is slightly less then the time delay of their switching, {\it i.e.} when the centre of the signal from $A$ is slightly past the centre of $B$ at the time of switching.  
 %\sout{behind the light cone where at the point where the signal from $A$ overlaps detector $B$.} 
 %\rbm{How do we see these light cone effects in the figures?}
 The specific effect of the signal depends on  the choice of configuration. 
 
 %\ljh{"point-towards"=the axis lies along a line connecting the centres of the two detectors}
 
In cases 1, 3, and 4, we find that the transition probability of the second detector depends significantly on the angle, $\varphi$, between the line connecting the centres of the two detectors and the semi-major axis of  detector  $A$, with the maximum occurring for $\varphi=\pi/2$ and $\varphi=3\pi/2$.  From the left column of Fig.~\ref{fig13BothSqueezed}, it can be seen that these angles correspond to the configuration where the vector $S$ is coincident with the semi-minor axis of $A$, indicating that the signal from $A$ is strongest along this direction.

Additionally, we find dependence of the transition probably of $B$ on the angle the semi-major axis of $B$ makes with the vector $\bd{S}$, which is most straightforwardly demonstrated by comparing cases 3 and 4.  In case 3, the semi-major axis of detector $B$ lies along $\bd{S}$, while in case 4, the semi-major axis of $B$ is perpendicular to $\bd{S}$.  We find that when detector $B$ is placed at the optimal distance from $A$, the transition probability of $B$ is larger in case 4 than in case 3, which suggests that the detector is most sensitive to signals that propagate in a direction perpendicular its semi-major axis.

In case 2, the orientation of $B$ is fixed   with its semi-major axis perpendicular to the semi-major axis of $A$, as shown in row two of the left column of Fig.~\ref{fig13BothSqueezed}.  The effect of this configuration in flat space (row two of the centre column of Fig.~\ref{fig13BothSqueezed}) removes any dependence of the angle $\varphi$ from the transition probability of detector $B$; it only depends on the distance between the centers of the two detectors, $S$.  This case most clearly demonstrates that the transition probably of $B$ is greatest when the separation between the centres of the detectors is slightly less then the time delay between their switching.  It also shows that the two orientation effects, the strength of the signal from detector $A$ and the sensitivity of detector $B$, are of equal importance.  By fixing the semi-major axis of $B$ to remain perpendicular to the semi-major axis of $A$, it will be perpendicular, and hence the most sensitive, to the signal from $A$ when the vector $\bd{S}$ is coincident with the semi-major axis of $A$ ($\varphi=0$ and $\varphi=\pi$).  At these angles, the signal from $A$ is the weakest.  This fixed angle also ensures that when detector $B$ is located where the signal from $A$ is the strongest ($\varphi=\pi/2$ and $\varphi=3\pi/2$) its semi-major axis will be coincident with the vector $\bd{S}$ and be the least sensitive to the signal from $A$.  Since both cases (and all other values of $\varphi$) give the same transition probability for detector $B$, we can conclude that neither orientation effect is more important in increasing the transition probability of detector $B$.

We can interpret the relative strength of the signal and sensitivity of the detector to the relative length scale of the detector as follows.  Recall that squeezing increases the sensitivity of the detector to field modes that  have large momenta in the squeezing direction, so we expect detectors to be more sensitive to signals that propagate along the direction of their semi-minor axis.  Likewise we expect the signal sent by the squeezed detector $A$ to contain more field modes with high momenta in the direction of squeezing, namely the semi-minor axis. 

\subsubsection{Spherical Spacetime}

Next, we consider the case where the two detectors are on the surface of the sphere,  illustrating the results in the right column of Fig.~\ref{fig13BothSqueezed} for the four orientations illustrated in the left column of the same figure, where we now interpret the figures to be a top down view of the north pole of the sphere.  Detector $A$ is located its centre is at the north pole of the sphere with its semi-major axis aligned with $\varphi=0$; therefore, the two angles describing the relative orientation of the two detectors, $\theta$ and $\varphi$, are the azimuthal and polar angle respectively.

Overall, we find a similar relationship between the transition probability of detector $B$ to its relative position and orientation to detector $A$ as we did in flat space.  The transition probability of $B$ is greatest when the distance between the centres of the detectors, $\theta$, is slightly less then the time delay between their switching times, which can be seen most clearly in case 2.

Again, in cases 1, 3, and 4, we note that when the detector $B$ is located at the optimal distance from $A$, the transition probably of $B$ is maximized for $\varphi=\pi/2$ and $\varphi=3\pi/2$.  These orientations correspond to the case where the line connecting the centroids of the  two detectors is coincident with the semi-minor axis of $A$.  Since the maximum occurs at this position, independent of the orientation of $B$, we conclude the field mediated signal from $A$ is the strongest in this direction.

By comparing cases 3 and 4, we are able to isolate the dependence of the transition probability of $B$ on its orientation relative to detector $A$.  We find, analogous to the flat space case, that when $B$ is located at the optimal value of $\theta$, the transition probability is larger in case 4.  In this case, $B$ is oriented so its semi-major axis lies along a constant value of $\theta$, \textit{i.e.\ }it is perpendicular to the direction of propagation of the signal from $A$, which suggests that $B$ is most sensitive to signal that propagate in this direction.

Case 2 illustrates that unlike in flat space, orienting detector $B$ so that its semi-major axis is perpendicular to the semi-major axis of detector $A$ \textit{before} it is parallel transported to the given value of $(\theta,\varphi)$ does not remove the $\varphi$ dependence.  This suggests that the effect of the anisotropy on the signal from $A$ and the effect of orientation on the sensitivity of $B$ do not cancel each other out the way they do in flat space, likely due the curvature of the sphere distorting the shape of detector $B$.

%We again find that the strongest signal from detector $A$ is sent in the direction of its semi-minor axis, meaning that the transition probability of detector $B$ will be the strongest when it is positioned in line 
%\rbm{Clarify!}
%with the semi-minor axis of $A$, here located at $\varphi=\pi/2$ and $\varphi=3\pi/2$.  We also find that that detector $B$ has the largest transition probability, when its semi-minor axis points towards detector $A$.  However, unlike   the flat space case, the two effects are not equal, as shown in \ref{fig:BothSqueezedSphere}b;  this is likely due to the curvature of sphere distorting the shape of detector $B$.

In both cases, we find that the signal from $A$ that propagates in the direction perpendicular to its semi-major axis is the strongest and is weakest in the direction of propagation parallel to its semi-major axis.  Similarly, detector $B$ is most sensitive to the signal when it is orientation so that its semi-minor axis is parallel to the direction of signal propagation and is the least sensitive to signals that with a direction of propagation parallel to its semi-major axis.  We understand both of these effects as resulting from the effective bandlimit of the shape of the detector.  These detectors are smaller in the direction of their semi-minor axis, say the $y$-axis, so are able to interact with field modes of higher frequency in that direction, here $k_y$.  Therefore, detectors will be more easily excited by signals propagating in this direction, since more of the signal will interact with the detector, and detectors that signal through the field are able to access more modes in this direction, which will lead to stronger signals.

\section{Conclusions}
\label{conc}

We have exactly calculated the density matrix describing the state of two UDW detectors, each with a squeezed Gaussian smearing function, that each couple to a conformal scalar field via Dirac-delta switching.  We considered this on  a $2+1$-dimensional spherical spacetime and 
on $2+1$ Minkowski spacetime 
for comparison.  For both spacetimes we implemented a conventional bandlimit on the field by imposing a hard cutoff of the angular momentum modes in the former case and the linear momentum modes in the latter case.

For a single unsqueezed UDW detector, we found for both spacetimes that  the results were analogous to those obtained previously in
$3+1$ Minkowski space \cite{Henderson:2020ucx}. 
There is an optimal size detector for bandlimit detection -- small, but not too small.  We also found that in Minkowski space, when the smearing function is a squeezed Gaussian, and the overall size of the detector is larger than the optimal size, higher squeezing significantly increases  sensitivity to the bandlimit.  However if the detector is smaller than the optimal size, squeezing slightly decreases its sensitivity.

In considering two detectors, while a no-go theorem \cite{Simidzija:2017kty} prevents them from becoming entangled through this interaction, they can signal to each other. When the detectors are on a sphere, and the first detector, detector $A$, is located at the north pole, we found that the signal caused by $A$ interacting with the field is reflection symmetric about $t=\pi$, that is the signal at  $(t,\theta,\varphi)$ is the same at $(t-\pi,\theta,\varphi)$, and this property is inherited from the Greens function.

Since a sphere is compact, the signal from $A$ does not dissipate and has a much more significant effect on the transition probability of the second detector $B$, particularly when it is located at the south pole.  The specific effect depends heavily on the switching time and the size of the detector.  Additionally, by tuning the switching time and size of the second detector appropriately, it can become significantly more sensitive to the bandlimit than can a single detector alone. This is in notable contrast to flat spacetime:  since the signal can dissipate in Minkowski space, the transition probability of a second detector does not depend heavily on the signal from the first detector, and as a result cannot be used to significantly increase bandlimit detector capabilities.

Finally, we explored the effect of squeezing on the response of the second detector, and found 
%the optimal orientation is the same for both Minkowski space and spherical spacetime. \rbm{Can we say more here?}
that for both Minkowski and spherical spacetime, a Gaussian detector is most sensitive to a signal when it is orientated so that its semi-major axis is perpendicular to the direction of propagation. Similarly, the strongest signal from a Gaussian detector propagates in the direction perpendicular to its semi-major axis.  Both effects are due to the detectors having the smallest length scale along their semi-minor axis, which leads to field interactions accessing field modes with higher momenta in that direction.  This optimization could be exploited to maximize communication between a pair of detectors, such as quantum collect calling \cite{jonsson:2015prl}, or to minimize signal jamming\cite{Sahu:prd2022}.

Our results can be naturally extended to $\text{AdS}_3$, which can be constructed as a conformal transformation of the sphere, and further extended to the BTZ black hole spacetime.

Overall, we exactly calculated the signalling between two detectors on a compact curved spacetime, and found, as a consequence of the compactness of a sphere, an order of magnitude improvement on local bandlimit detection.  We conjecture that a similar effect may be observed in Minkowski space in a cavity with periodic boundary conditions, and may be harnessed in an experimental setup to better detect a UV cutoff of a field, or the shape of a qubit detector \cite{McKay:2017pra}.

%  Ideas for future work:
%  \begin{itemize}
%      \item Bandlimit  detection with (or without) delta switching with a qutrit, or qudit.  This should be able to ''hold'' more  information before becoming saturated to the maximally mixed state.  This could mean the the detector could be made to be more sensitive to at smaller sizes and more able to detect a higher bandlimit.
%      \item Conformal transformations to AdS$_3$ and BTZ blackholes, reference papers like: https://arxiv.org/pdf/1809.06862.pdf and https://arxiv.org/pdf/1712.10018.pdf 
%      \item Emma's paper related to superconducting circuits?
%  \end{itemize}
% \rbm{Need to clean up the concluding section. Do we have a strong punchline?}
% \ahs{The two main punchlines,
% First time Two UDW detectors were used non perturbatively on a curved spacetime. Second, the at least one order of magnitude improvement in bandlimit detection due to the focusing of the signal. And saying something about cavities with reflecting boundary conditions if this were to be done experimentally.}
%%%%%%%%%%%%%%%%%% Acknowledgments %%%%%%%%%%%%%%%%%%%%
\acknowledgments
This work was supported in part by the Natural Sciences and Engineering Research Council of Canada, and by the Asian Office of Aerospace Research and Development Grant No.\ FA2386-19-1-4077.

%%%%%%%%%%%%%%%%%  Appendices %%%%%%%%%%%%%%%%%%%%%%%%%
\newpage
\onecolumngrid
\appendix

\section*{Appendix A: Showing that $\Ket{\al_{\ell m}}$ is a coherent state}
\label{sec:AppA}

 In this section we will show that
\aln{
	\Ket{\al_{\ell m}} &= \hat{\mc{D}}_{\ell m}\Ket{0}  = \exp\lb(\sum_{\ell,m}\lb(\al_{\ell m}\oad_{\ell m}-\al_{\ell m}^*\oa_{\ell m}\rb)\rb)\Ket{0}
}
is a coherent state, i.e.\ that $\Ket{\al_{\ell m}}$ is an eigenstate of the annihilation operator $\oa_{ij}$.

Using the canonical commutation relation,
$\lb[\oa_{ij},\oad_{\ell m}\rb] = \de_{i\ell}\de_{jm}$, we can show that
\aln{
\oa_{ij} \lb(\sum_{\ell,m}\lb(\al_{\ell m}\oad_{\ell m}-\al_{\ell m}^*\oa_{\ell m}\rb)\rb)  
	%&\qquadd = \sum_{\ell,m} \bigg(\al_{\ell m}\lb(\lb[\oa_{ij},\oad_{\ell m}\rb]+\oad_{\ell m}\oa_{ij}\rb)-\al_{\ell m}^*\oa_{\ell m}\oa_{ij}\bigg) \nn\\
	& = \al_{ij} + \lb(\sum_{\ell,m} \lb(\al_{\ell m}\oad_{\ell m} - \al_{\ell m}^*\oa_{\ell m}\rb)\rb)\oa_{ij}
	\label{eq:CoherPart1}
}
yielding
\aln{
	\lb[\oa_{ij},\sum_{\ell,m}\lb(\al_{\ell m}\oad_{\ell m}-\al_{\ell m}^*\oa_{\ell m}\rb)\rb]  
	 = \al_{ij} \in \bb{C}
	\label{eq:CoherPart2}
}
from which we find
\aln{
	[\oa_{ij},\hat{\mc{D}}_{\ell m}] = \lb[\oa_{ij},\sum_{\ell,m}\lb(\al_{\ell m}\oad_{\ell m}-\al_{\ell m}^*\oa_{\ell m}\rb)\rb]\exp\lb(\sum_{\ell,m}\lb(\al_{\ell m}\oad_{\ell m}-\al_{\ell m}^*\oa_{\ell m}\rb)\rb) = \al_{ij} \hat{\mc{D}}_{\ell m}\; .
}
Hence
\aln{
	\oa_{ij}\Ket{\al_{\ell m}} =  \oa_{ij}\hat{\mc{D}}_{\ell m}\Ket{0}  
	= [\oa_{ij},\hat{\mc{D}}_{\ell m}]\Ket{0} + \hat{D}_{\ell m}\oa_{ij}\Ket{0}  
= \al_{ij}\Ket{\al_{\ell m}} 
}
showing that  $\Ket{\al_{\ell m}}$ is an eigenstate of the annihilation operator $\oa_{ij}$ and so is a coherent state.

\section*{Appendix B: Derivation of the spherical harmonics coefficients $g_{\ell m}$ for a Gaussian detector on $S^2 \times R$}
\label{sec:AppB}
The spherical harmonics coefficients $g_{\ell m}$ for the symmetric FB distribution   \eqref{FBSymmetric} are given by 
\aln{
g_{\ell m} &= \Braket{g(\kappa,\beta,\theta),Y_{\ell}^m} = \int_{0}^{\pi}\int_{0}^{2 \pi} \frac{e^{\kappa\cos(\theta)}}{C(\kappa,0)} {Y_{\ell}^m}^* \sin(\theta) d \theta d \varphi  =\frac{N_{\ell m}}{C(\kappa, 0)} \int_{0}^{\pi} e^{\kappa \cos \theta} P_{\ell}^{m}(\cos \theta) \sin \theta d \theta  \nn \\
&\int_{0}^{2 \pi} e^{-i m \varphi} d \varphi =2 \pi \delta_{m, 0} \frac{N_{\ell 0}}{C(\kappa, 0)} \int_{0}^{\pi} e^{\kappa \cos \theta} P_{\ell}^{0}(\cos \theta) \sin \theta d \theta =\delta_{m, 0} \sqrt{\frac{2 \ell+1}{4 \pi}} \frac{2 \pi}{C(\kappa, 0)} \int_{0}^{\pi} e^{\kappa \cos \theta} P_{\ell}^{0}(\cos \theta) \sin \theta d \theta}
and using \cite{alem2015spherical}
\eqn{e^{\kappa \cos \theta}=\sqrt{\frac{\pi}{2 \kappa}} \sum_{n=0}^{\infty}(2 n+1) I_{2 n+1 / 2}(\kappa) P_{n}^{0}(\cos \theta)}
we can rewrite the $\theta$ integral as
\aln{
g_{\ell m} &=\sqrt{\frac{2 \ell+1}{4 \pi}} \frac{2 \pi \delta_{m, 0}}{C(\kappa, 0)} \sqrt{\frac{\pi}{2 \kappa}} \sum_{n=0}^{\infty}(2 n+1) I_{n+1 / 2}(\kappa)  \int_{0}^{\pi} P_{n}^{0}(\cos \theta) P_{\ell}^{0}(\cos \theta) \sin \theta d \theta \nonumber \\
&=\sqrt{\frac{2 \ell+1}{4 \pi}} \frac{2 \pi \delta_{m, 0}}{C(\kappa, 0)} \sqrt{\frac{\pi}{2 \kappa}} \sum_{n=0}^{\infty}(2 n+1) I_{n+1 / 2}(\kappa) \frac{2}{2 n+1} \delta_{\ell, n}  =\sqrt{\frac{2 \ell+1}{4 \pi}} \frac{4 \pi \delta_{m, 0}}{C(\kappa, 0)} \sqrt{\frac{\pi}{2 \kappa}} I_{\ell+1 / 2}(\kappa) \nonumber \\
&=\delta_{m, 0} \sqrt{\frac{2 \ell+1}{4 \pi}} \frac{4 \pi \sqrt{\kappa}}{2 \pi \sqrt{2 \pi} I_{1 / 2}(\kappa)} \sqrt{\frac{\pi}{2 \kappa}} I_{\ell+1 / 2}(\kappa)  
\equiv \delta_{m, 0} \sqrt{\frac{2 \ell+1}{4 \pi}} \frac{I_{\ell+1 / 2}(\kappa)}{I_{1 / 2}(\kappa)}
\nn \\
&= \delta_{m,0} g_{\ell}}
For the squeezed FB distribution (equation \ref{FB dist}), the spherical harmonics coefficients are given by

\aln{
g_{\ell m} &= \int_{0}^{\pi}\int_{0}^{2 \pi} \frac{e^{\kappa \cos \theta+\beta \sin ^{2} \theta \cos 2 \varphi}}{C(\kappa,\beta)} {Y_{\ell}^m}^* \sin(\theta) d \theta d \varphi \nn \\
&=\frac{N_{\ell m}}{C(\kappa, \beta)} \int_{0}^{\pi} e^{\kappa \cos \theta} P_{\ell}^{m}(\cos \theta) \sin \theta d \theta \times \int_{0}^{2 \pi} e^{\beta \sin ^{2} \theta \cos (2 \varphi)-i m \varphi} d \varphi \nn \\
&=\frac{2 \pi N_{\ell m}}{C(\kappa, \beta)} \int_{0}^{\pi} e^{\kappa \cos \theta} P_{\ell}^{m}(\cos \theta) \sin \theta d \theta\left( I_{m / 2}\left(\beta \sin ^{2} \theta\right)\right)}

Now let $\cos{(\theta)} = z$, which allows us to rewrite the integral above as
\eqn{
g_{\ell m}= 2 \pi \frac{N_{\ell m}}{C(\kappa, \beta)} \int_{-1}^{1} e^{\kappa z} P_{\ell}^{m}(z) I_{m / 2}\left(\beta \left( 1 - z^2 \right) \right)dz}

\section*{Appendix C: The position independence of the non-signaling part of the transition probability.}
\label{sec:AppC}
 In this appendix  we provide details of the simplification of the non-signaling part of the transition probability.

The transition probabilities of the two detectors are
\aln{
	P_A &= \frac{1}{2}\lb[1-\exp\lb(-\al_A\rb)\rb] \\
	P_B &= \frac{1}{2}\lb[1-\exp\lb(-\al_B\rb)\cos(2\Tht)\rb]
}
where $\Tht$ is the field commutator between detectors $A$ and $B$ defined in Eq.~\eqref{eq:SphereCommNoLocation}.
We refer to $\al_D$ $(D\in\{A,B\})$ as the non-signalling part of the transition probability because it does not depend on the the properties of the other detector.

In general
\eqn{
	\al_{D} = \sum_{\ell}\sum_{m=-\ell}^{\ell} \Abs{\al_{\ell,m}^D}^2
	\label{eq:alpha}
}
where
\eqn{
	\al_{\ell,m}^D = -\frac{\iu\la_D\et_D}{\sqrt{2\ell+1}}\ \ec^{\iu(\ell+1/2)T_D} f_{\ell,m}^D
	\label{eq:alphalm}
}
and $f_{\ell,m}^D$ are the spherical harmonic coefficients of the spacial profile of the detector $D$.  Recall that these coefficients include the information about the location of the detector. Let $g_{\ell,m}^D$ be the spherical harmonic coefficients of detector $D$ if it were located at the north pole and oriented along the $\vph=0$ axis, so that
\eqn{
	f_{\ell,m}^D = \sum_{n=-\ell}^{\ell} D_{m,n}^\ell(\al,\be,\ga)g_{\ell,n}^D
	\label{eq:fellmApC}
}
where $D^{\ell}_{m,n}(\alpha,\beta,\gamma)$ are the Wigner-$d$ functions,  Now all the information about the position and orientation of the detector is encoded in the Euler angles $\al$, $\be$, and $\ga$.

Substituting Eq.~\eqref{eq:alphalm} and Eq.~\eqref{eq:fellmApC} into the non-signalling part [Eq.~\eqref{eq:alpha}] yields:
\aln{
	\al_{D} &= \sum_{\ell}\sum_{m=-\ell}^{\ell} \Abs{-\frac{\iu\la_D\et_D}{\sqrt{2\ell+1}}\ \ec^{\iu(\ell+1/2)T_D} \sum_{n=-\ell}^{\ell} D_{m,n}^{\ell}(\al,\be,\ga)g_{\ell,n}^D}^2 \nn\\
	&= \la_D^2\et_D^2 \sum_{\ell} \frac{1}{2\ell+1} \sum_{m=-\ell}^{\ell} \Abs{\sum_{n=-\ell}^{\ell} D_{m,n}^{\ell}(\al,\be,\ga)g_{\ell,n}^D}^2 \nn\\
	%
	%&= \la_D^2\et_D^2 \sum_{\ell} \frac{1}{2(\ell+1/2)} \sum_{m=-\ell}^{\ell} \lb(\sum_{n=-\ell}^{\ell} D_{m,n}^{\ell}(\al,\be,\ga)g_{\ell,n}^D\rb)\lb(\sum_{n'=-\ell}^{\ell} {D_{m,n'}^\ell}^*(\al,\be,\ga) {g_{\ell,n'}^D}^*\rb) \nn\\
	%
	&= \la_D^2\et_D^2 \sum_{\ell} \frac{1}{2\ell+1} \sum_{n=-\ell}^{\ell} \sum_{n'=-\ell}^{\ell} g_{\ell,n}^D {g_{\ell,n'}^D}^* \lb(\sum_{m=-\ell}^{\ell} D_{m,n}^\ell(\al,\be,\ga) {D_{m,n'}^\ell}^*(\al,\be,\ga)\rb).
	\label{eq:alphalm2}
}

The Wigner-$d$ functions obey the orthogonality relations  \cite{pagaran_fritzsche_gaigalas_2006}
\gthr{
	\sum_{m=-\ell}^{\ell} D_{m,n}^\ell(\al,\be,\ga) {D_{m,n'}^{\ell *}}(\al,\be,\ga) = \de_{n,n'} \label{eq:OR1}\\
	\sum_{n=-\ell}^{\ell} {D_{m,n}^{\ell *}}(\al,\be,\ga) D_{m',n}^\ell(\al,\be,\ga) = \de_{m,m'} \label{eq:OR2},
} which we will now use to simplify Eq.~\eqref{eq:alphalm2}:
\aln{
	\al_D &= \la_D^2\et_D^2 \sum_{\ell} \frac{1}{2\ell+1} \sum_{n=-\ell}^{\ell} \sum_{n'=-\ell}^{\ell} g_{\ell,n}^D {g_{\ell,n'}^{D *}} \lb(\de_{n,n'}\rb) %\nn\\
	%
	%&= \la_D^2\et_D^2 \sum_{\ell} \frac{1}{2(\ell+1/2)} \sum_{n=-\ell}^{\ell} g_{\ell,n}^D {g_{\ell,n}^D}^* \nn\\
	%
	%&= \la_D^2\et_D^2 \sum_{\ell} \frac{1}{2(\ell+1/2)} \sum_{n=-\ell}^{\ell} g_{\ell,n}^D {g_{\ell,n}^D}^* \nn\\
	%
	= \la_D^2\et_D^2 \sum_{\ell} \frac{1}{2\ell+1} \sum_{n=-\ell}^{\ell} \Abs{g_{\ell,n}^D}^2
	\label{eq:genalpha}
}
which does not depend the position and orientation of the detector.

Additionally, in the case of identical detectors ($\la_A=\la_B$, $\et_A=\et_B$ and $g_{\ell,m}^A=g_{\ell,m}^B$), then
\eqn{
	\al_A = \la_D^2\et_D^2 \sum_{\ell} \frac{1}{2\ell+1} \sum_{m=-\ell}^{\ell} \Abs{g_{\ell,m}^A}^2 = \la_B^2\et_B^2 \sum_{\ell} \frac{1}{2\ell+1} \sum_{m=-\ell}^{\ell} \Abs{g_{\ell,m}^B}^2 = \al_B.
}

\section*{Appendix D: The smeared field commutator in the case of unsqueezed detectors.}
\label{sec:AppD}
{In this section we provide details of the simplification of the smeared field commutator, Eq.~\eqref{eq:SphereCommNoLocation} in the case of regular detectors.}

 If a detector is centred at the north pole of the sphere, and its smearing function is regular, i.e.\ it only depends on the angle $\theta$, then its spherical harmonics coefficients can be written as
\eqn{
    g_{\ell,m} = \delta_{m,0}g_{\ell}.
}
 If a regular detector is moved away from the north pole, then its spherical harmonics coefficients can now be written in terms of Wigner-$d$ functions
\aln{
    f_{\ell,m} = \sum_{n=-\ell}^{\ell} D_{m,n}^\ell(\al,\be,\ga) \lb(\delta_{n,0}g_\ell\rb) = D^{\ell}_{m,0}(\alpha,\beta,\gamma)g_{\ell}.
}

 Now consider two regular detectors, one centred at the north pole   and the second centred at $(\theta,\varphi)=(\theta_B,\varphi_B)$.  The smeared field commutator [Eq.~\eqref{eq:SphereCommNoLocation} between can be simplified as:
\begin{align}
\left[\hat{\mathcal{Y}}_{A}, \hat{\mathcal{Y}}_{B}\right] &= -\lambda_{A} \lambda_{B} \eta_{A} \eta_{B} \sum_{\ell, m} \frac{1}{2\ell+1 }\left[\left(f_{\ell}^{A} \delta_{m, 0}\right)^{*}\left(D_{m, 0}^{\ell}\left(\varphi_{B}, \theta_{B}, \gamma\right) g_{\ell}^{B}\right) \mathrm{e}^{\mathrm{i}(\ell+1 / 2)\left(T_{B}-T_{A}\right)}\right. \nn\\
&\qquad\qquad\qquad
\qquad\qquad\qquad \left.-\left(f_{\ell}^{A} \delta_{m, 0}\right)\left(D_{m, 0}^{\ell}\left(\varphi_{B}, \theta_{B}, \gamma\right) g_{\ell}^{B}\right)^{*} \mathrm{e}^{-\mathrm{i}(\ell+1 / 2)\left(T_{B}-T_{A}\right)}\right] \nn\\
&=-\lambda_{A} \lambda_{B} \eta_{A} \eta_{B} \sum_{\ell=0}^{\infty} \frac{1}{2\ell+1}\left[f_{\ell}^{A^{*}} g_{\ell}^{B} D_{0,0}^{\ell}\left(\varphi_{B}, \theta_{B}, \gamma\right) \mathrm{e}^{\mathrm{i}(\ell+1 / 2)\left(T_{B}-T_{A}\right)}\right. \nn\\
&\qquad\qquad\qquad
\qquad\qquad\qquad \left.-f_{\ell}^{A} g_{\ell}^{B^{*}} D_{0,0}^{\ell}{}^{*}\left(\varphi_{B}, \theta_{B}, \gamma\right) \mathrm{e}^{-\mathrm{i}(\ell+1 / 2)\left(T_{B}-T_{A}\right)}\right] \nn\\
&=-\lambda_{A} \lambda_{B} \eta_{A} \eta_{B} \sum_{\ell=0}^{\infty} \frac{1}{2\ell+1}\left[f_{\ell}^{A^{*}} g_{\ell}^{B}\left(\sqrt{\frac{4 \pi}{2 \ell+1}} Y_{\ell, 0}^{*}\left(\theta_{D}, \varphi_{D}\right)\right) \mathrm{e}^{\mathrm{i}(\ell+1 / 2)\left(T_{B}-T_{A}\right)}\right. \nn\\
&\qquad\qquad\qquad
\qquad\qquad\qquad \left.-f_{\ell}^{A} g_{\ell}^{B^{*}}\left(\sqrt{\frac{4 \pi}{2 \ell+1}} Y_{\ell, 0}\left(\theta_{D}, \varphi_{D}\right)\right) \mathrm{e}^{-\mathrm{i}(\ell+1 / 2)\left(T_{B}-T_{A}\right)}\right] \nn\\
&=-\lambda_{A} \lambda_{B} \eta_{A} \eta_{B} \sum_{\ell=0}^{\infty} \frac{1}{2\ell+1}\left[f_{\ell}^{A^{*}} g_{\ell}^{B}\left(\sqrt{\frac{4 \pi}{2 \ell+1}} \sqrt{\frac{2 \ell+1}{4 \pi}} P_{\ell}\left[\cos \left(\theta_{B}\right)\right]\right) \mathrm{e}^{\mathrm{i}(\ell+1 / 2)\left(T_{B}-T_{A}\right)}\right. \nn\\
&\qquad\qquad\qquad
\qquad\qquad\qquad \left.-f_{\ell}^{A} g_{\ell}^{B^{*}}\left(\sqrt{\frac{4 \pi}{2 \ell+1}} \sqrt{\frac{2 \ell+1}{4 \pi}} P_{\ell}\left[\cos \left(\theta_{B}\right)\right]\right) \mathrm{e}^{-\mathrm{i}(\ell+1 / 2)\left(T_{B}-T_{A}\right)}\right] \nn\\
&=-\lambda_{A} \lambda_{B} \eta_{A} \eta_{B} \sum_{\ell=0}^{\infty} \frac{1}{2\ell+1}\left[f_{\ell}^{A^{*}} g_{\ell}^{B} P_{\ell}\left[\cos \left(\theta_{B}\right)\right] \mathrm{e}^{\mathrm{i}(\ell+1 / 2)\left(T_{B}-T_{A}\right)}\right. \nn\\
&\qquad\qquad\qquad
\qquad\qquad\qquad \left.-f_{\ell}^{A} g_{\ell}^{B^{*}} P_{\ell}\left[\cos \left(\theta_{B}\right)\right] \mathrm{e}^{-\mathrm{i}(\ell+1 / 2)\left(T_{B}-T_{A}\right)}\right].
\end{align}

\twocolumngrid

%%%%%%%%%%%%%%%%% References  %%%%%%%%%%%%%%%%% 
\bibliographystyle{unsrt}
\bibliography{references2} 

\end{document}